\documentclass[aps,prd,twocolumn,groupedaddress,nofootinbib]{revtex4-1}
\usepackage{graphicx}
\usepackage{here}
\usepackage[dvips]{epsfig}
\def\beq{\begin{equation}}
\def\eeq{\end{equation}}
\def\bea{\begin{eqnarray}}
\def\eea{\end{eqnarray}}
\def\bq{\begin{quote}}
\def\eq{\end{quote}}

\def\nnb{\nonumber}
\def\ga{\left(}
\def\dr{\right)}

\def\rar{\rightarrow}

\def\nnb{\nonumber}

\def\la{\langle}
\def\ra{\rangle}
\def\nin{\noindent}
\def\ba{\vspace*{-0.2cm}\begin{array}}
\def\ea{\end{array}\vspace*{-0.2cm}}

\def\als{\alpha_s}

\def\gg2{\la\alpha_s G^2 \ra}
\def\gg3{g^3f_{abc}\la G^aG^bG^c \ra}
\def\ggg4{\la\als^2G^4\ra}

\def\qq{\la\bar{q}q\ra}

\def\lv{\mathcal{L}_v}
\def\lp{\mathcal{L}_+}
\def\lid{\mbox{Li}_2}
\def\ln{\mbox{Log}}
\def\gg{\lag g^{2}_{s} G^2 \rag}
\def\ggg{\lag g^{3}_{s}G^3\rag}

\usepackage{amsmath}
\usepackage{slashed}
\usepackage{color}

\begin{document}

\title{$Z_c$
-like spectra  from QCD Laplace sum rules at NLO}
\author{R.M. Albuquerque}
\affiliation{Faculty of Technology, Rio de Janeiro State University (FAT,UERJ), Brazil}
\email[Email address:~] {raphael.albuquerque@uerj.br}
\author{S. Narison
}
\altaffiliation{ICTP-Trieste consultant for Madagascar - Corresponding author.}
\affiliation{Laboratoire Univers et Particules de Montpellier (LUPM) \\
CNRS-IN2P3 and University of Montpellier  
Case 070, Place Eug\`ene
Bataillon, 34095 - Montpellier, France\\
and\\
Institute of High-Energy Physics of Madagascar (iHEPMAD)\\
University of Ankatso,
Antananarivo 101, Madagascar}
\email[Email address:~] {snarison@yahoo.fr} 

\author{D. Rabetiarivony}
\email[Email address:~] {rd.bidds@gmail.com}

\affiliation{Institute of High-Energy Physics of Madagascar (iHEPMAD)\\
University of Ankatso,
Antananarivo 101, Madagascar}

\date{\today}
\begin{abstract}
\noindent
We present a global analysis of the observed $Z_c,~Z_{cs}$ and future $Z_{css}$-like spectra  using  (inverse) Laplace Sum Rule (LSR) within stability criteria. Integrated compact QCD expressions of the LO spectral functions  up to dimension-six condensates are given.  Next-to-Leading Order (NLO) factorized perturbative contributions are included.  
We re-emphasize the importance to include PT radiative corrections (though numerically small) for heavy quark sum rules in order to justify the (ad hoc) definition and value of the heavy quark mass used frequently  at LO in the literature. We also demonstrate that, contrary to a qualitative large $N_c$-counting,  the two-meson scattering contributions to the four-quark spectral functions are numerically negligible confirming the reliability of the LSR predictions. 
Our results are summarized in Tables\,\ref{tab:res} to \ref{tab:lsr-rad}. The $Z_{c}(3900)$ and $Z_{cs}(3983)$ spectra are well reproduced  by the ${\cal T}_c(3900)$ and ${\cal T}_{cs}(3973)$ {\it tetramoles} (superposition of quasi-degenerated molecules and tetraquark states having the same quantum numbers and with almost equal couplings to the currents). The $Z_{c}$(4025) or $Z_{c}$(4040) state can be fitted with the $D^*_0D_1$ molecule having a mass  4023(130) MeV while the $Z_{cs}$ bump around 4.1 GeV can be likely due to the $D^*_{s0}D_1\oplus D^*_{0}D_{s1}$ molecules. The $Z_c(4430)$ could be a radial excitation of the $Z_{c}(3900)$ weakly coupled to the current, while all strongly coupled ones are in the region  $(5634\sim 6527)$ MeV. The double strange {\it tetramole} state ${\cal T}_{css}$ which one may identify with the future $Z_{css}$ is predicted to be at 4064(46) MeV. It is remarkable to notice the regular mass-spliitings of the tetramoles due to $SU(3)$ breakings\,:  $ M_{{\cal T}_{cs}}-M_{{\cal T}_c}\approx M_{{\cal T}_{css}}-M_{{\cal T}_{cs}}\simeq (73\sim 91)$\,MeV. 
\end{abstract}
\maketitle
 \section{Introduction}
 Beyond the successful quark model of Gell-Mann\,\cite{GELL}  and Zweig\,\cite{ZWEIG}, Rossi and Veneziano have introduced the four-quark states within the string model\,\cite{ROSSI} in order to describe baryon-antibaryon scattering, while Jaffe\,\cite{JAFFE1} has introduced them  within the bag models for an attempt to explain the complex structure of the $I=1,0$ light scalar mesons (see also\,\cite{ISGUR,ACHASOV,THOOFT}).
 
In earlier papers, QSSR has been used to estimate  the $I=0$ light scalar mesons ($\sigma, f_0,$) masses and widths\,\cite{LATORRE} assumed to be four-quark states.  However, the true nature of these states remains still an open question as they can be well interpreted as glueballs/gluonia\,\cite{VENEZIA,SNG,OCHS,MENES3}. 

After the recent discovery of many exotic XYZ states beyond the quark model found in different accelerator experiments\,\footnote{For a recent review, see e.g.\,\cite{WU}.}, there was a renewed interest on the four-quarks and molecule states for attempting to explain the properties of these new exotic states\,\footnote{For reviews, see e.g. \,\cite{MOLEREV,MAIANI,RICHARD,ROSSI,SWANSON,DOSCH2,ZHU,QIANG,BRAMBILLA}.}. 

 In previous works\,\cite{HEP18,SU3,MOLE16}, we have systematically extracted the couplings and masses of the $XYZ$ states using QCD spectral sum rules (QSSR) \`a la Shifman-Vainshtein-Zakharov (SVZ)\,\cite{SVZa,ZAKA}\,\footnote{For reviews, see e.g\,\cite{SNB1,SNB2,SNB3,IOFFEb,RRY,DERAF,BERTa,YNDB,PASC,DOSCH,COL}.} where the N2LO factorized perturbative and QCD condensates up to dimension-six in the Operator Product Expansion (OPE) corrections have been included\,\footnote{For a recent review on the uses of QSSR for exotic hadrons, see e.g\,\cite{MOLEREV} where different LO results (LO) are quoted.}. In so doing, we have used the inverse Laplace transform (LSR)\,\cite{BELLa,BECCHI,SNR} version of  QSSR within stability criteria where we have emphasized the importance of the PT corrections for giving a meaning on the input heavy quark mass value which plays an important role in the analysis, though these corrections are numerically small, within the $\overline{MS}$-scheme. 
 
 More recently, we have applied the LSR for interpreting the new states around (6.2-6.9) GeV found by the LHCb-group\,\cite{LHCb1} to be a doubly/fully hidden-charm molecules $(\bar QQ) (Q\bar Q)$ and $( \bar Q \bar Q)(QQ)$ tetraquarks states\,\cite{4Q}, while the new states found by the same group from the $DK$ invariant mass\,\cite{LHCb3} have been interpreted by a $0^+$ and $1^-$ tetramoles (superposition of almost degenerate molecules and tetraquark states having the same quantum numbers and couplings) slightly mixed with their radial excitations\,\cite{DK}. 
\begin{figure}[hbt]
\vspace*{-0.25cm}
\begin{center}
\includegraphics[width=6.5cm]{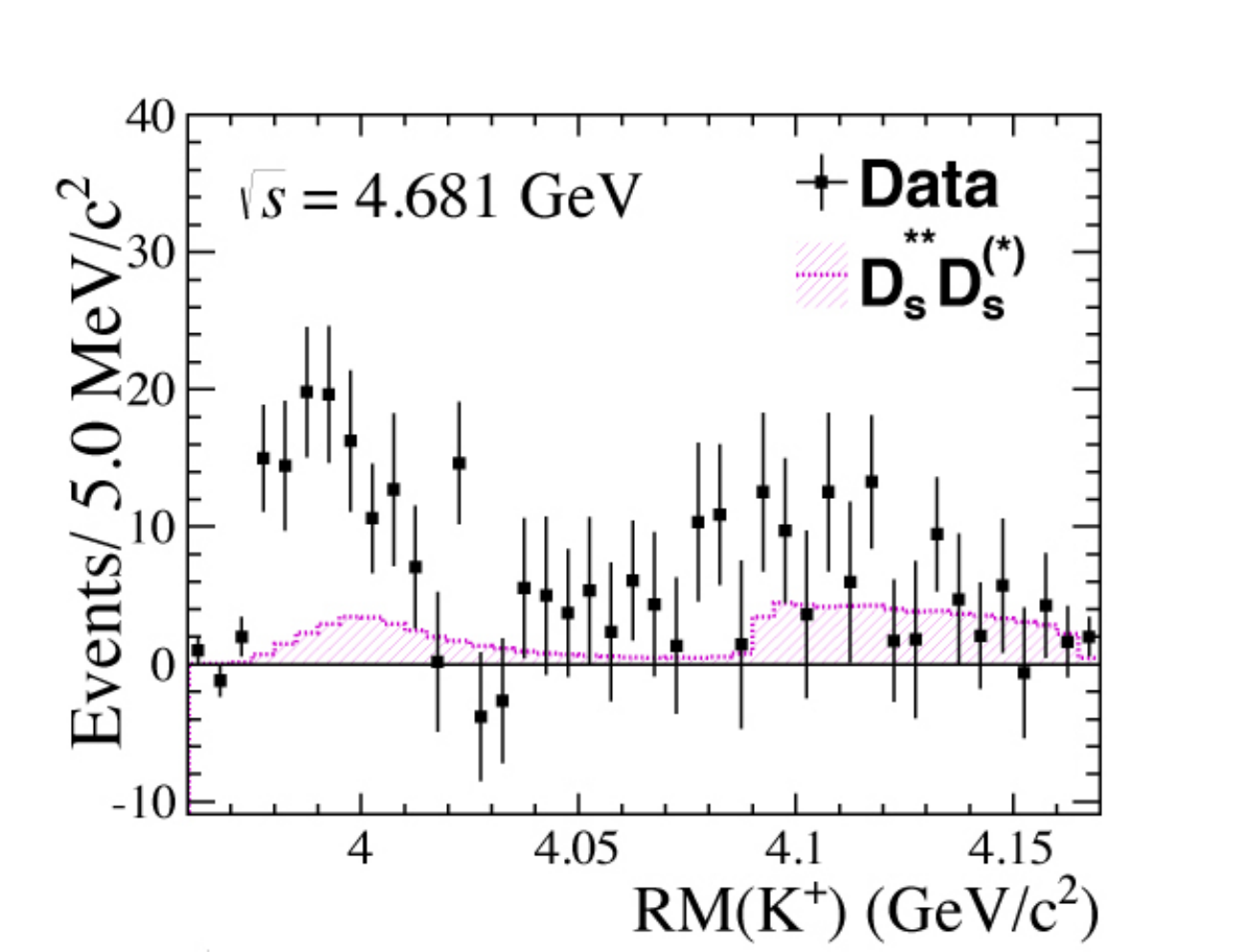}
\vspace*{-0.25cm}
\caption{\footnotesize  $K^+$ recoil-mass at $\sqrt {s}=4.681$ GeV after background subtractions from BESIII\,\cite{BES}.} 
\label{fig:bes}
\end{center}
\vspace*{-0.5cm}
\end{figure}
 
In this paper, we pursue the analysis using LSR by studying the recent data from BESIII  where the $K^+$ recoil or invariant $D^*_sD \oplus D^*D_s$ mass (see Fig.\,\ref{fig:bes})\,\cite{BES} is a good $Z_{cs}(1^+)$-like state candidate. A narrow peak is experimentally found at (in units of MeV):
\beq
M=(3982.5^{+1.8}_{-2.6}\pm 2.1), ~~\Gamma = (12.8^{+5.3}_{-4.4}\pm 3.0).
\eeq
\section{The Inverse Laplace sum rules}
\subsection{The QCD molecule and tetraquarks currents}
    We shall be concerned with the QCD local  currents ${\cal O}^\mu_H(x)$ of dimension-six given in Table\,\ref{tab:current} for $1^+$ axial-vector molecules and tetraquarks where $q\equiv u,s$. 
$b$ is a free mixing parameter where its optimal value was found to be zero\,\cite{MOLE16,SU3}. 
\begin{table}[hbt]
{\scriptsize
\setlength{\tabcolsep}{1.7pc}
    {
    \begin{center}
  \begin{tabular}{ll}
\hline
\hline
Molecules &Currents  \\
\hline
$D^*_qD$ & $(\bar c\gamma_\mu q)(\bar u\,i\gamma_5c)$  \\
$D^*D_q$ & $(\bar u\gamma_\mu c)(\bar c\,i\gamma_5q)$  \\
$D^*_{q0}D_1$ & $(\bar c q)(\bar u\gamma_\mu\gamma_5c)$  \\
$D^*_{0}D_{q1}$ & $(\bar uc)(\bar c\gamma_\mu\gamma_5q)$  \\
$D^*_sD_s$ & $(\bar c\gamma_\mu s)(\bar s\,i\gamma_5c)$  \\
$D^*_{s0}D_{1s}$ & $(\bar c s)(\bar s\gamma_\mu\gamma_5c)$  \\
\\
\hline\hline
Tetraquarks &  Currents \\
\hline
$A_{cq}$ &$\epsilon_{ijk}\epsilon_{mnk}\big{
[}(q^T_i\,C\gamma_5\,c_j)(\bar u_m \gamma_\mu C\, \bar c_n^T)$\\
&$\,+b\,(q^T_i\,C\,c_j)(\bar u_m\gamma_\mu \gamma_5 C\, 
\bar c_n^T)\big{]}$\\
$A_{css}$ &$\epsilon_{ijk}\epsilon_{mnk}\big{
[}(s^T_i\,C\gamma_5\,c_j)(\bar s_m\gamma_\mu C\, \bar c_n^T)$\\
&$\,+b\,(s^T_i\,C\,c_j)(\bar s_m\gamma_\mu\gamma_5 C\, 
\bar c_n^T)\big{]}$\\

\hline\hline
\end{tabular}
\end{center}
\vspace*{-0.5cm}
{\scriptsize
 \caption{\label{tab:current}$(1^+)$ molecules and tetraquarks currents ($q\equiv u,s$).}  }
} 
}
\end{table}
The appropriate $1^+$ hadron ${\cal H}$ couples to the current as :
\beq
\hspace*{-0.6cm}\la 0|{\cal O}^\mu_{\cal H}(x)\vert {\cal H}\ra= f_{\cal H}M_{\cal H}^5\epsilon^\mu
\eeq
where  $f_{\cal H}$ is the hadron decay constant analogue to $f_\pi$ and $\epsilon^\mu$ is the  axial-vector polarisation. 
In general, the four-quark operators mix under renormalization and acquire anomalous dimensions\,\cite{SNTARRACH}. In the present case where the interpolating currents are
constructed from bilinear (pseudo)scalar currents,  the anomalous dimension can be transfered to  the decay constants as\,:
\beq
f_{\cal H}(\mu)=\hat f_{\cal H} \ga -\beta_1a_s\dr^{2/\beta_1}(1-k_f\,a_s) ,
\eeq
where : $\hat f_{\cal H}$ is the renormalization group invariant coupling and $-\beta_1=(1/2)(11-2n_f/3)$ is the first coefficient of the QCD $\beta$-function for $n_f$ flavours. $a_s\equiv (\alpha_s/\pi)$ is the QCD coupling. $k_f=1.014 $ for $n_f=4$ flavours. 
\subsection{Form of the sum rules}
We shall work with the  Finite Energy version of the QCD Inverse Laplace sum rules (LSR) and their ratios :
\bea
 {\cal L}^c_n(\tau,\mu)&=&\int_{t_0}^{t_c}\hspace*{-0cm}dt~t^n~e^{-t\tau}\frac{1}{\pi} \mbox{Im}~\Pi^{(1)}_{\cal H}(t,\mu)~,\nnb\\
 {\cal R}^c_n(\tau)&=&\frac{{\cal L}^c_{n+1}} {{\cal L}^c_n},
\label{eq:lsr}
\eea
 where $m_c$ is the charm quark mass, $\tau$ is the LSR variable, $n=0,...$ is the degree of moments, $t_0$ is the quark/hadronic threshold. $t_c$ is the threshold of the ``QCD continuum" which parametrizes, from the discontinuity of the Feynman diagrams, the spectral function  ${\rm Im}\,\Pi^{(1)}_{\cal H}(t,m_c^2,\mu^2)$   where  $\Pi^{(1)}_{\cal H}(t,m_c^2,\mu^2)$ is the  transverse scalar correlator corresponding to a spin one hadron\,: 
 \bea
\hspace*{-0.6cm} \Pi^{\mu\nu}_{\cal H}(q^2)&=&i\int \hspace*{-0.15cm}d^4x ~e^{-iqx}\la 0\vert {\cal T} {\cal O}^\mu_{\cal H}(x)\ga {\cal O}^\nu_{\cal H}(0)\dr^\dagger \vert 0\ra \nnb\\
&\equiv& -\ga g^{\mu\nu}-\frac{q^\mu q^\nu}{q^2}\dr\Pi^{(1)}_{\cal H}(q^2)+\frac{q^\mu q^\nu}{q^2} \Pi^{(0)}_{\cal H}(q^2)
 \label{eq:2-pseudo}
 \eea
\section{QCD two-point function\label{sec:twopoint}}
Using the SVZ\,\cite{SVZa} Operator Product Expansion (OPE), we give in the Appendix the QCD expressions to lowest order (LO) of the two-point correlators associated to the currents given in Table\,\ref{tab:current} up to dimension $d=6$ condensate contributions. 
 \subsection{LO perturbative (PT) and $1/N_c$ counting}
Motivated by the criticisms raised in Ref.\,\cite{LUCHA0} based on the large $N_c$-limit 
which state that the non-factorized contribution of the two-point correlator starts at order $\alpha_s^2$ and that the non-resonant (scattering states) dominate the sum rules, we check explicitly these statements  for finite $N_c$here and in the following subsection C, which we complete in Section X by comparing the $D^*D$ molecule resonance and the non-resonating $D^*$ and $D$ channels contributions to the sum rule.
 
\begin{figure}[hbt]
\begin{center}
\centerline {\hspace*{-6cm} \bf a) }
\includegraphics[width=3.cm]{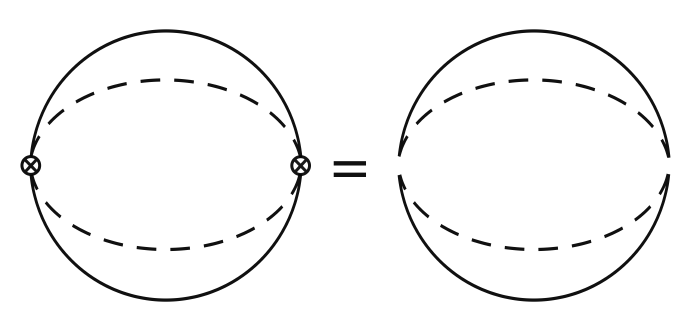}
\centerline {\hspace*{1.5cm} \hspace*{1.5cm } $N_c^2$ {or} \bf $N_c^2(1-1/N_c)$}
\centerline {\hspace*{-6cm} \bf b) }
\includegraphics[width=4.cm]{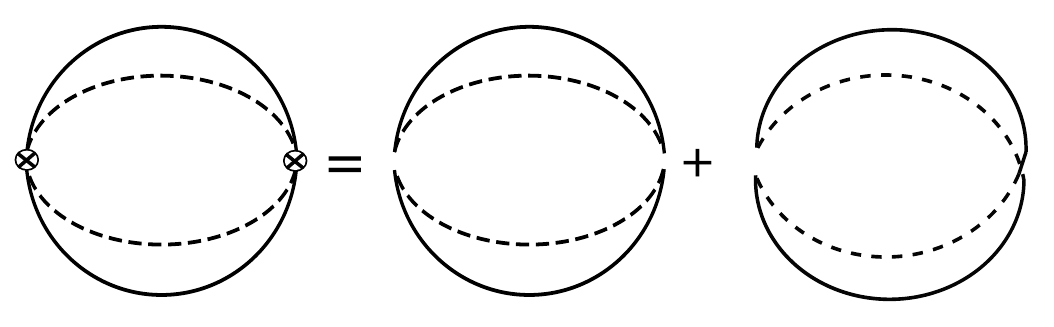}
\centerline {\hspace*{-0.2cm} \bf  \hspace*{1.5cm } $N_c^2$  \hspace*{0.2cm} +  \hspace*{0.2cm} $N_c$}\\
\caption{\footnotesize  LO PT contributions to the spectral function: a) factorised; b) factorised $\oplus$ non-factorised.  } 
\label{fig:pert}
\end{center}
\end{figure}

 The LO perturbative contributions are given by the diagrams in Fig.\,\ref{fig:pert}. Explicit evaluations of the Trace
 appearing in the two-point function indicates that the lowest order (LO)  PT contribution behaves like $N_c^2$  (or $2N^2_c(1-1/N_c)$ if the current has an $\epsilon$-tensor like the one in Table\,\ref{tab:current} where the $1/N_c$ term arises from the $\epsilon$-contraction) as expected from large $N_c$. 
 
\begin{figure}[hbt]
\vspace*{-0.5cm}
\begin{center}
\includegraphics[width=1.5cm]{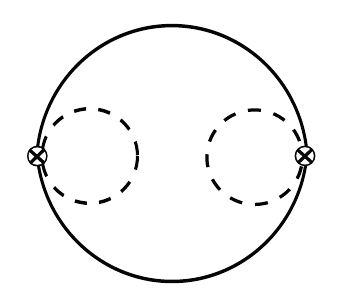}
\centerline {\hspace*{-0.2cm} \bf $N_c$}
\vspace*{-0.3cm}
\caption{\footnotesize  LO eye diagram.  } 
\label{fig:eye}
\end{center}
\vspace*{-0.75cm}
\end{figure}

 However, non-factorised contribution appears at LO both from PT and condensate contributions  when one has two or more  identical quark flavours because one has more possibilities to do the Wick's contraction\,\footnote{Eye-diagram of the type in Fig.\,\ref{fig:eye} will not contribute in our analysis as it leads to non-open charm final states.}. Moreover, some care has to be taken when applying the large $N_c$ analysis to the case
 of baryons and tetraquark states with string junctions\,\cite{ROSSI}\,\footnote{We thank G. Veneziano for discussions on this point.}. 

 \subsection{The LO $d\leq 6$ condensates contributions}
\begin{figure}[hbt]
\vspace*{-0.25cm}
\begin{center}
\includegraphics[width=2.5cm]{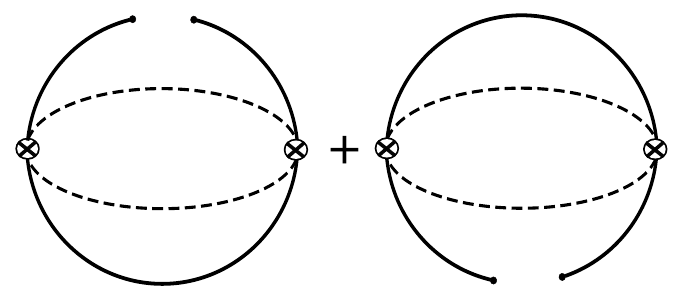}
\vspace*{-0.3cm}
\caption{\footnotesize  $\la \bar qq\ra$ quark condensate.  } 
\label{fig:qq}
\end{center}
\end{figure}

 The QCD condensates entering in the analysis are the light quark condensate $\qq$ and the $SU(3)$-breaking parameter $\kappa\equiv \la\bar ss\ra/\la\bar qq\ra$,  the gluon condensates $ \la
\alpha_sG^2\ra
\equiv \la \alpha_s G^a_{\mu\nu}G_a^{\mu\nu}\ra$ 
and $ \la g^3G^3\ra
\equiv \la g^3f_{abc}G^a_{\mu\nu}G^b_{\nu\rho}G^c_{\rho\mu}\ra$, 
the mixed quark-gluon condensate $g\la\bar qGq\ra
\equiv {\la\bar qg\sigma^{\mu\nu} (\lambda_a/2) G^a_{\mu\nu}q\ra}=M_0^2\la \bar qq\ra$ 
and the four-quark 
 condensate $\rho\la\bar qq\ra^2$, where
 $\rho\simeq (3\sim 4)$ indicates the deviation from the four-quark vacuum 
saturation. 
Their different contributions within the SVZ expansion to LO are shown in Figs.\,\ref{fig:qq} to\,\ref{fig:qq2}. 

Unlike often used in the literature, we have not included higher dimension  $d\geq 8$ condensates contributions\,\footnote{Some classes of  $d=8$ contributions are given in\,\cite{MOLE16,SU3}.} due to our poor knowledge of their size.  
Indeed, a violation of the vacuum saturation for the four-quark condensates\,\cite{SNTAU,JAMI2a,JAMI2c,LNT,LAUNERb} has been already noticed in different light quark channels.  In addition, the mixing of different four-quark condensates under renormalization\,\cite{SNTARRACH} does not also favour the vacuum saturation estimate. On the other, the inaccuracy of a simple dilute gas instanton estimate\,\cite{SVZa,SHURYAK1} has been also observed from the phenomenological estimate of high-dimension gluon condensates\,\cite{SNH10,SNH11,SNH12}.

\vspace*{-0.5cm}
\begin{figure}[hbt]
\vspace*{-0.25cm}
\begin{center}
\includegraphics[width=5.5cm]{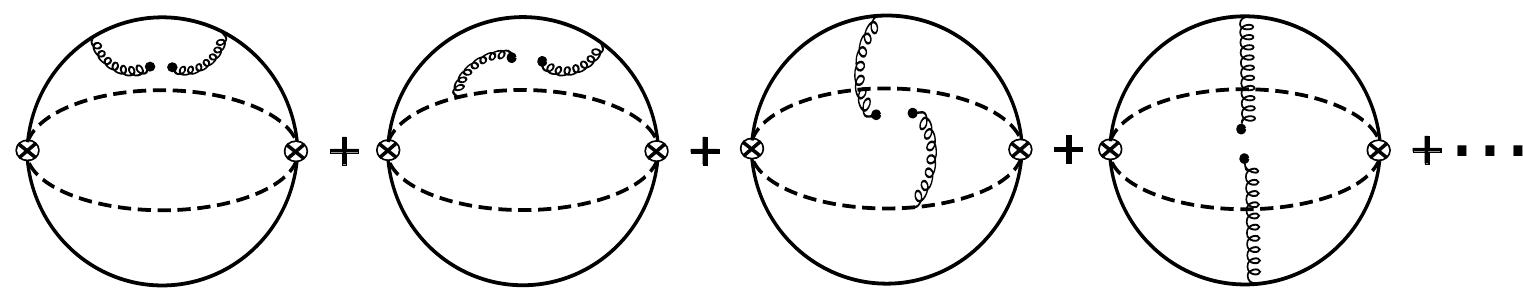}
\vspace*{-0.3cm}
\caption{\footnotesize  $\la\alpha_s G^2\ra$ gluon condensate.  } 
\label{fig:gg}
\end{center}
\end{figure}
\begin{figure}[hbt]
\vspace*{-0.5cm}
\begin{center}
\includegraphics[width=6cm]{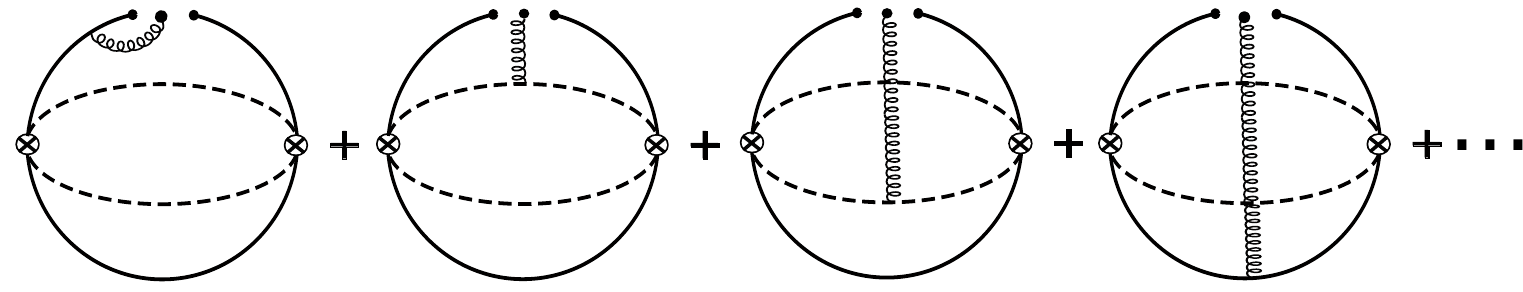}
\vspace*{-0.3cm}
\caption{\footnotesize  $g\la \bar qG q\ra$ mixed quark-gluon condensate.  } 
\label{fig:mix}
\end{center}
\end{figure}
\begin{figure}[hbt]
\vspace*{-0.25cm}
\begin{center}
\includegraphics[width=6cm]{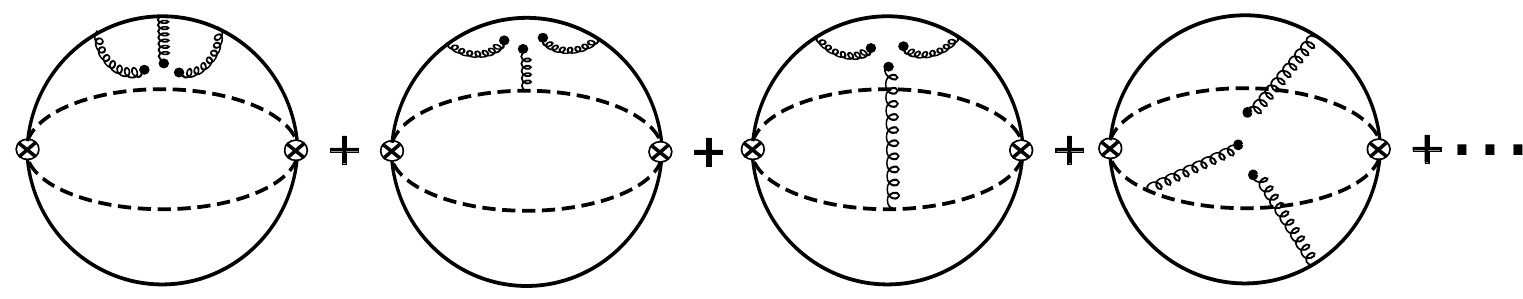}
\vspace*{-0.3cm}
\caption{\footnotesize  $\la g^3G^3\ra$ triple gluon condensate.  } 
\label{fig:ggg}
\end{center}
\end{figure}
\begin{figure}[hbt]
\vspace*{-0.25cm}
\begin{center}
\includegraphics[width=1.2cm]{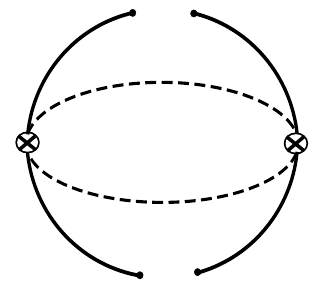}
\vspace*{-0.3cm}
\caption{\footnotesize  $\rho\la \bar \psi\psi\ra^2$ four-quark condensate. } 
\label{fig:qq2}
\vspace*{-0.5cm}
\end{center}
\end{figure}
\vspace*{-1cm}
\subsection{Convolution representation and Matching}
\begin{figure}[hbt]
\vspace*{-0.25cm}
\begin{center}
\hspace*{0.5cm}\includegraphics[width=6cm]{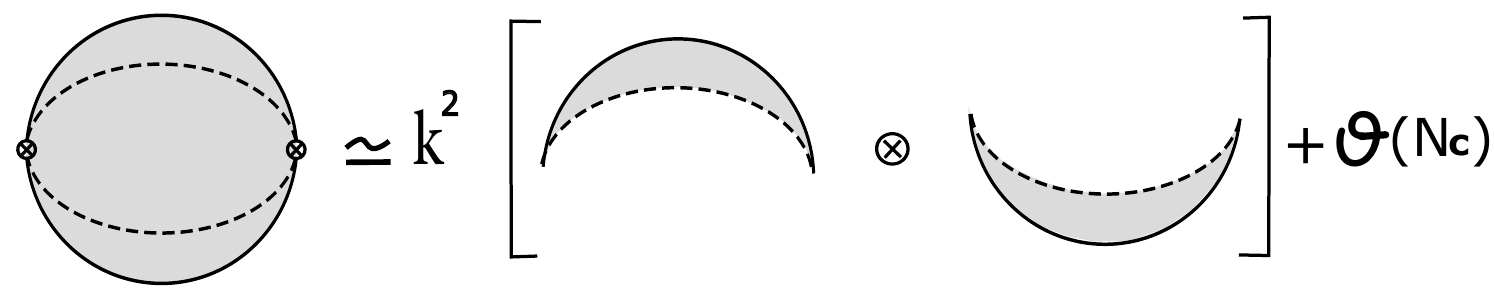} 
\vspace*{-0.2cm}
\caption{\footnotesize The four-quark  spectral function as a convolution of two quark bilinear  ones (see  Eq.\,\ref{eq:convolution}). The black region means perturbative $\oplus$ non-perturbative contributions.} 
\label{fig:conv}
\end{center}
\vspace*{-0.75cm}
\end{figure}

We have explicitely proved in our previous papers\,\cite{HEP18,SU3,MOLE16,4Q,DK} that the non-factorized contribution to the four-quark correlator which appears at lowest order $\alpha_s^0$ of PT QCD to order in $1/N_c$ (but not to order $\alpha_s^2$ as claimed by\,\cite{LUCHA0}),  gives numerically negligible contribution to the sum rule.
Therefore, we can consider that the molecule /tetraquark two-point spectral function is well approximated by the convolution of the two ones built from 
two quark bilinear currents (factorization) as illustrated in Fig.\,\ref{fig:conv} where the diagrams in the two sides of Fig.\,\ref{fig:conv} are of the order $N_c^2$.  

In order to fix the matching factor $k^2$, we consider the example of the $D^*D$ molecule current where the QCD expression is given in the Appendix. The  bilinear currents and the corresponding spectral functions entering in the RHS of Fig.\,\ref{fig:conv} are : 
\bea\label{eq:qqcurrent}
J^{P,S}(x)&\equiv&\bar c [i\gamma_5,1] c ~\rar ~ \frac{1}{\pi}{\rm Im}\,\psi^{P,S}(t)\sim \frac{3}{8\pi^2} \nnb\\
J^{V,A}(x)&\equiv&\bar c [\gamma_\mu,\gamma_\mu\gamma_5] c ~ \rar ~ \frac{1}{\pi}{\rm Im}\,\psi^{V,A}(t)\sim \frac{1}{4\pi^2}~,
\eea
in the limit where $m_c^2\ll t$. 
In this way, we obtain, for a spin 1 state, the convolution integral\,\cite{PICH,SNPIVO,HAGIWARA}:
\bea
\hspace*{-0.1cm}\frac{1}{ \pi}{\rm Im}\, \Pi_{\cal H}(t)&&= \theta [t-(\sqrt{t_{10}}+\sqrt{t_{20}})^2]\ga  \frac{k}{4\pi}\dr^2\hspace*{-0.15cm} t^2 \hspace*{-0.15cm}\int_{t_{10}}^{(\sqrt{t}-\sqrt{t_{20}})^2}\hspace*{-1.2cm}dt_1\times\nnb\\
&&\hspace*{-1.6cm}\int_{t_{20}}^{(\sqrt{t}-\sqrt{t_1})^2}  \hspace*{-1.2cm}dt_2~2\,\lambda^{3/2}\ga\frac{t_1}{t},\frac{t_2}{t}\dr\frac{1}{\pi}{\rm Im} \,\psi^{S,P}(t_1) \frac{1}{\pi} {\rm Im}\, \psi^{A,V}(t_2) : 
\nnb\\
\eea
with the phase space factor:
\beq
{\footnotesize 
\lambda\ga\frac{t_1}{t},\frac{t_2}{t}\dr=\ga 1-\frac{\ga \sqrt{t_1}- \sqrt{t_2}\dr^2}{ t}\dr \ga 1-\frac{\ga \sqrt{t_1}+ \sqrt{t_2}\dr^2}{ t}\dr.}
\label{eq:convolution}
\eeq
$\sqrt{t_{10}}$ and $\sqrt{t_{20}}$  are the quark / hadronic thresholds and $m_c$ is the on-shell / pole perturbative charm quark mass. 

The appropriate $k$-factor which matches this convolution representation with the direct perturbative calculation of the molecule spectral function given in the appendix is\,\footnote{For the tetraquark case, one should add a factor (4/3).} :
\beq
k^2= \frac{5}{3\times 2^6}~,
\label{eq:match}
\eeq 
which comes from the dynamics of the Feynman diagram calculations and which is missed in a standard large $N_c$-results of\,\cite{LUCHA}. 
Related phenomenology will be discussed in Section\,\ref{sec:scatt}.

\subsection{NLO PT corrections to the Spectral functions}

\begin{figure}[hbt]
\vspace*{-0.25cm}
\begin{center}
\hspace*{0.5cm}\includegraphics[width=2.5cm]{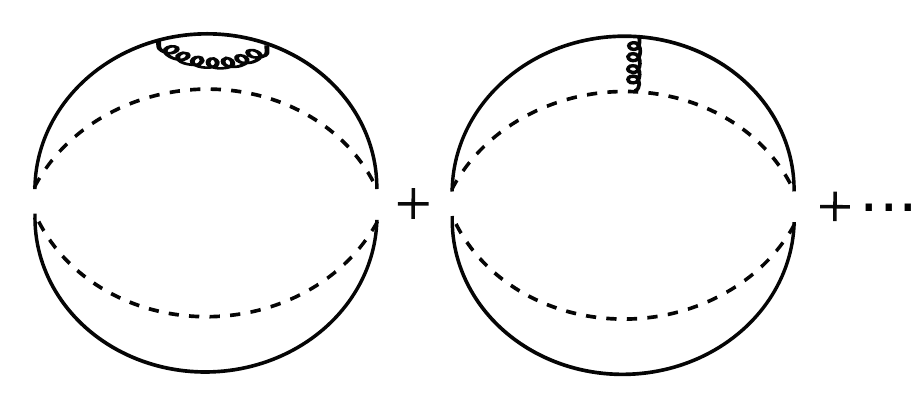}
\vspace*{-0.25cm}
\caption{\footnotesize  NLO factorised PT contribution to the spectral function. } 
\label{fig:as}
\end{center}
\vspace*{-0.5cm}
\end{figure}

We extract the next-to-leading (NLO) perturbative (PT) corrections by approximating  the molecule /tetraquark two-point spectral function with the convolution of the two ones built from 
two quark bilinear currents (factorization) illustrated in Fig.\,\ref{fig:as}. 

The NLO perturbative expressions of the bilinear unequal masses (pseudo)scalar and (axial)-vector spectral functions are known in the literature \cite{BROAD,RRY,SNB1,SNB2,CHET1,PIVOSU3}. 


\subsection{From the On-shell to the $\overline{MS}$-scheme}
We transform the pole mass $m_c$ to the running mass $\overline m_c(\mu)$ using the known relation  in the
$\overline{MS}$-scheme to order $\alpha_s^2$\,\cite{SNB1,SNB2} 
\beq
m_c = \overline{m}_c(\mu)\Big{[}
1+\frac{4}{3} a_s\nnb\\
+a_s\ln{\frac{\mu^2}{m_c^2}}+{\cal O}(a_s^2)\Big{]}.
\label{eq:pole}
\eeq
 In the following, we shall use $n_f$=4 total number of flavours for the numerical value of $a_s\equiv\alpha_s/ \pi$. 
\section{QCD input parameters}
 \vspace*{-1cm}
\nin
{\scriptsize
\begin{table}[hbt]
\setlength{\tabcolsep}{0.45pc}
    {\scriptsize
  \begin{tabular}{llll}
&\\
\hline
\hline
Parameters&Values&Sources& Ref.    \\
\hline
$\alpha_s(M_Z)$& $0.1181(16)(3)$&$M_{\chi_{0c,b}-M_{\eta_{c,b}}}$&
\cite{SNparam,SNparam2,SNm20} \\
$\overline{m}_c(m_c)$ [MeV]&$1266(6)$ &$D, B_c \oplus$&
\cite{SNm20,SNparam,SNbc20,SNFB13}\\
&&$ {J/\psi}, \chi_{c1},\eta_{c}$ &\\
$\hat \mu_q$ [MeV]&$253(6)$ &Light  &\,\cite{SNB1,SNp15} \\
$\hat m_s$ [MeV]&$114(6)$ &Light &\,\cite{SNB1,SNp15} \\
$\kappa\equiv\la \bar ss\ra/\la\bar dd\ra$& $0.74(6)$&Light-Heavy&\cite{SNB1,SNp15,HBARYON1}\\
$M_0^2$ [GeV$^2$]&$0.8(2)$ &Light-Heavy&\,\cite{SNB1,DOSCH,JAMI2a}\\
&&&\cite{JAMI2c,HEIDa,HEIDc,SNhl} \\
$\la\alpha_s G^2\ra$ [GeV$^4$]& $6.35(35) 10^{-2}$&Light-Heavy &
 \cite{SNparam,SNm20}\\
${\la g^3  G^3\ra}/{\la\alpha_s G^2\ra}$& $8.2(1.0)$[GeV$^2$]&${J/\psi}$&
\cite{SNH10,SNH11,SNH12}\\
$\rho \alpha_s\la \bar qq\ra^2$ [GeV$^6$]&$5.8(9) 10^{-4}$ &Light,$\tau$-decay&\cite{DOSCH,SNTAU,JAMI2a,JAMI2c,LNT,LAUNERb}\\
\hline\hline
\end{tabular}}
 \caption{QCD input parameters estimated from QSSR (Moments, LSR and ratios of sum rules) used here. 
 }  
\label{tab:param}
\vspace*{-0.5cm} 
\end{table}
} 
\subsection{QCD coupling $\alpha_s$}
We shall use from the $M_{\chi_{0c}}-M_{\eta_{c}}$ mass-splitting sum rule\,\cite{SNparam}: 
 \bea&&\hspace*{-1cm} 
 \alpha_s(2.85)=0.262(9) \rar\alpha_s(M_\tau)=0.318(15)\nnb\\
& \rar&\alpha_s(M_Z)=0.1183(19)(3)
\eea
which is more precise than the one from  $M_{\chi_{0b}}-M_{\eta_{b}}$\,\cite{SNparam}\,: 
\bea &&\hspace*{-1cm} 
 \alpha_s(9.50)=0.180(8) \rar\alpha_s(M_\tau)=0.312(27)\nnb\\
&&\rar\alpha_s(M_Z)=0.1175(32)(3).
 \eea
 These lead to the mean value quoted in Table\,\ref{tab:param}, which is
 in complete agreement with the world average\,\cite{PDG}\,:
\beq
\alpha_s(M_Z)=0.1181(11)~.
\eeq
\subsection{Quark masses}
We shall use the recent determinations of  the running masses $\overline m_s(\mu)$ and $\overline m_{c}(\overline m_{c})$ quoted in Table\,\ref{tab:param} and the corresponding value of $\alpha_s$ evaluated at the scale $\mu$ obtained using the same sum rule approach. 
\subsection{QCD  condensates\label{sec:cond}}
Their values are quoted in Table\,\ref{tab:param}. One should notice that taking into account the anomalous dimension of the $\la\bar qq\ra$ condensate, $\alpha_s\la\bar qq\ra^2$ has a very smooth $1/\log{(\tau\Lambda^2)}^{1/25}$ behaviour for $n_f=4$ flavours such that, contrary to simple minded, its contribution is not suppressed by $1/\log{(\tau\Lambda^2)}$ for light mesons\,\cite{LNT,LAUNERb} and $\tau$-decays\,\cite{SNTAU}  permitting its reliable phenomenological estimate while  its extraction from light baryons\,\cite{DOSCH,JAMI2a,JAMI2c} occurs without the $\alpha_s$ factor like in the case of the four-quark currents discussed here.


\section{The spectral function\label{sec:spectral}}
\subsection{The minimal duality ansatz}
   In the present case, with no complete data on the spectral function, we use the minimal duality ansatz:
 \bea
\hspace*{-0.65cm} \frac{1}{\pi}{\rm Im} \Pi_{\cal H}\hspace*{-0.1cm}&\simeq&\hspace*{-0.1cm} f_{\cal H}^2 M_{\cal H}^{8} \delta(t-M_{\cal H}^2) +\Theta(t-t_c) \frac{1}{\pi}{\rm Im} \Pi^{QCD}_{\cal H}(t),
 \eea
 for parametrizing the molecule spectral function. $M_{\cal H}$ and $f_{\cal H}$ are the lowest ground state mass and coupling analogue to $f_\pi$. The ``QCD continuum"  is the imaginary part of the QCD correlator (as mentioned after Eq.\,\ref{eq:lsr}) from the  threshold $t_c$ which is assumed to smear all higher states contribution. Then, it insures that both sides of the sum rules have the same large $t$ asymptotic behaviour which leads to the Finite Energy Sum Rule (FESR) in Eq.\,\ref{eq:lsr}. Within a such parametrization, one  obtains: 
 \beq
  {\cal R}^{c}_n\equiv {\cal R}\simeq M_{\cal H}^2~,
  \label{eq:mass}
  \eeq
 indicating that the ratio of moments is a useful tool for extracting the mass of the hadron ground state\,\cite{SNB1,SNB2,SNB3}. The corresponding value of $t_c$ approximately  corresponds to the mass of the 1st radial excitation. However, one should bear in mind that a such parametrization cannot distinguish two nearby resonances but instead will consider them as one ``effective resonance". 
 
 This simple model has been tested successfully in different channels where complete data are available (charmonium, bottomium and $e^+e^-\to I=1$ hadrons)\,\cite{SNB1,SNB2,BERTa}.   It was shown that, within the model, the sum rule  reproduces quite well the integrated data while the masses of the lowest ground state mesons ($J/\psi,~\Upsilon$ and $\rho$) have been predicted within a good accuracy. 
    
In the extreme case of the pseudoscalar Goldstone pion, the sum rule using the spectral function parametrized by this simple model and the more complete one by Chiral Perturbation Theory (ChPT)\,\cite{BIJNENS} lead to  similar values of the sum of light quark masses $(m_u+m_d)$ indicating the efficiency of this simple parametrization\,\cite{SNB1,SNB2}. 

An eventual violation of the quark-hadron duality (DV)\,\cite{SHIF,PERIS} tested from hadronic $\tau$-decay data\,\cite{PERIS,SNTAU,PICHROD} is negligible here thanks to the double exponential suppression of this contribution in the Laplace sum rule (see e.g \,\cite{4Q} for details). 
 
The uses of this model for the tetraquarks and molecule states are also quite successful compared to the recent data (see e.g \,\cite{4Q,DK} and references therein). Then, we (a priori) expect to extract with a good accuracy the  masses and  couplings of the  mesons within the approach. 
 
In order to minimize the effects of radial excitations smeared by the QCD continuum, 
 we shall work with the lowest  moment $ {\cal L}^c_0 $ and ratio of moments $ {\cal R}^{c}_0$ for extracting the meson masses and couplings $f_{\cal H}$. Moments with $n<0$ will not be considered due to their sensitivity on the non-perturbative contributions at zero momentum. 

However, once we have fixed the ground state parameters, we attempt to extract 
the mass and coupling of the first radial excitation by using a :``two resonance" +($\Theta(t-t_{c1})$ ``QCD continuum " parametrization where $t_{c1}$ is above the $t_c$-value obtained for the ground state. 
  \vspace*{-0.5cm}
\subsection{Optimization Criteria}
 As $\tau,~ t_c$ and $\mu$ are free external parameters , we shall use stability criteria (minimum sensitivity on the variation of these parameters) to extract the hadron masses and couplings.  Results based on these stability criteria have lead to successful predictions in the current literature (see\,\cite{SNB1,SNB2,SNB3}  and original papers). 

 \vspace*{-0.25cm}
\section{Revisiting $f_{D*D}$ and $M_{D^*D}$}
We start by revisiting and checking the results obtained in\,\cite{MOLE16} where they have included the factorized contributions to N2LO of perturbative series.  In this example, we show explictly our strategy for extracting the mass and coupling. The same strategy will be used in some other channels discussed later in this paper. This example is also a test of the efficiency of the method by confronting the prediction with the\,: $Z_c(3900)$\,\cite{BELLE1, BES0}, $Z_c(4020)$\,\cite{BES2}, $Z_c(4025)$\,\cite{BES3}, $Z_c(4050)$\,\cite{BELLE3}, $Z_c(4226), Z_c(4257)$ by BESIII\,\cite{BES5} and $Z_c(4430)$\,\cite{BELLE4,LHCb5} found earlier, where one can notice that only the $Z_c(3900)$ and  $Z_c(4430)$ have been retained as established in the Meson Summary Table of Particle Data Group (PDG)\,\cite{PDG}.
\begin{figure}[hbt]
\vspace*{-0.25cm}
\begin{center}
\centerline {\hspace*{-7.5cm} \bf a) }
\includegraphics[width=7cm]{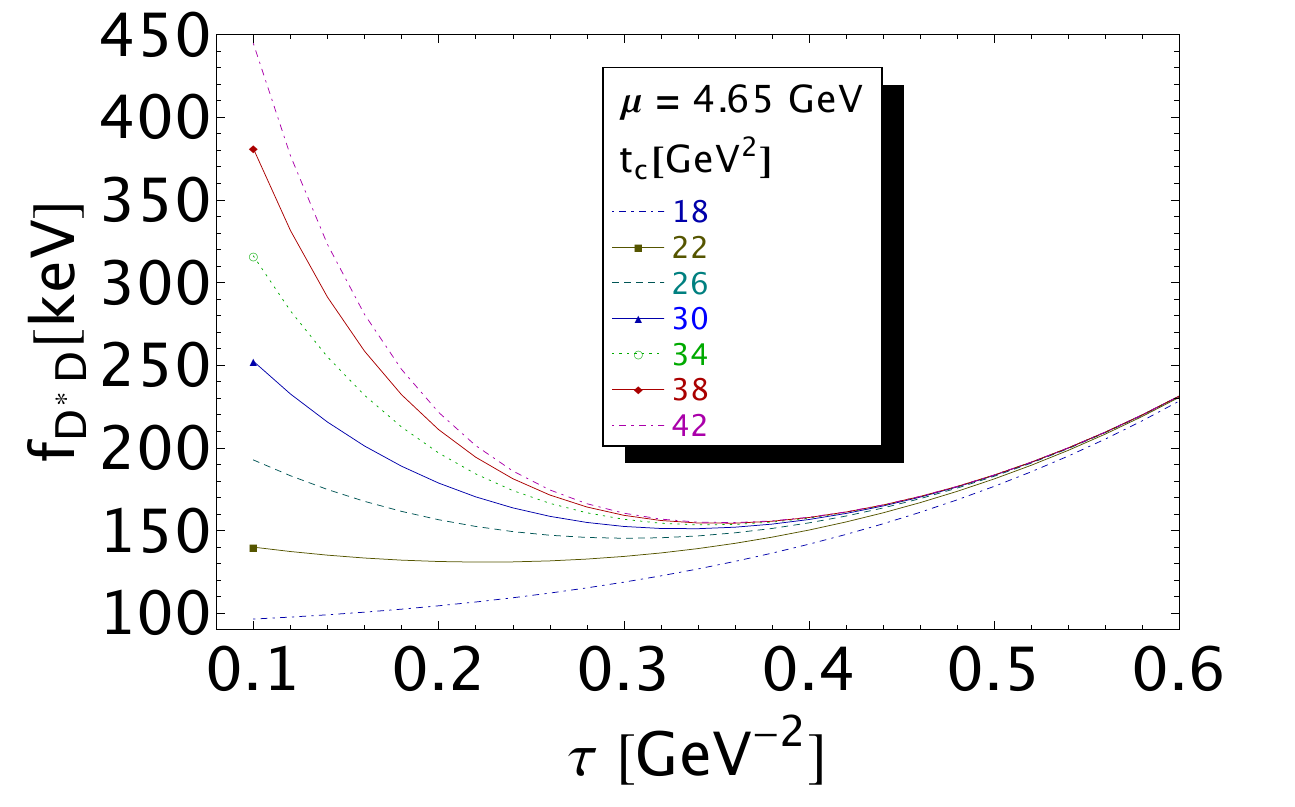}
\centerline {\hspace*{-7.5cm} \bf b) }
\includegraphics[width=6.5cm]{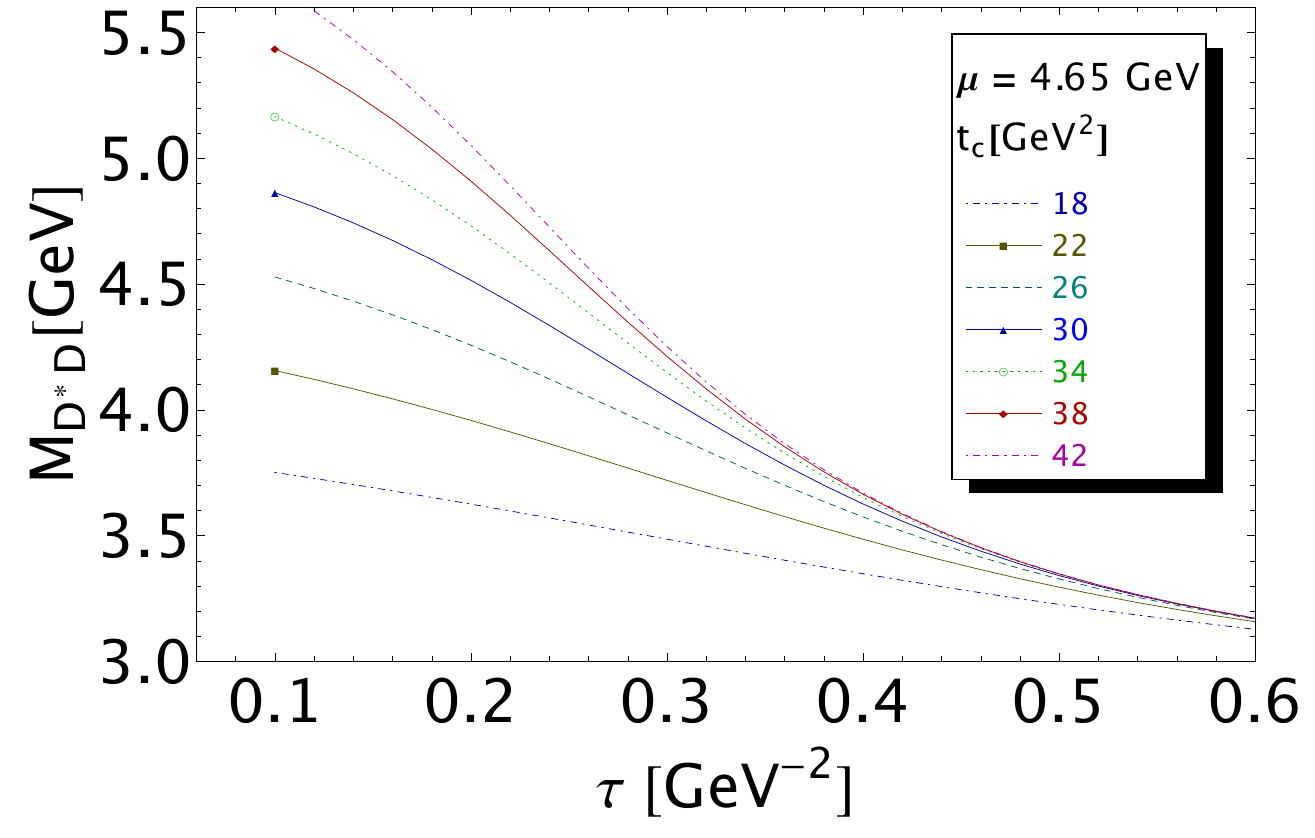}
\vspace*{-0.5cm}
\caption{\footnotesize  $f_{D^*D}$ and $M_{D^*D}$ as function of $\tau$ at NLO for different values of $t_c$, for $\mu$=4.65 GeV and for values of $\overline m_{c}(\overline m_{c})$ given in Table\,\ref{tab:param}.} 
\label{fig:dstard}
\end{center}
\vspace*{-1cm}
\end{figure} 
\subsection{$\tau$ and $t_c$-stabilities} 
We show in Fig.\,\ref{fig:dstard} the $\tau$-behaviour of the coupling and of the mass for different values of $t_c$ and fixing the value of the subtraction constant $\mu$ at 4.65 GeV (see next subsection) where a $\mu$ stability $(4.5\pm 0.5)$ GeV has been found  in\,\cite{MOLE16}. 
From this figure, one can see that the coupling presents minimas in $\tau$ and the mass inflexion points.  

-- In a first step,  we use the experimental mass $M_{Z_c}=3900$ MeV for
extracting the value of the coupling $f_{D^*D}$ shown in Fig.\,\ref{fig:dstard}a). 

-- In a 2nd step, we take the value of $\tau$ at the minimum of the coupling and use it for extracting the value of $M_{D^*D}$ from Fig.\,\ref{fig:dstard}b).

-- In a 3rd step, we take the common range of $t_c$ where both curves present stabilities in $\tau$. In the present case, this value ranges from $t_c= 22$ GeV$^2$ (beginning of $\tau$-stability) to $t_c= 38$ GeV$^2$ (beginning of $t_c$-stability)
where the range is given in Table\,\ref{tab:lsr-param}.  For the mean $t_c=$ 30 GeV$^2$, it is : $\tau_R\simeq 0.38$(resp. 0.34) GeV$^{-2}$ for LO (resp. NLO) QCD expression. 

-- The errors given in Table\,\ref{tab:res} and the resulting values of $f_{D^*D}$ and $M_{D^*D}$ in Table\,\ref{tab:summary} are the mean of the ones from the previous two extremal values of $t_c$. 

\subsection{$\mu$-stability\label{sec:mu}} 
\begin{figure}[hbt]
\vspace*{-0.25cm}
\begin{center}
\centerline {\hspace*{-7.5cm} \bf a) }
\includegraphics[width=7cm]{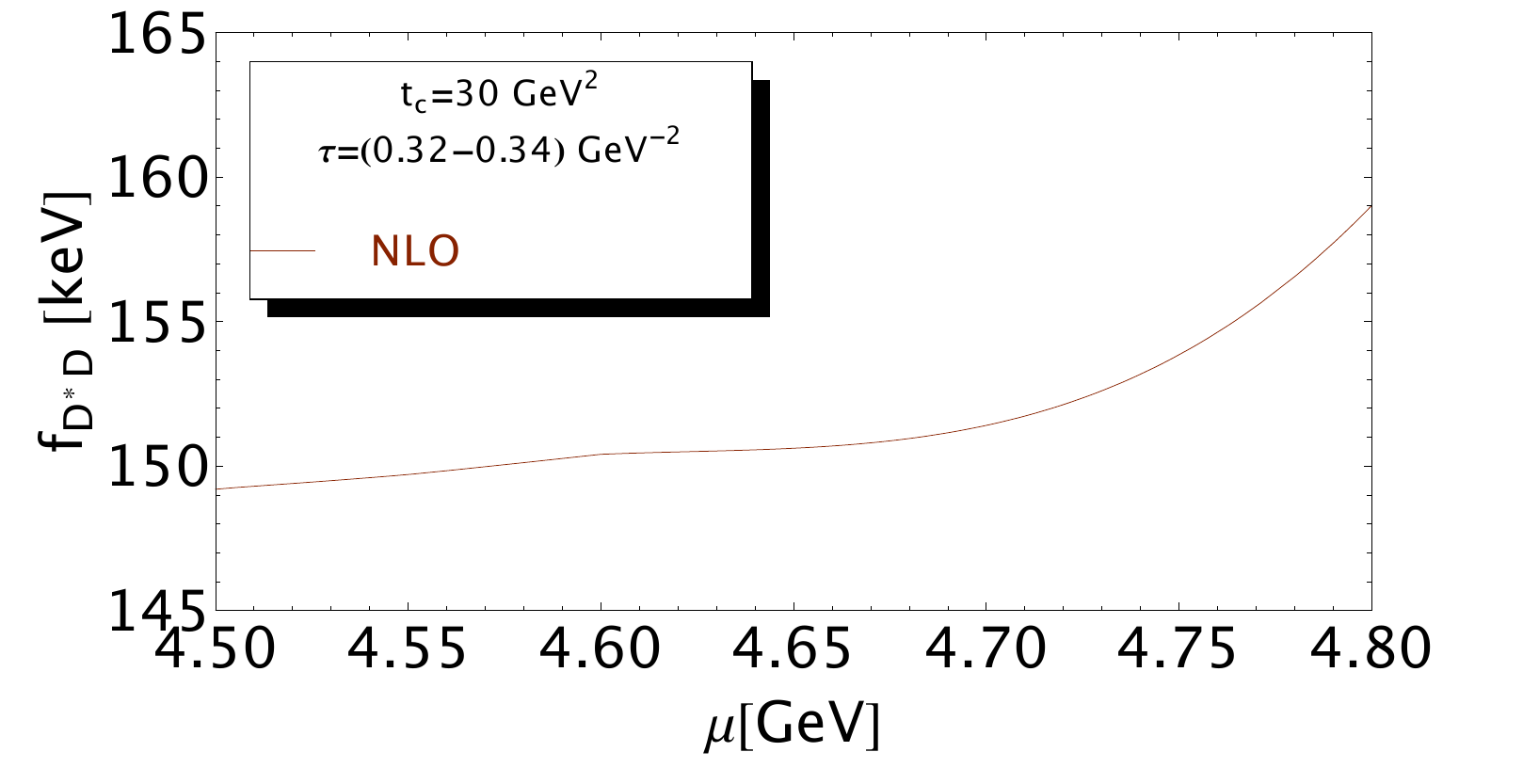}
\centerline {\hspace*{-7.5cm} \bf b) }
\includegraphics[width=7cm]{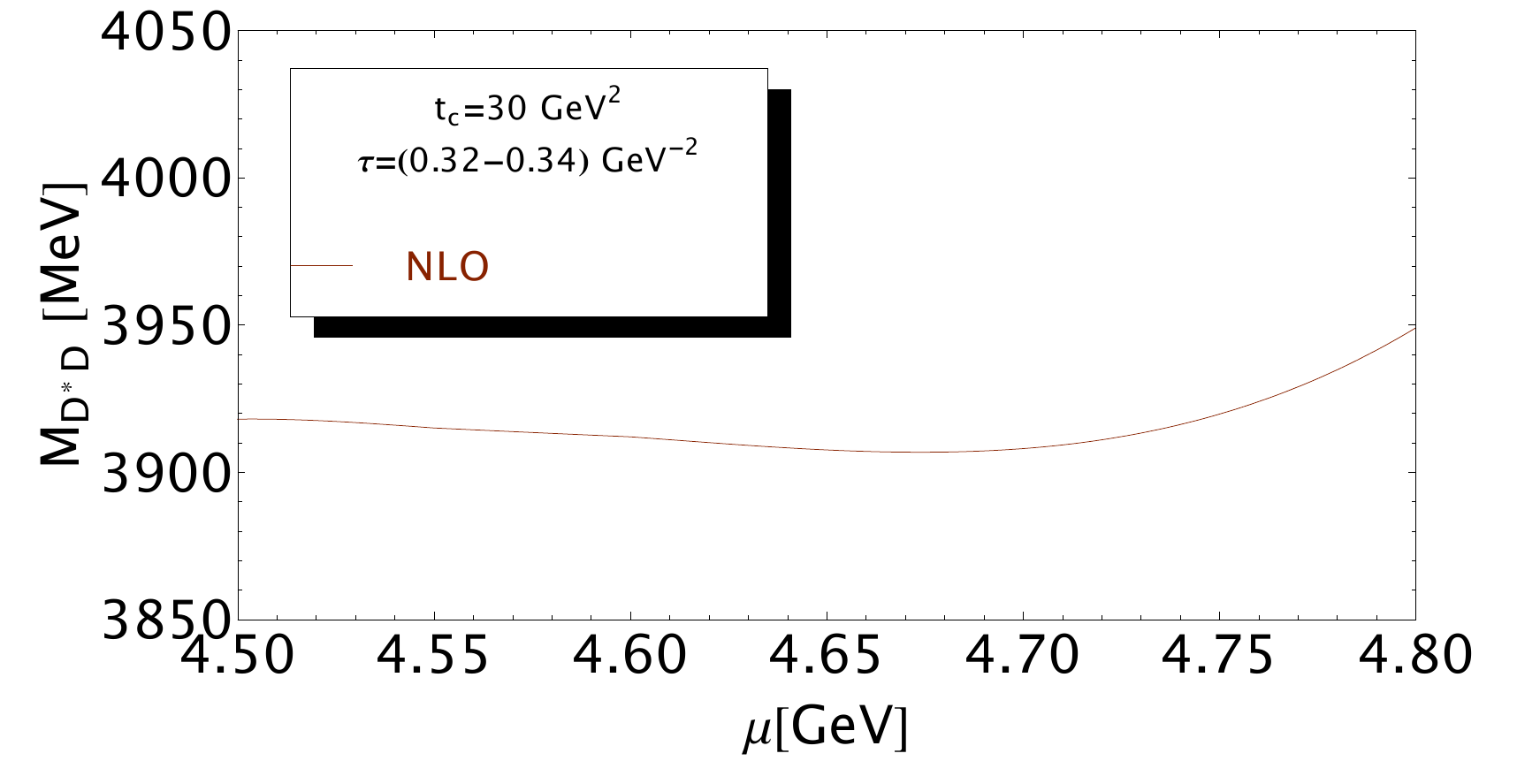}
\vspace*{-0.5cm}
\caption{\footnotesize  $f_{D^*D}$ and $M_{D^*D}$ as function of $\mu$ at NLO and for $t_c=30$ GeV$^2$.} 
\label{fig:dstard-mu}
\end{center}
\vspace*{-1cm}
\end{figure}

 For doing the analysis, we shall fix $t_c=30$ GeV$^2$ which is the mean of the two extremal values delimiting the stability region and take $M_{D^*D}=3900$ MeV for fixing the coupling. The analysis is shown in Fig.\,\ref{fig:dstard-mu}. 
 One can see a $\mu$-stability for :
 \beq
 \mu=(4.65\pm 0.05) ~{\rm GeV}~,
 \label{eq:mu}
 \eeq
 at which we shall evaluate the results quoted in Table\,\ref{tab:res}. 
 
  This value of $\mu$ from a more refined analysis is more precise than the conservative one $(4.5\pm 0.5)$ GeV quoted in\,\cite{MOLE16}. The results of the analysis are given in Tables\,\ref{tab:res} and \,\ref{tab:summary}. 
 
 One should note that the resummed QCD expression of the sum rule which obeys an ``Homogeneous" Renormalization Group Equation (RGE) is obtained by putting $\mu^2=1/{\tau}$ in the QCD expression of the sum rule and where the parameters having anomalous dimension $\gamma$ run as $1/(\log{\tau\Lambda^2})^{\gamma/\beta}$ (see e.g.\,\cite{SNR}).  We have often used this choice in the past (see\,\cite{SNB1,SNB2,SNB3}) which corresponds here to the value:  
 \beq
 \mu=1/\sqrt{\tau_0}\simeq 1.6 ~{\rm GeV},
 \eeq
 at the $\tau$-minimum for $t_c$=30 GeV$^2$.  However, this value is outside the $\mu$-stability region obtained previously and then does not correspond to the optimal choice of $\mu$.  This is the reason why we have abandoned this choice $\mu^2=1/{\tau}$.  This result does not support the argument of Ref.\,\cite{WANG1} based on the observation that, when the term of the type $(\alpha_s(\tau)/\alpha_s(\mu))^{{\gamma}/{\beta_1}}$ disappears for $\mu=1/\sqrt{\tau}$, one would obtain the best choice of $\mu$. Indeed, the PT series behaves obviously much better in the $\mu$-stability region where the value of $\mu$ is about 3 times higher at which the radiative corrections are more suppressed. 
 
 Moreover, the physical meaning of the relation between $\mu$ with the so-called bound energy or virtuality  $ \mu^2={M_Z^2-4M_c^2}$ used in his different papers (see e.g.\,\cite{WANG1,WANGMU}) remains unclear to us where $M_c$ is the PT constituent or pole charm quark mass taken by the author to be about 1.84 GeV which corresponds to $\mu\simeq $ 1.3 GeV. 
 
  Indeed,  one may expect from this formula that the difference between the resonance and PT quark constituent masses  has a non-perturbative origin which (a priori) has nothing to do with the scale $\mu$ where the PT series and the Wilson coefficients of the condensates are evaluated.

However, one may also consider $\mu$ as a scale separating the calculable PT Wilson coefficients and the NPT non-calculable condensates in the OPE\,\cite{SVZa} though one has to bear in mind that the $d\leq 6$ condensates appearing in the present analysis are renormalization group invariant ($\mu$ independent like $m_c\la\bar qq\ra,~m^2_c\la\bar qq\ra^2,$ ) or have a weak dependence on $\mu$ ($\la\alpha_s G^2\ra,\la g\bar qGq\ra$)\,\cite{SNB1}\,(Part VII page 285) such that the truncation of the PT series does not affect much their values. This feature indicates that the separation of the condensates from the PT Wilson coefficients are not ambiguous while the size of the non-perturbative condensate is almost independent on the scale at which the PT series is truncated. 
\subsection{NLO and Truncation of the perturbative series\label{sec:ho}} 
We have mentioned in previous works that the inclusion of the NLO perturbative corrections is important
for justifying the choice of heavy quark mass definition used in the analysis  where an {\it ad hoc} value of the $\overline {MS}$ running mass is frequently used in the literature while the spectral function has been evaluated using an on-shell renormalization where the pole (on-shell) quark mass naturally enters into the LO expression. 

\begin{figure}[hbt]
\vspace*{-0.25cm}
\begin{center}
\centerline {\hspace*{-7.5cm} \bf a) }
\includegraphics[width=7cm]{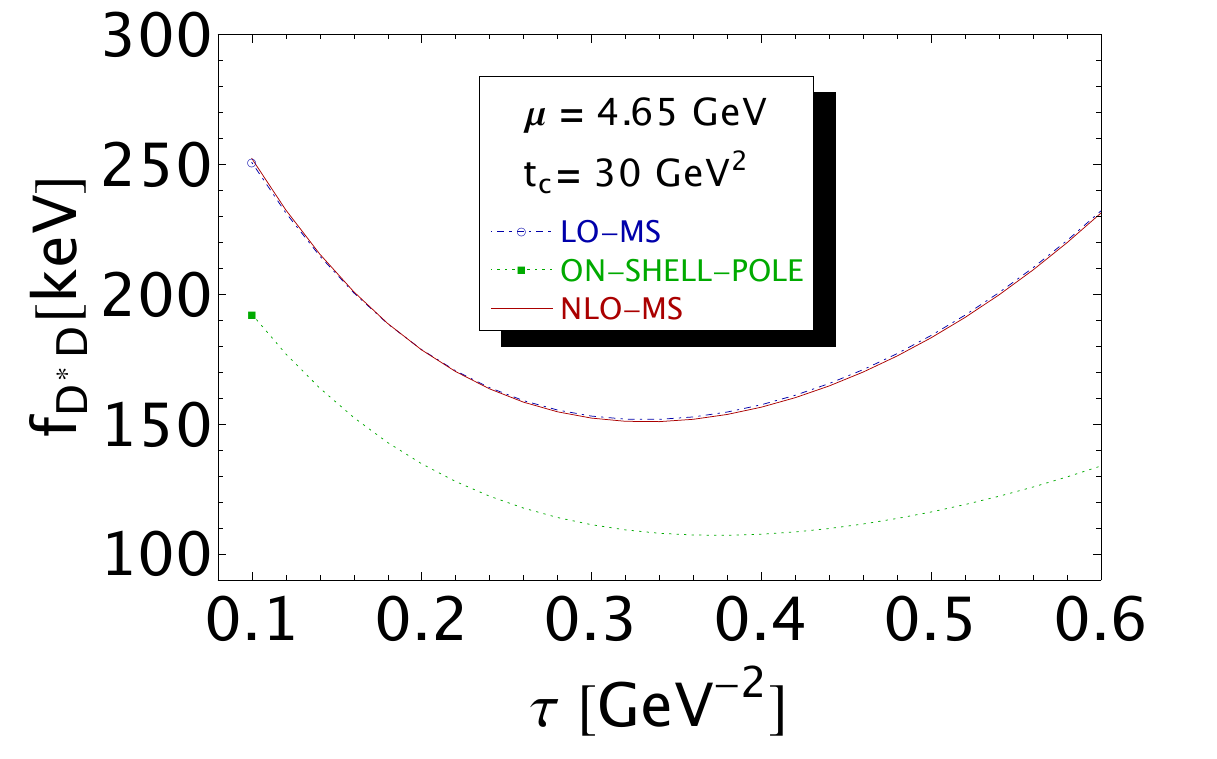}
\vspace{0.25cm}
\centerline {\hspace*{-7.5cm} \bf b) }
\includegraphics[width=6.5cm]{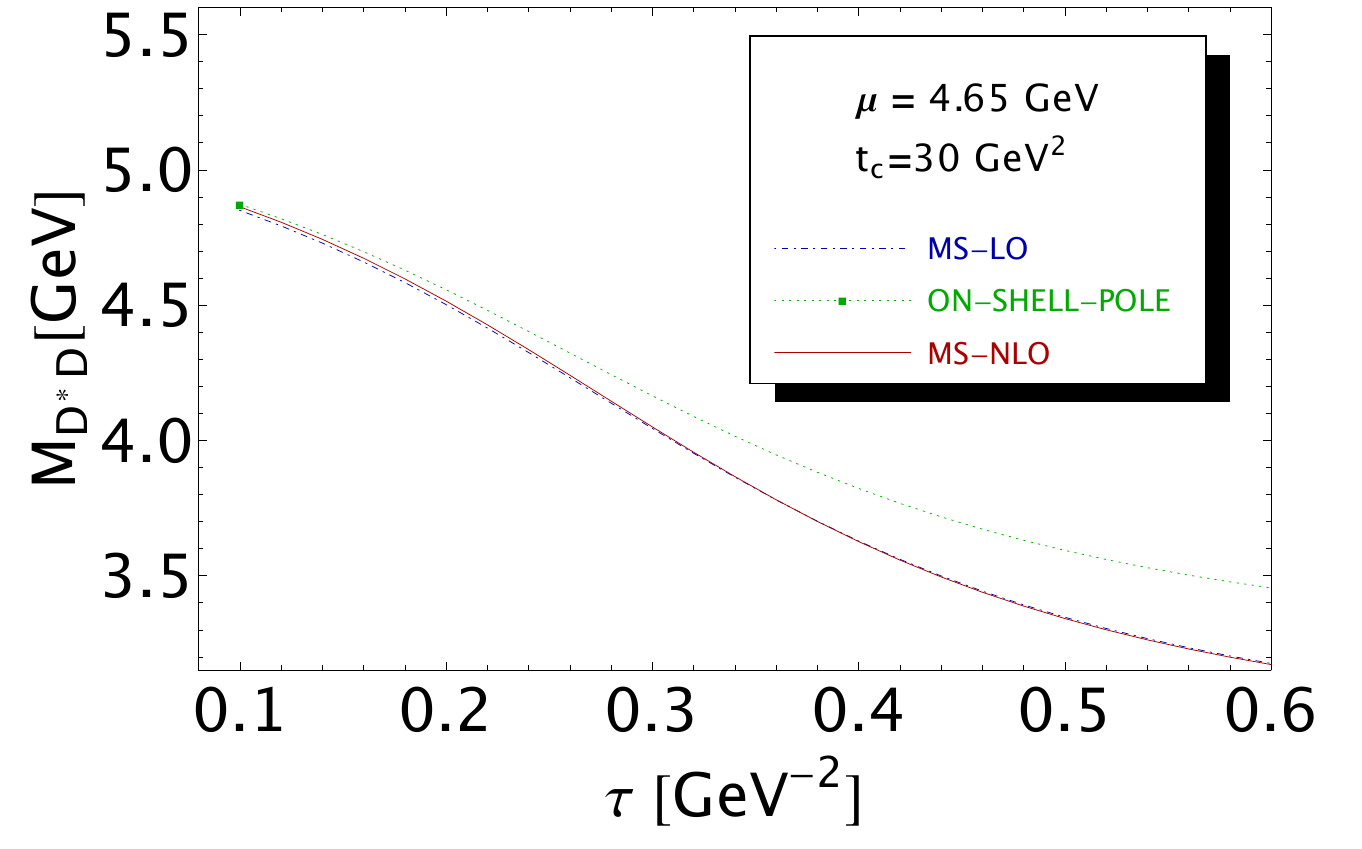}
\vspace*{-0.5cm}
\caption{\footnotesize  $f_{D^*D}$ and $M_{D^*D}$ as function of $\tau$ at LO and NLO  for $t_c=30$ GeV$^2$ and $\mu=4.65$ GeV   for different definitions of the charm quark mass. We use $m_c(pole)=1.5$ GeV and the running mass given in Table\,\ref{tab:param}.} 
\label{fig:lo}
\end{center}
\vspace*{-0.25cm}
\end{figure}

 We show the analysis for the mass and coupling in Fig.\,\ref{fig:lo} at fixed value of the continuum threshold $t_c$ and subtraction constant $\mu$, where one can find that the use of the pole mass at LO decreases by 30\% at the minimum (resp. increases by  0.5\% at the inflexion point) the value of the coupling (resp. mass) obtained using the  $\overline{MS}$ running mass at LO  while the NLO correction is relatively small within the $\overline {MS}$-scheme. 
 
 The smallness of radiative corrections for the ratio of moments demonstrates (a posteriori) why the use of the $\overline {MS}$ running mass at LO leads to a surprisingly good prediction for the mass.
  \begin{figure}[hbt]
\vspace*{-0.25cm}
\begin{center}
\centerline {\hspace*{-7.5cm} \bf a) }
\includegraphics[width=6.cm]{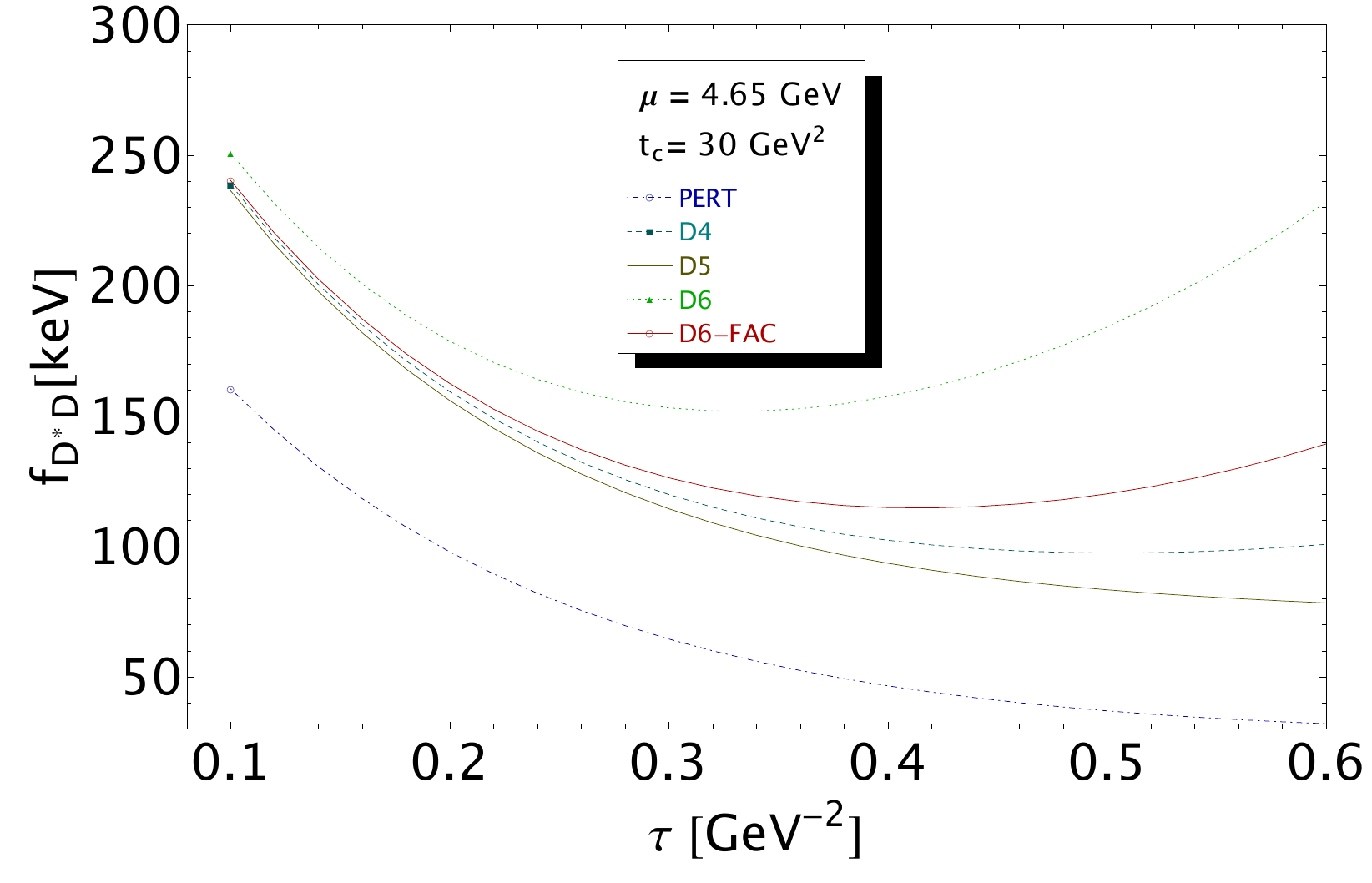}
\vspace{0.25cm}
\centerline {\hspace*{-7.5cm} \bf b) }
\includegraphics[width=6.cm]{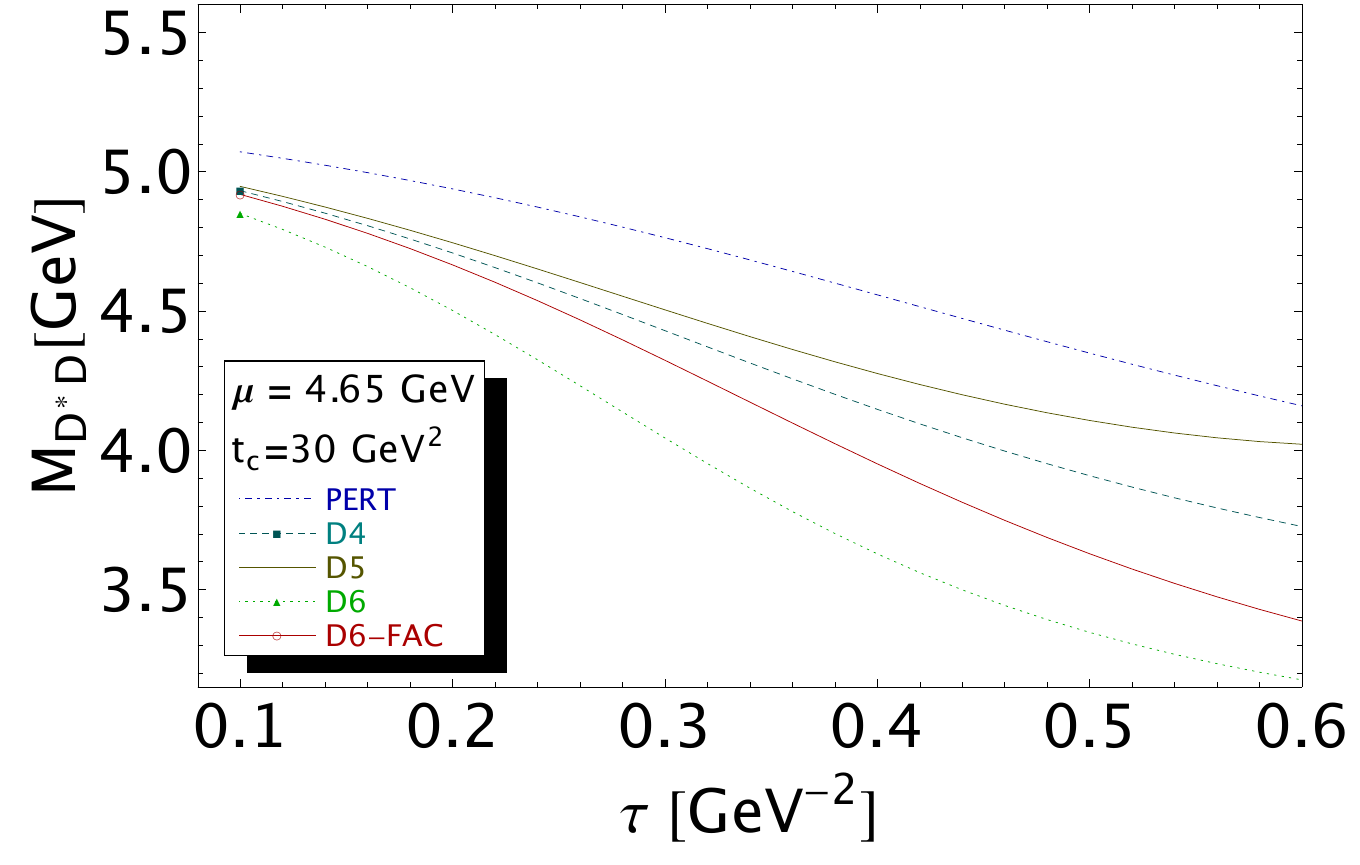}
\vspace*{-0.5cm}
\caption{\footnotesize  $f_{D^*D}$ and $M_{D^*D}$ as function of $\tau$ at NLO for $t_c=30$ GeV$^2$ and $\mu=4.65$ GeV and for different truncations of the OPE.  $D4\equiv$ perturbative $\oplus$ dimension-4; $D5\equiv D4\oplus$ dimension-5; $D6\equiv D5\oplus$ dimension-6. The values of $\overline m_{c}(\overline m_{c})$ and the condensates are given in Table\,\ref{tab:param}. FAC means that we use the vacuum saturation assumption for the estimate of the four-quark condensates.}
\label{fig:cond}
\end{center}
\vspace*{-0.25cm}
\end{figure} 

Assuming that the PT series grow geometrically\,\cite{SZ,CNZa,CNZb,ZAKa}, one can deduce from the previous analysis an estimate of higher order PT contributions given in Table\,\ref{tab:res} which can be compared with the ones in \,\cite{MOLE16,SU3}. 
\subsection{QCD condensates and truncation of the OPE\label{sec:ope}}
 We show in Fig.\,\ref{fig:cond} the contributions of the QCD condensates for different truncations of the OPE. Fixing $\tau\approx 0.34$ GeV$^{-2}$ where the final value of the coupling presents $\tau$-stability and the mass inflexion points, one obtains for $t_c$=30 GeV$^2$ :
\bea
f_{D^*D}&\simeq& 56~{\rm keV}\big{[} 1+0.98  - 0.13 + 0.86\big{]}\nnb\\
&\simeq&151~{\rm keV},\nnb\\
M_{D^*D}&\simeq&4683~{\rm MeV}\big{[}1-0.08 +0.02-0.12\big{]}\nnb\\
&\simeq& 3864~{\rm MeV}~.
\label{eq:violation}
\eea

One can notice the important role of the dimension\,-4 and -6 condensates which are dominated by the chiral condensates $\la\bar qq\ra$ and $\rho\la\bar qq\ra^2$ in the extraction of the coupling and mass where their strengths are more pronounced for the coupling. 

We estimate the sytematic errors due to the truncation of the OPE from the size of the dimension-6 condensate contributions rescaled by the factor $m^2_c\tau/3$ where 1/3 comes from the LSR exponential form of the sum rule. It can be compared with the contributions of the known $d=8$ $\la\bar qq\ra \la \bar q Gq\ra$ condensate obtained in\,\cite{MOLE16,SU3} but bearing in mind that this is only a part of the complete $d$=8 condensate ones where the validity of the vacuum saturation used for its estimate is questionable. 

The quoted errors in Table\,\ref{tab:res} are the mean from the two extremal values of $t_c$ delimiting the $(\tau,t_c)$ stability region. We expect that the systematics for the $m_s\not=0$ cases are similar as confirmed  explicitly in Table\,\ref{tab:res}.


\subsection{Vacuum saturation of the four-quark condensate}
The four-quark condensates have been demonstrated to mix under renormalization\,\cite{SNTARRACH} and \cite{SNB1}\,(Part VII page 285) at finite $N_c$ while its vacuum saturation estimate is only valid in the large $N_c$-limit. A such estimate from light mesons\,\cite{LNT,LAUNERb}, light baryons\,\cite{DOSCH,JAMI2a,JAMI2c} and $\tau$-decays\,\cite{SNTAU} has been shown to underestimate the actual value of the four-quark condensates by a factor 3-4 (see also the comments in Subsection\,\ref{sec:cond}).  For completeness,  we also show in Fig.\,\ref{fig:cond} the effect of this estimate, where we see that the position of the minimum and inflexion point are shifted at $\tau\approx 0.4$ GeV$^{-2}$. At this value and for $t_c=30$ GeV$^2$, one obtains an effect of +38\% for the coupling and --7\% for the mass leading to :
\beq
f_{D^*D}\vert_{FAC}\simeq  115 ~{\rm keV},~~~~
M_{D^*D}\vert_{FAC}\simeq   3951~{\rm MeV},
\eeq
which can be compared with the one in Eq.\,\ref{eq:violation}. 
\subsection{Results of the analysis}
The sizes of the errors from different sources and the final results are collected inTables\,\ref{tab:res} and\,\ref{tab:summary}\,:
\beq 
f_{D^*D}=140(15)~{\rm keV},~~~~~M_{D^*D}=3912(61)\,{\rm MeV}~. 
\eeq
One can notice that the results are in perfect agreement with the previous ones in\,\cite{MOLE16}. The slight difference in the error calculation is due to the fact that, in Ref.\,\cite{MOLE16}, we have estimated the error by choosing $t_c=38$ GeV$^2$ but not considering the one due to $t_c=22$ GeV$^2$. 

The result for the mass is  in a very good agreement with the BELLE \,\cite{BELLE1} and BESIII\,\cite{BES0} data $Z_c(3900)$ MeV which we shall discuss later on. 
\vspace*{-0.25cm}
\begin{table*}[hbt]
\setlength{\tabcolsep}{0.15pc}
\catcode`?=\active \def?{\kern\digitwidth}
    {\scriptsize
  \begin{tabular*}{\textwidth}{@{}l@{\extracolsep{\fill}}|ccccccccc   c ccccccc  l}
\hline
\hline
 Observables\,&$ \Delta t_c$&$\Delta\tau$&$ \Delta\mu $ &$\Delta \alpha_s$& $\Delta PT$ &$\Delta m_s$ & $\Delta m_c$&$\Delta \bar\psi\psi$&$\Delta \kappa$&$\Delta G^2 $&$\Delta M_0^2$&$\Delta\bar\psi\psi^2$&$\Delta G^3 $&$\Delta OPE$&$\Delta M_G$& $\Delta f_G$&$\Delta M_{(G)_1}$&Values\\
\hline
\hline
Coupling $f_{G}$ [keV] \\
\cline{0-0} 
{\it Molecule} $(c\bar{d})(\bar{c}u)$\\
$D^*D$&11.0&0.10&0.50&2.10&0.03&--&1.25&1.90&--&0.0&1.15&5.60&0.0&6.20&3.12&--&--&140(15)\\
$D^{*}_{0}D_1$&9.80&0.20&0.14&0.80&17.0&--&0.70&4.30&--&0.14&1.60&7.60&0.06&7.30&2.08&--&--&96(23)\\
{\it Tetraquark} $(\bar{c}\bar{d})(cu)$\\
$A_{cd}$&12.6&0.10&0.60&2.90&0.47&--&1.60&2.80&--&0.0&1.80&6.60&0.0&8.16&3.11&--&--&173(17) \\
\hline
 {\it Molecule} $(c\bar{s})(\bar{c}u)$\\
$D^{*}_{s}D$ &12.4&0.10&0.50&2.40&0.07&0.0&2.0&1.90&3.20&0.0&0.90&4.30&0.0&4.80&1.93&--&--&130(15)\\
$D^*D_{s}$ &12.2&0.10&0.50&2.40&0.11&0.07&1.40&1.70&3.20&0.02&1.30&4.40&0.03&7.0&2.19&--&--&133(16)\\
$D^{*}_{s0}D_1$&9.0&0.20&0.45&0.53&18.0&0.48&0.85&4.10&1.12&0.36&1.65&6.60&0.33&6.90&1.97&--&--&86(23)\\
$D^{*}_{0}D_{s1}$&8.60&0.20&0.15&0.86&17.5&0.20&0.75&3.60&1.60&0.02&1.80&6.50&0.07&6.90&1.95&--&--&89(22)\\
{\it Tetraquark} $(\bar{c}\bar{s})(cu)$\\
$A_{cs}$&13.2&0.12&0.60&2.80&0.01&0.20&1.60&1.90&3.70&0.0&2.0&5.40&0.0&6.70&2.34&--&--&148(17) \\
\hline
 {\it Molecule} $(c\bar{s})(\bar{c}s)$\\
$D^{*}_{s}D_s$&9.60&0.11&0.40&2.40&1.40&0.10&1.30&2.0&5.90&0.0&1.10&3.50&0.05&2.40&3.35&--&--&114(13)\\
$D^{*}_{s0}D_{s1}$&9.60&0.20&0.13&0.80&5.90&0.41&0.75&4.35&1.73&0.05&1.58&5.50&0.08&3.50&2.31&--&--&79(14)\\
{\it Tetraquark} $(\bar{c}\bar{s})(cs)$\\
$A_{css}$&10.9&0.14&0.50&2.80&0.90&0.25&1.40&2.10&6.90&0.0&2.0&4.40&0.10&5.20&2.40&--&--&114(15) \\
\hline
\hline
Mass $M_{G}$ [MeV]\\
\cline{0-0}
{\it Molecule} $(c\bar{d})(\bar{c}u)$\\
$D^*D$&15.0&40.0&2.0&6.30&0.0&--&2.80&13.5&--&0.0&5.30&11.0&0.0&39.0&--&--&--&3912(61)\\
$D^{*}_{0}D_1$&1.60&105&4.50&9.50&2.50&--&7.10&50.0&--&1.80&12.0&40.0&0.30&39.5&--&--&--&4023(130)\\
{\it Tetraquark} $(\bar{c}\bar{d})(cu)$\\
$A_{cd}$&8.30&38.0&1.90&7.50&1.90&--&2.90&10.0&--&0.06&4.90&7.10&0.0&39.5&--&--&--&3889(58) \\
\hline
 {\it Molecule} $(c\bar{s})(\bar{c}u)$\\
$D^{*}_{s}D$ &5.70&37.8&2.50&7.50&0.23&2.50&3.0&8.0&15.0&0.0&7.0&10.0&0.0&25.4&--&--&--&3986(51)\\
$D^*D_{s}$ &1.50&37.6&2.0&7.0&0.03&2.70&2.30&8.80&17.0&0.06&10.0&6.20&0.18&33.8&--&--&--&3979(56)\\
$D^{*}_{s0}D_1$&13.1&105&4.25&6.25&1.30&3.75&6.50&20.0&44.0&0.20&18.0&20.9&0.23&57.5&--&--&--&4064(133)\\
$D^{*}_{0}D_{s1}$&18.8&102&4.0&8.80&0.28&2.80&5.30&30.3&51.2&0.06&12.6&19.3&0.16&52.8&--&--&--&4070(133)\\
{\it tetraquark} $(\bar{c}\bar{s})(cu)$\\
$A_{cs}$&8.50&38.8&2.0&6.90&0.03&2.20&2.40&7.60&16.0&0.07&7.50&6.70&0.0&33.5&--&--&--&3950(56) \\
\hline
{\it Molecule} $(c\bar{s})(\bar{c}s)$\\
$D^{*}_{s}D_s$&0.30&42.5&2.40&6.0&12.0&6.20&2.10&6.70&15.3&0.05&9.10&13.8&0.05&25.7&--&--&--&4091(57)\\
$D^{*}_{s0}D_{s1}$&0.10&108&2.75&4.30&0.05&5.0&10.3&15.0&30.5&0.25&15.0&30.3&0.38&50.1&--&--&--&4198(129)\\
{\it Tetraquark} $(\bar{c}\bar{s})(cs)$\\
$A_{css}$&1.60&44.0&15.0&15.0&0.06&16.0&13.0&23.0&16.0&0.15&12.3&17.0&0.08&43.7&--&--&--&4014(77) \\
\hline\hline
\end{tabular*}
{\scriptsize
 \caption{Sources of errors and predictions from LSR at NLO for the couplings and masses of the molecules and tetraquark ground states. The errors from the QCD input parameters are from Table\,\ref{tab:param}. We take $|\Delta \tau|= 0.01$ GeV$^{-2}$ and  $\Delta \mu=0.05$ GeV.  The quoted errors are the mean from the two extremal values of $t_c$ delimiting the stability region quoted in Table\,\ref{tab:lsr-param}. 
 }
 \label{tab:res}
}
}
\end{table*}
\begin{table*}[hbt]
\setlength{\tabcolsep}{0.pc}
\catcode`?=\active \def?{\kern\digitwidth}
{\scriptsize
\begin{tabular*}{\textwidth}{@{}l@{\extracolsep{\fill}}|cccccc  cc  ccc}
\hline
\hline
      States         & \multicolumn{11}{c}{ ($1^+$) ground states}\\ 

\hline
Parameters          & \multicolumn{1}{c}{$D^*D$} 
       & \multicolumn{1}{c}{$D^{*}_{s}D$} 
                 & \multicolumn{1}{c}{{$D^*D_{s}$}} 
                 & \multicolumn{1}{c}{{$D^{*}_{s}D_{s}$ }} 
              & \multicolumn{1}{c}{$D^{*}_{0}D_{1}$} 
       & \multicolumn{1}{c}{$D^{*}_{s0}D_{1}$} 
                 & \multicolumn{1}{c}{{$D^{*}_{0}D_{s1}$}} 
                 & \multicolumn{1}{c}{{$D^{*}_{s0}D_{s1}$}} 
                  & \multicolumn{1}{c}{$A_{cd}$} 
       & \multicolumn{1}{c}{$A_{cs}$}
            & \multicolumn{1}{c}{{$~A_{css}$}} 
                                \\
\hline
$t_c$ [GeV$^2$]&22 -- 38&22 -- 38&22 -- 38&24 -- 40&28 -- 40&28 -- 44&28 -- 44&28 -- 44&22 -- 38&22 -- 8&24 -- 40 \\
$\tau$ [GeV]$^{-2} 10^2$&23 ; 35&22 ; 36&24 ; 38&24 ; 36&21 ; 30&20 ; 30&23 ; 31&20 ; 30&25 ; 37&25 ; 38&27 ; 38 \\
\hline\hline
\end{tabular*}
}
 \caption{Values of the set of the LSR parameters $(t_c,\tau)$ at the optimization region for the PT series up to NLO and for the OPE truncated at the dimension-six condensates and for $\mu=4.65$ GeV.}
  \vspace*{-0.25cm}
\label{tab:lsr-param}
\end{table*}

\section{Revisiting $f_{D^*_0D_1,A_{cd}}$ and $M_{D^*_0D_1,A_{cd}}$}
Here, we also revisit the estimate of the $D^*_0D_1, A_{cd}$ masses and couplings done in\,\cite{MOLE16}. 
We repeat exactly the same procedure as in the previous section.
\subsection{$f_{D^*_0D_1}$ and $M_{D^*_0D_1}$}
  The $\tau$ and $t_c$-behaviours of the coupling and mass are similar to the previous case and will not be shown here. The $\tau$-stability region ranges  from $\tau=0.21$ GeV$^{-2}$ for $t_c=$ 28 GeV$^2$  (beginning of $\tau$-stability) to 0.30 GeV$^{-2}$ for $t_c$= 40 GeV$^2$ (beginning of $t_c$-stability). One can notice that the $\tau$-stability starts at a larger value of $t_c$ than the one $t_c=22$ GeV$^2$ for the case of $D^*D$ which will imply a larger value of $M_{D^*_0D_1}$ than of $M_{D^*D}$. 

\begin{figure}[hbt]
\vspace*{-0.25cm}
\begin{center}
\centerline {\hspace*{-7.5cm} \bf a) }
\includegraphics[width=7cm]{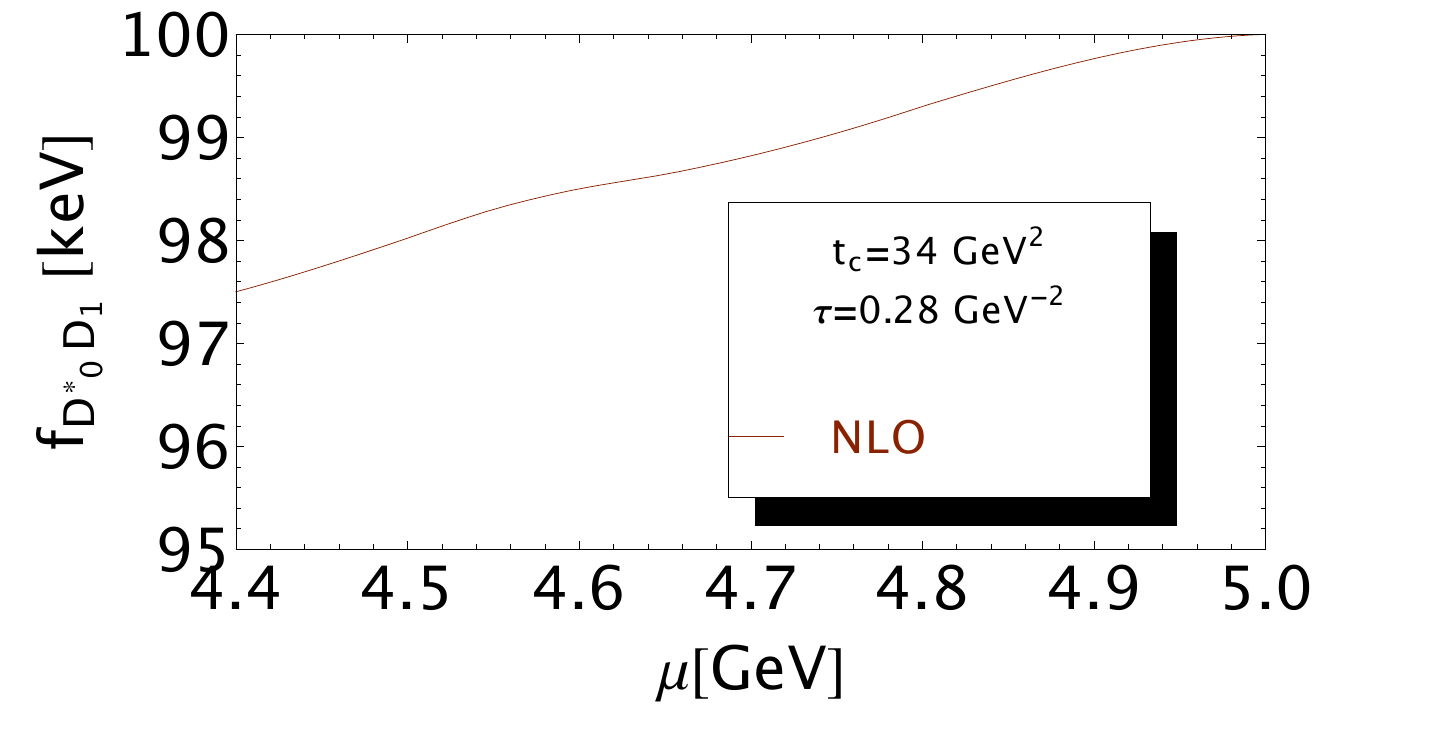}
\centerline {\hspace*{-7.5cm} \bf b) }
\includegraphics[width=7cm]{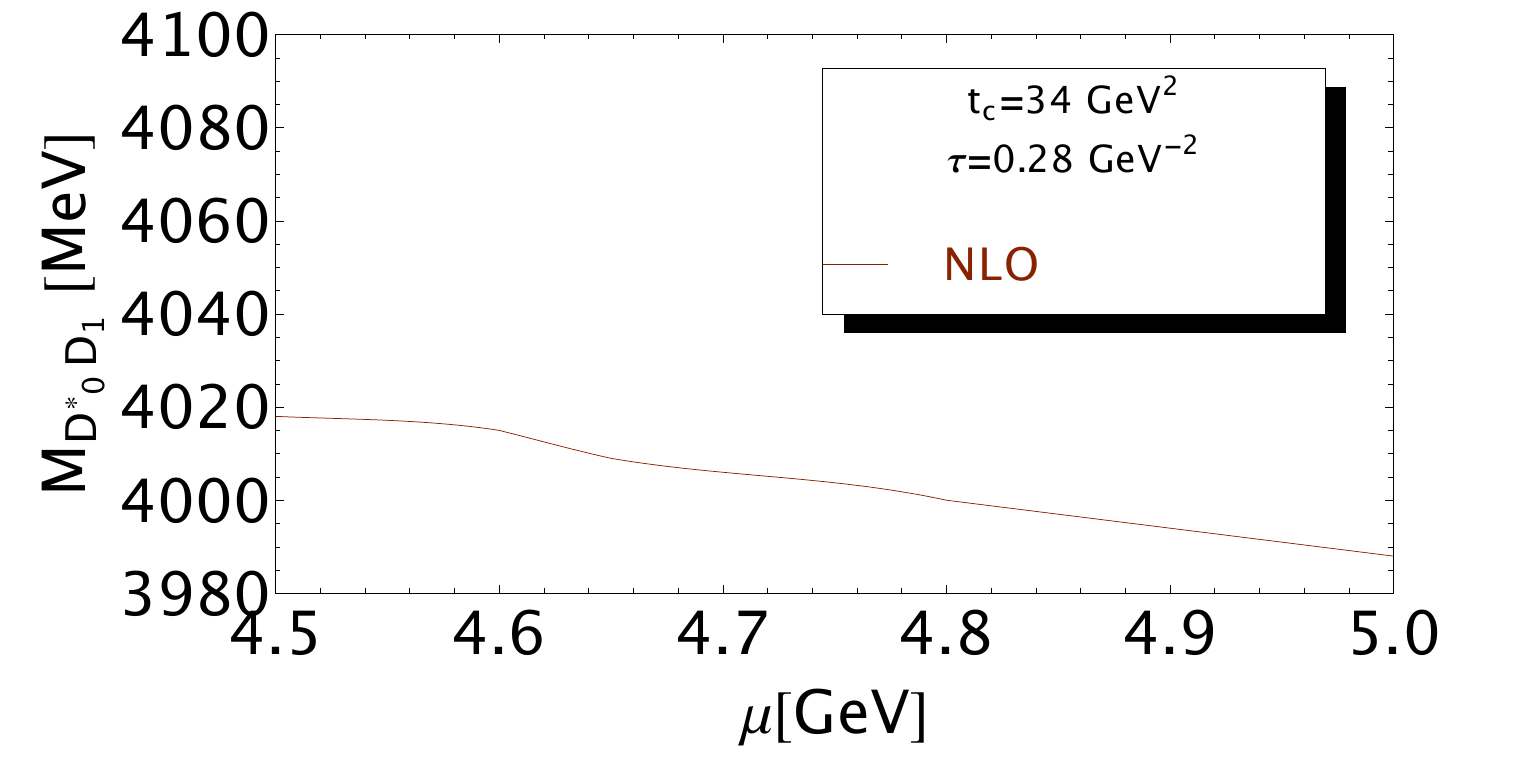}
\vspace*{-0.5cm}
\caption{\footnotesize  $f_{D^*_0D_1}$ and $M_{D^*_0D_1}$ as function of $\mu$ at NLO and for $t_c=30$ GeV$^2$.} 
\label{fig:dstar0d1-mu}
\end{center}
\vspace*{-1cm}
\end{figure}

 The $\mu$-stability is shown in Fig.\,\ref{fig:dstar0d1-mu} where the optimal value is the same as in Eq.\,\ref{eq:mu}. The result\,:
\beq
f_{D^*_0D_1}=96(23)~{\rm keV} ,~~~~M_{D^*_0D_1}=4023(130)~{\rm MeV},
\eeq
differs with the one given in\,\cite{MOLE16,SU3} (see Table\,\ref{tab:summary}) originated from the unprecise value of the $\tau$ used there for extracting the optimal value which affect in a sensible way the value of the mass in this channel. 
\subsection{$f_{A_{cd}}$ and $M_{A_{cd}}$}
 The $\tau$ and $t_c$-behaviours of the $A_{cd}$-mass and coupling are similar to the one in Fig\,\ref{fig:dstard}. 
The $\tau$-stability starts for $t_c$=22 GeV$^2$ at $\tau=0.25$ GeV$^{-2}$ while the $t_c$-stability starts at $t_c=$ 38 GeV$^2$ where $\tau=0.36$ GeV$^{-2}$. 

 The $\mu$-behaviours are the same as the ones of $D^*_0D_1$ where the $\mu$-stability is also at 4.65 GeV as in Eq.\,\ref{eq:mu}. 

 The estimates of the errors and the results for the coupling and mass :
\beq
f_{A_{cd}}=173(17)~{\rm keV}~~~~~~~~M_{A_{cd}} = 3889(58)~{\rm MeV}
\eeq
are collected in Tables\,\ref{tab:res} and\,\ref{tab:summary}. 
  The mass value is in good agreement with the one 3888(130) MeV obtained in Ref.\,\cite{MOLE16} but with a smaller error due to a better localisation of the $\tau$ and $\mu$ stability points. The $A_{cd}$ mass also coincides with the observed $Z_c(3900)$. The different sets of $(t_c,\tau)$ used to get the previous optimal results are summarized in Table\,\ref{tab:lsr-param}.
  \vspace*{-0.5cm}
\section{The $(1^+)$  $(c\bar s)(\bar cu)$ and $(\bar c\bar s)(cu)$states }
\subsection{New estimate of $f_{D^*_sD}$ and $M_{D^*_sD}$}

In this section, we present a new estimate of the $D^*_sD$ molecule mass and coupling.

\begin{figure}[hbt]
\vspace*{-0.25cm}
\begin{center}
\centerline {\hspace*{-7.5cm} \bf a) }
\includegraphics[width=6.5cm]{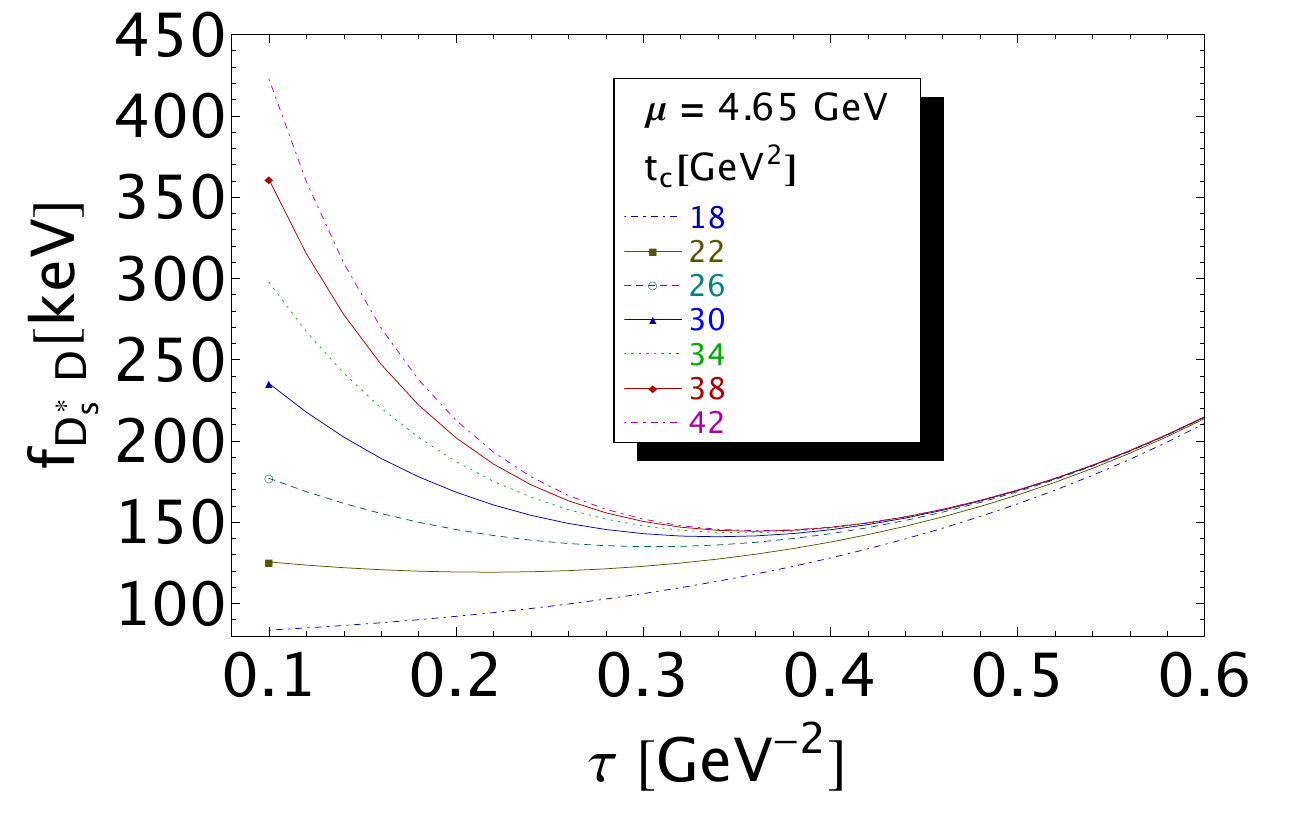}
\vspace{0.25cm}
\centerline {\hspace*{-7.5cm} \bf b) }
\includegraphics[width=6.5cm]{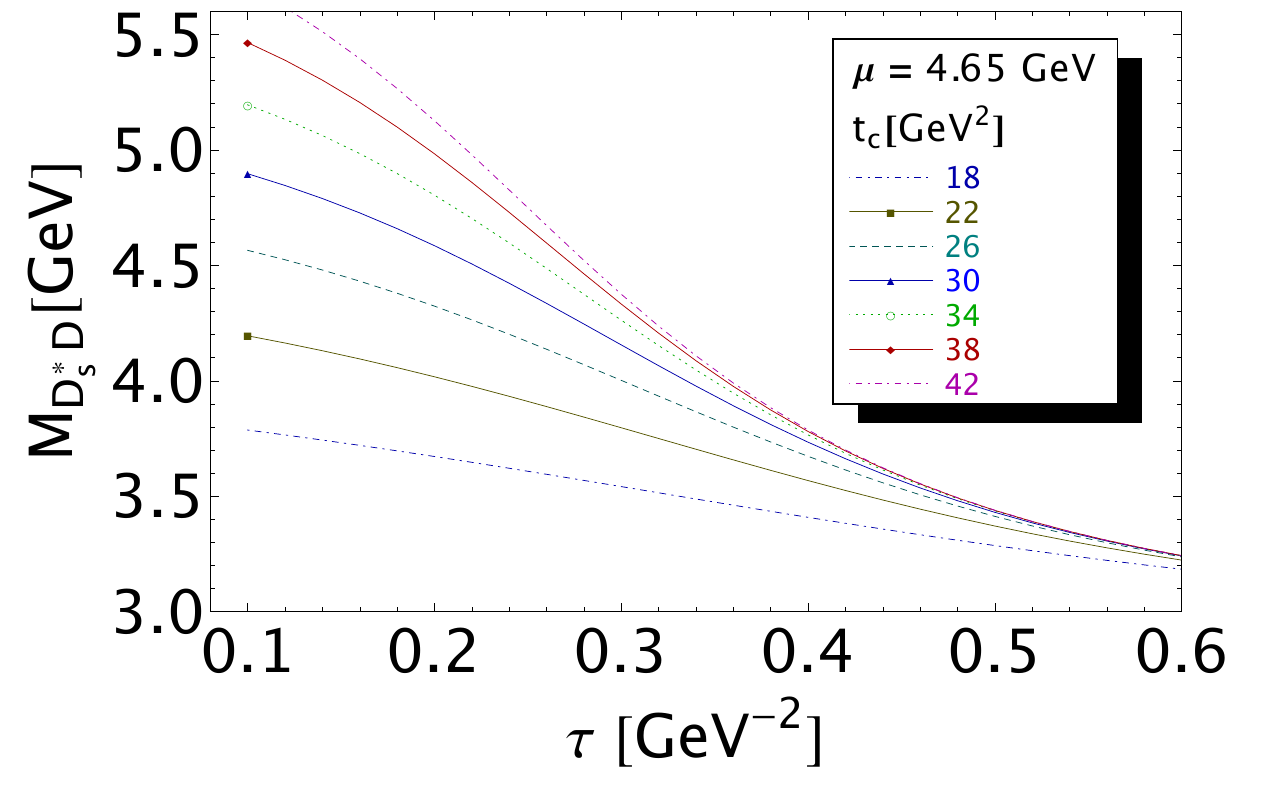}
\vspace*{-0.5cm}
\caption{\footnotesize  $f_{D^*_sD}$ and $M_{D^*_sD}$ as function of $\tau$ at NLO for different values of $t_c$, for $\mu$=4.65 GeV and for values of $\overline m_{c}(\overline m_{c})$ given in Table\,\ref{tab:param}.} 
\label{fig:dstarsd}
\end{center}
\vspace*{-0.25cm}
\end{figure} 
\begin{figure}[hbt]
\vspace*{-0.25cm}
\begin{center}
\centerline {\hspace*{-7.5cm} \bf a) }
\includegraphics[width=7.3cm]{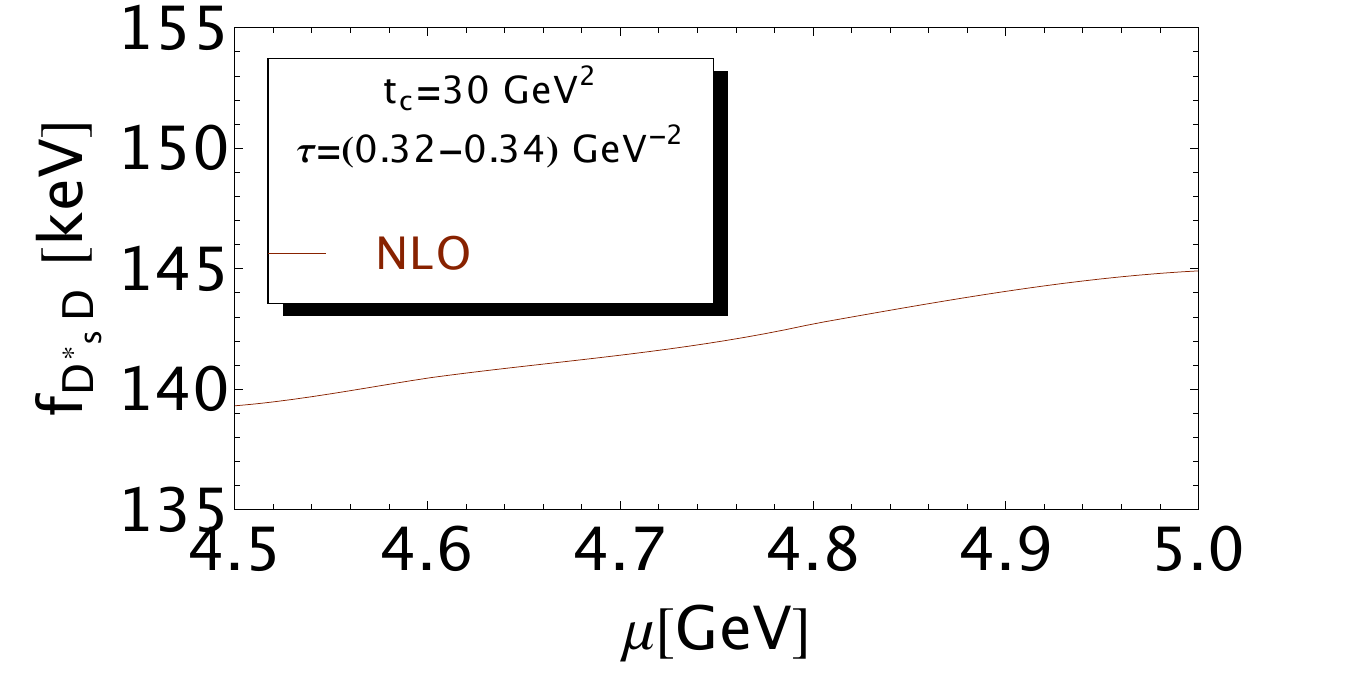}
\centerline {\hspace*{-7.5cm} \bf b) }
\includegraphics[width=7cm]{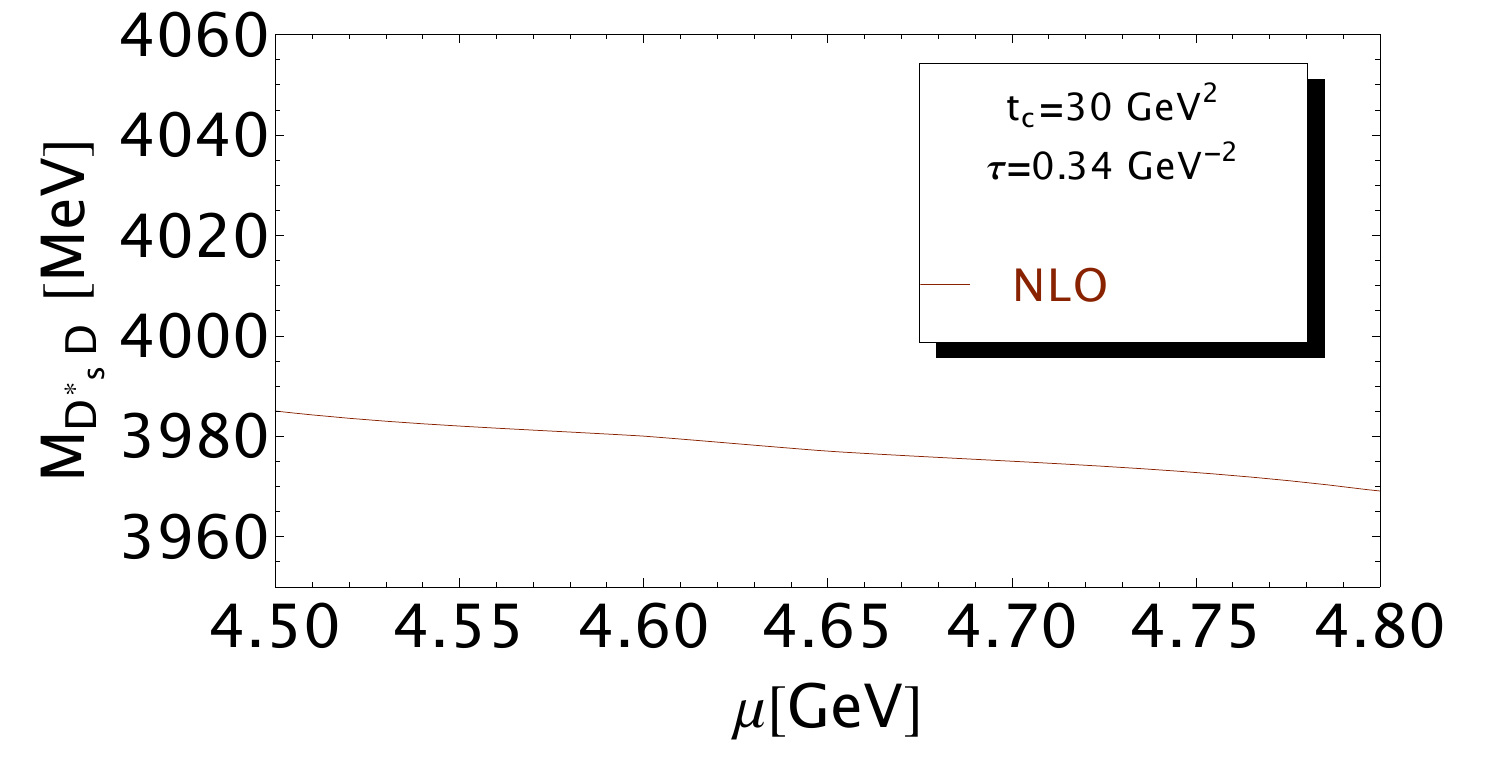}
\vspace*{-0.5cm}
\caption{\footnotesize  $f_{D^*_sD}$ and $M_{D^*_sD}$ as function of $\mu$ at NLO and for $t_c=30$ GeV$^2$.} 
\label{fig:dstarsd-mu}
\end{center}
\vspace*{-0.25cm}
\end{figure} 
\begin{figure}[hbt]
\vspace*{-0.25cm}
\begin{center}
\centerline {\hspace*{-7.5cm} \bf a) }
\includegraphics[width=7cm]{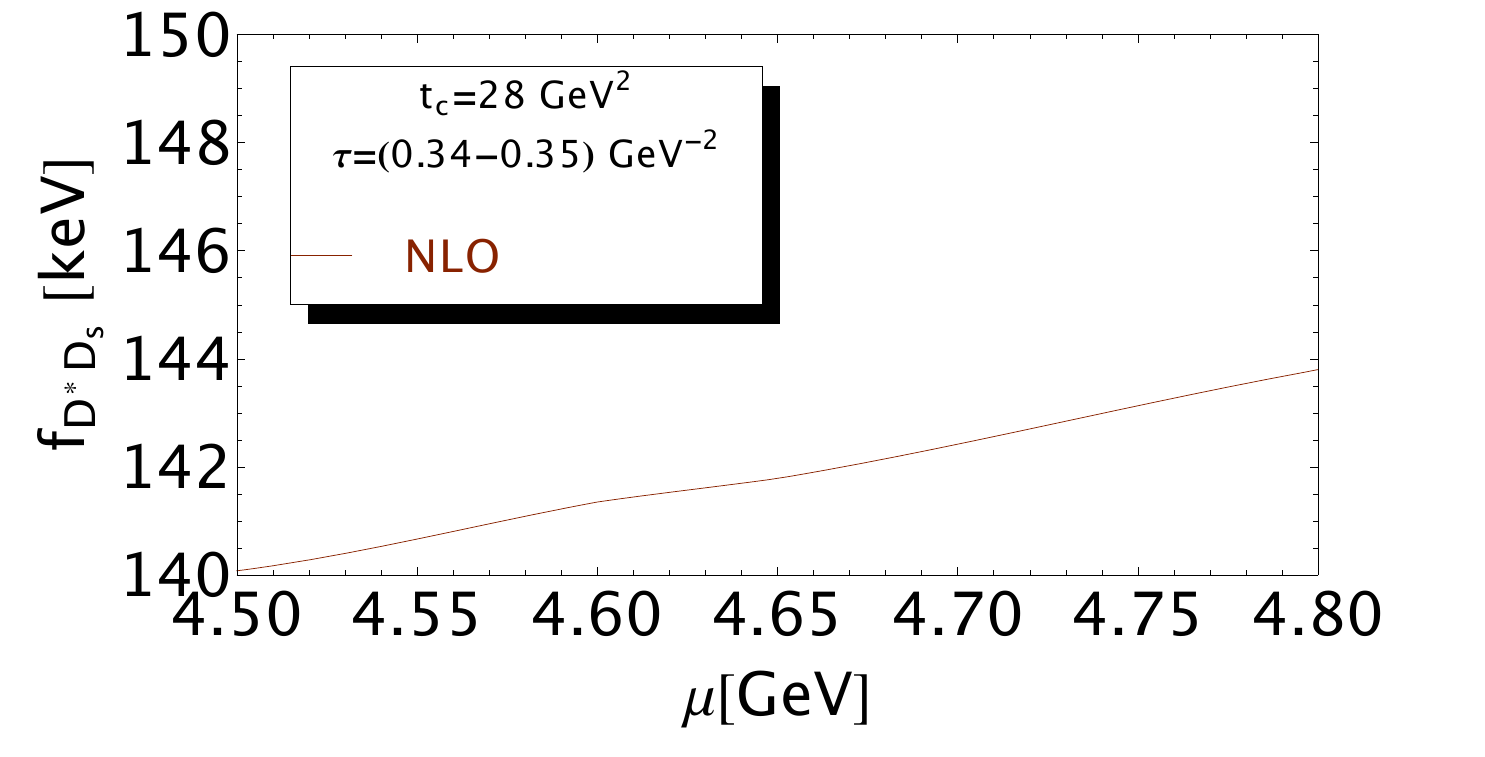}
\centerline {\hspace*{-7.5cm} \bf b) }
\includegraphics[width=7cm]{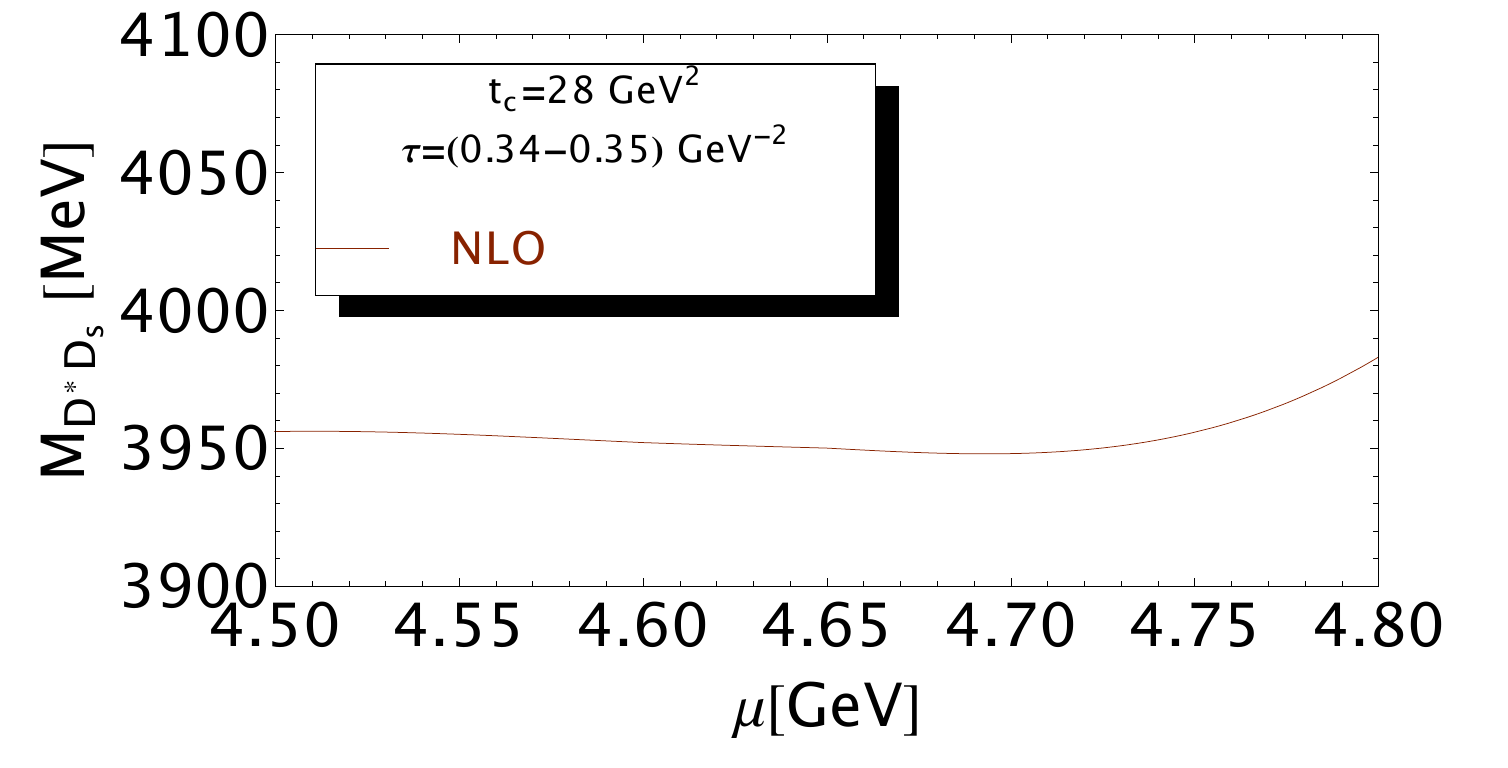}
\vspace*{-0.5cm}
\caption{\footnotesize  $f_{D^*D_s}$ and $M_{D^*D_s}$ as function of $\mu$ at NLO and for $t_c=28$ GeV$^2$.} 
\label{fig:dstards-mu}
\end{center}
\vspace*{-0.25cm}
\end{figure} 
   The $\tau$ and $t_c$-behaviours of the coupling and mass are also similar to the previous cases as shown in Fig.\,\ref{fig:dstarsd}. 

  The $\tau$-stability region ranges  from $\tau=0.22$ GeV$^{-2}$ for $t_c=$ 22 GeV$^2$ (beginning of $\tau$-stability) until 0.36 GeV$^{-2}$ for 38 GeV$^2$ (beginning of $t_c$-stability).

 The $\mu$-stability is shown in Fig.\,\ref{fig:dstarsd-mu} where the optimal value is the same as in Eq.\,\ref{eq:mu}.

 The sources of the errors and the results are quoted in Table\,\ref{tab:res} and \ref{tab:summary}  .  One can notice that the values of the $\tau,~t_c$ and $\mu$-stabilities are about the same as in the case of $D^*D$ indicating a good $SU(3)$ symmetry for the sum rule parameters. The obtained values quoted in Tables\,\ref{tab:res} and\,\ref{tab:summary}\,:
\beq
f_{D^*_sD}= 130(15)~{\rm keV},~~~~M_{D^*_sD}=3986(51)~{\rm MeV}
\eeq
also indicate small $SU(3)$ breakings of chiral  symmetry which is about $-7.1\%$ for the coupling  and $+1.9\%$ for the mass. 
\subsection{New estimate of $f_{D^*D_s}$ and $M_{D^*D_s}$}
A similar analysis is done for extracting $f_{D^*D_s}$ and $M_{D^*D_s}$. The $\tau$ and $t_c$ behaviours of the results are similar to the previous ones. The $\mu$-behaviour is  is shown in Fig. 18.  We obtain :
\beq
f_{D^*D_s}=133(16)~{\rm keV},~~~~M_{D^*D_s}=3979(56)\,{\rm MeV},
\eeq
where the values of the parameters are about the same as the ones of $D^*_sD$ as intuitively expected. They are quoted in Tables\,\ref{tab:res} and\,\ref{tab:summary}. 
\subsection{The $D^*_{s0}D_{1}$ and $D^*_0D_{s1}$ molecules}
Similar analysis leads to (see Tables\,\ref{tab:res} and\,\ref{tab:summary})\,:
\bea
&\hspace*{-0.3cm}f_{D^*_{s0}D_1}=86(23)~{\rm keV},~M_{D^*_{s0}D_1}=4064(133)\,{\rm MeV},\nnb\\
&\hspace*{-0.3cm}f_{D^*_{0}D_{s1}}=89(23)~{\rm keV},~M_{D^*_{0}D_{s1}}=4070(133)\,{\rm MeV},
\eea
where one can notice that the two molecule states are almost degenerated and have the same couplings to the currents.
\subsection{The $A_{cs}$ tetraquark}
We pursue the previous analysis for the $A_{cs}$ tetraquark. The behaviours of the different curves are similar to the previous ones and will not be shown. The result quoted in Tables\,\ref{tab:res} and\,\ref{tab:summary} is:
\beq
f_{A_{cs}}=148(17)~{\rm keV},~~~~~M_{A_{cs}}=3950(56)\,{\rm MeV}~, 
\eeq
where one can notice that it is almost degenerated to the $D^*D_s$ and $D^*_sD$ and has almost the same couplings. 
\section{The $(1^+)$  $(c\bar s)(\bar cs)$ and $(\bar c\bar s)(cs)$ states }
In this section, we revisit and improve our previous estimate of the masses and couplings of these above-mentioned  states and give a new estimate of the $D^*_sD_s$ radial excitation mass and coupling. 
\subsection{The ${D^*_sD_s}$ molecule}
\begin{figure}[hbt]
\begin{center}
\centerline {\hspace*{-7.5cm} \bf a) }
\includegraphics[width=6.5cm]{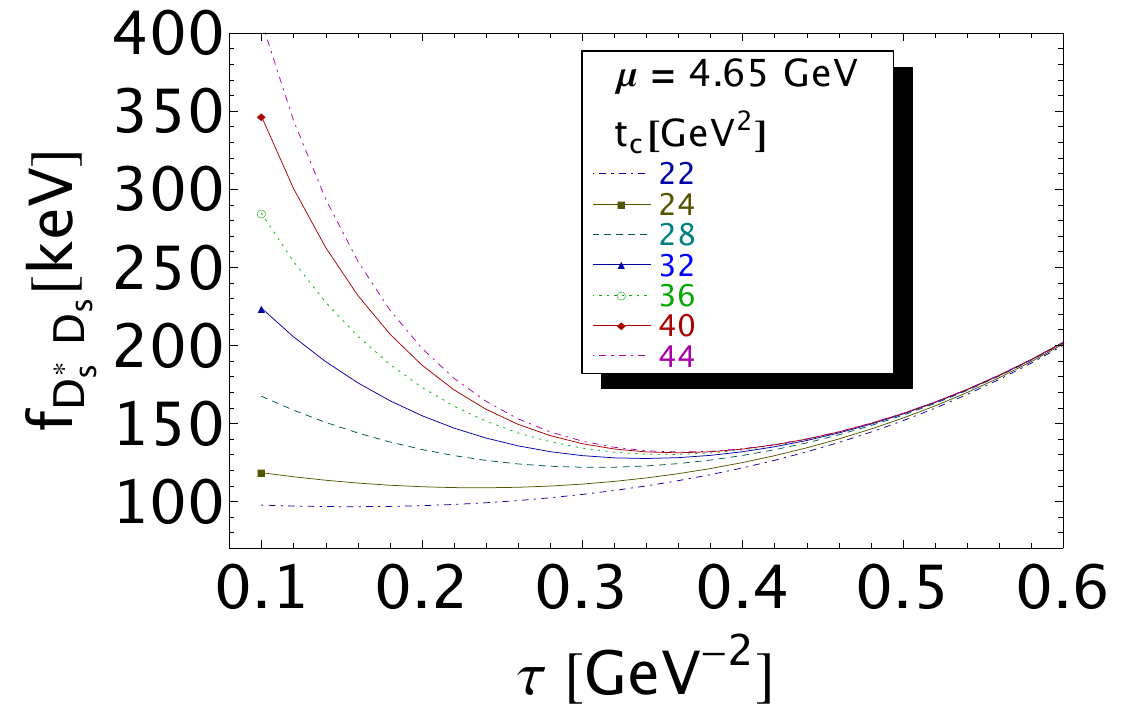}
\centerline {\hspace*{-7.5cm} \bf b) }
\hspace*{0.75cm}\includegraphics[width=6.5cm]{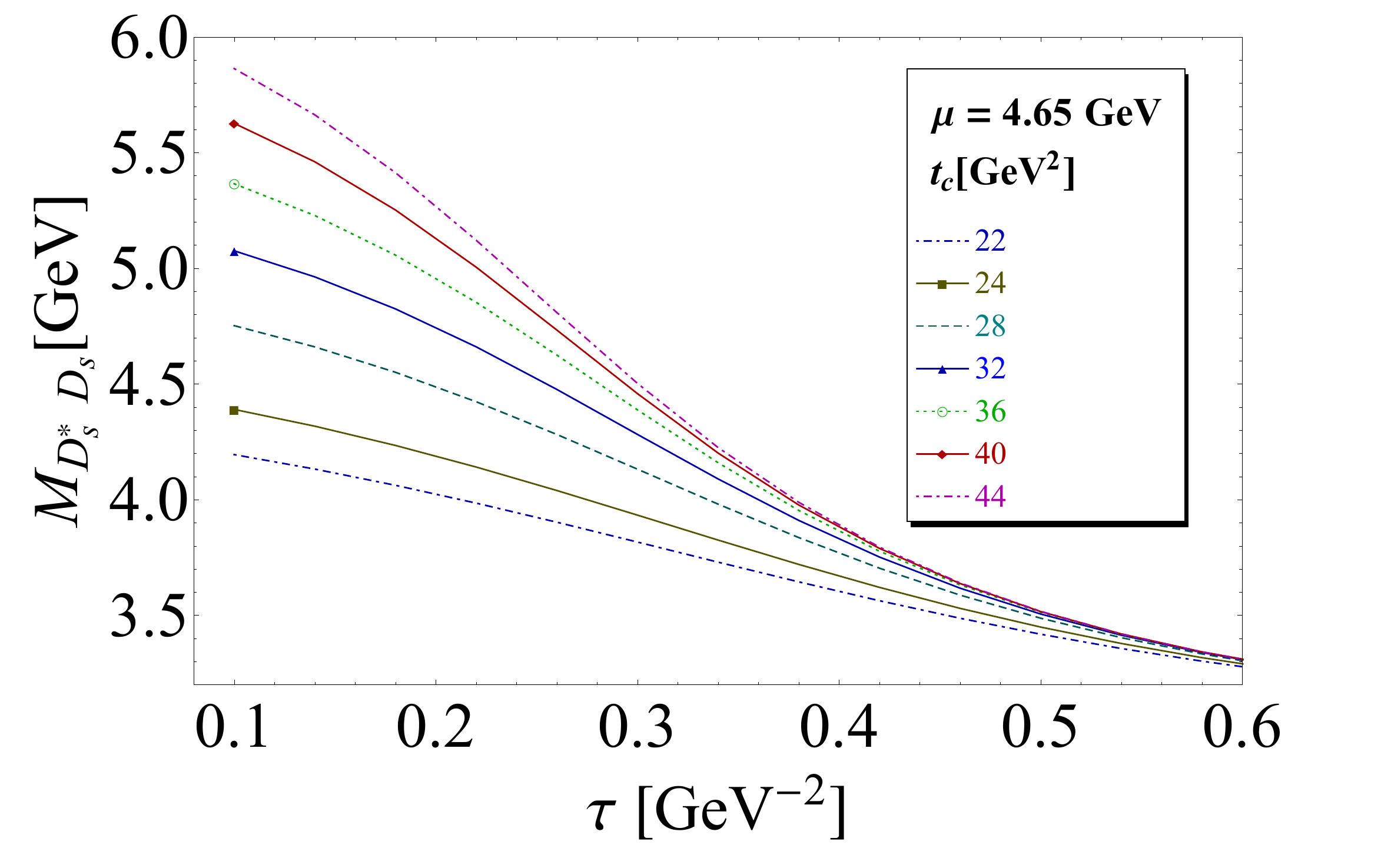}
\vspace*{-0.5cm}
\caption{\footnotesize  $f_{D^*_sD_s}$ and $M_{D^*_sD_s}$ as function of $\tau$ at NLO for different values of $t_c$, for $\mu$=4.65 GeV and for values of $\overline m_{c}(\overline m_{c})$ given in Table\,\ref{tab:param}.} 
\label{fig:dstarsds}
\end{center}
\vspace*{-0.25cm}
\end{figure}

\begin{figure}[hbt]
\vspace*{-0.25cm}
\begin{center}
\centerline {\hspace*{-7.5cm} \bf a) }
\includegraphics[width=7cm]{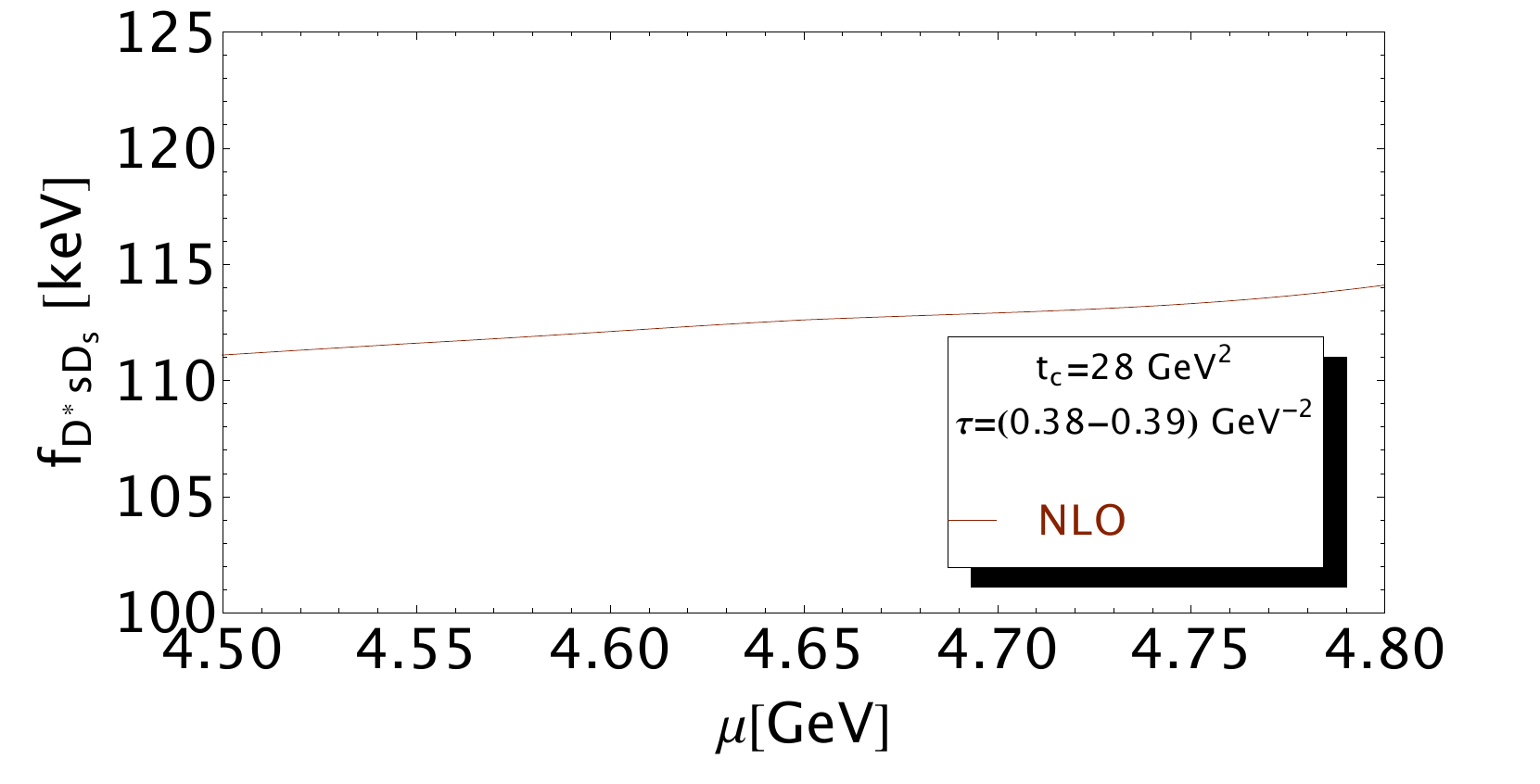}
\vspace{0.25cm}
\centerline {\hspace*{-7.5cm} \bf b) }
\includegraphics[width=7cm]{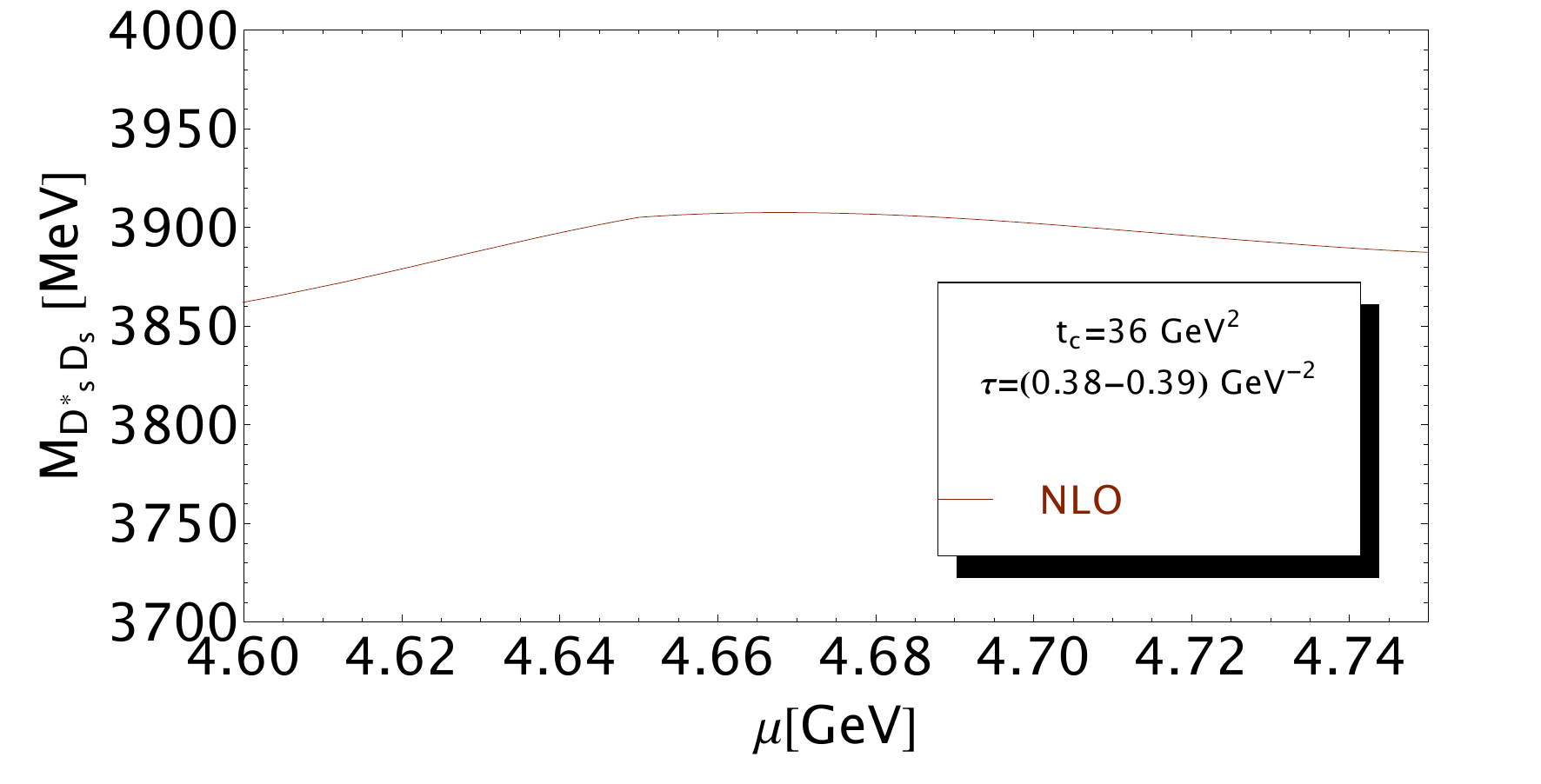}
\vspace*{-0.5cm}
\caption{\footnotesize  $f_{D^*_sD_s}$ and $M_{D^*_sD_s}$ as function of $\mu$ at NLO and for $t_c=36$ GeV$^2$.} 
\label{fig:dstarsds-mu}
\end{center}
\vspace*{-0.25cm}
\end{figure}

The analysis  of the $\tau$ and $t_c$ behaviours is shown in Fig.\,\ref{fig:dstarsds}. The $\mu$-behaviour is shown in Fig.\,\ref{fig:dstarsds-mu}. These behaviours are similar to the previous ones.  The $(\tau,t_c)$ stabilities are obtained for $t_c$ inside the range 24 to 40 GeV$^2$. We deduce (see Tables\,\ref{tab:res} and\,\ref{tab:summary})\,:
\beq 
f_{D^*_sD_s}=114(13)~{\rm keV},~~~M_{D^*_sD_s}=4091(57)\,{\rm MeV}, 
\eeq
One can notice the (almost) similar effect of $SU(3)$ breakings than in the previous cases. The mass increases by 105 MeV compared to the one of the $D^*_sD$ state which is about the one 74 MeV from $D^*D$ to $D^*_sD$. 

\subsection{The $D^*_{s0}D_{s1}$ molecule and $A_{css}$ tetraquark}
In this section, we revise and check the results obtained in\,\cite{SU3}. The behaviours of the $\tau,t_c$ and $\mu$-behaviours of the masses and couplings are similar to the previous cases and will not be shown.

 For the $D^*_{s0}D_{s1}$ molecule, the set of $(\tau,t_c)$-stabilities are obtained from (0.20,28) to (0.30,44) in units of (GeV$^{-2}$, GeV$^2$), where one notice the sensitivity of the results in the change of $\tau$ which is quantified by the large error
induced by the variation of $\tau$ as shown in Table\,\ref{tab:res}. One obtains from Tables\,\ref{tab:res} and\,\ref{tab:summary} :
\beq 
f_{D^*_{s0}D_{s1}}=79(14)~{\rm keV},~~~M_{D^*_{s0}D_{s1}}=4198(129)\,{\rm MeV}, 
\eeq
which agree within the errors  with the previous results in\,\cite{SU3}. 

 For the $A_{css}$ tetraquark, $\tau$-stability at 0.27 GeV$^{-2}$ starts for $t_c=24$ GeV$^2$ while the $t_c$-stability for $\tau=0.38$ GeV$^{-2}$
starts for $t_c=40$ GeV$^2$. These values are about the same as the ones for $A_{cd}$ and $A_{cs}$ indicating like in the case of molecule states a good $SU(3)$ symmetry for the values of the LSR parameters. The optimal results are given in Tables\, \ref{tab:res} and \,\ref{tab:summary} which read:
\beq 
f_{A_{css}}=114(16)~{\rm keV}~,~~~~~M_{A_{css}}=4014(86)\,{\rm MeV}~.
\label{eq:acss}
\eeq

Compared to the previous non-strange case where the molecule $D^*D$ is quasi-degenerated with the tetraquark $A_{cd}$, which leads us to conclude that the observed state is a tetramole ${\cal T}_c$, in this channel, we find the tetraquark $A_{css}$ is almost degenerated (within the errors) with the molecule $D^*_sD_s$ and has the same couplings implying that it can also be a tetramole ${\cal T}_{css}$ state with a coupling and mass:
\beq 
f_{{\cal T}_{css}}=114(10)~{\rm keV}~,~~~~~M_{{\cal T}_{css}}=4064(46)\,{\rm MeV}~.
\label{eq:facss}
\eeq
 The $D^*_{s0}D_{s1}$ molecule is about 100 MeV slightly higher but has a weaker coupling to the corresponding current. 
 \section{Two-meson scattering states\,\label{sec:scatt}}
To study this contribution, we take the example of the $D^*D$ molecule\,\footnote{Our conclusion remains valid for some other molecules and tetraquarks states.}.

We saturate the 
RHS of Fig.\,\ref{fig:conv} by the two non-resonant (scattering) states $D^*$ and $D$ which we shall compare with the one due to the molecule $D^*D$ (LHS) appropriately matched with the $k^2$-factor. 

Then, one obtains the two-resonance scattering LSR moment :
\bea
{\cal L}^c_0({D^*\oplus D}) &\equiv& 2 \ga\frac{k}{4\pi}\dr^2 f^2_{D^*}\ga\frac{f_{D}}{m_c}\dr^2 {\cal I}(t,M_D^2,M_{D^*}^2)\nnb\\
&\simeq& 1.16\times 10^{-6} {\rm GeV}^{10}, 
\eea
where ${\cal I}(t,M_D^2,M_{D^*}^2)$ is the integral (see Eq.\,\ref{eq:qqcurrent}):
\beq
{\cal I}\equiv\int_{(M_D+M_{D^*})^2}^{t_c} \hspace*{-1.4cm}dt\,t^2\,{\rm e}^{-t\tau}\int_{{M_D}^2}^{(\sqrt{t}-M_{D^*})^2} \hspace*{-1.4cm} dt_1\hspace*{0.6cm}
\int_{M_{D^*}^2}^{(\sqrt{t}-\sqrt{t_1})^2}\hspace*{-1.2cm}dt_2 \,\lambda^{3/2}\ga\frac{M_D^2}{t},\frac{M_{D^*}^2}{t}\dr,
\eeq
to be compared with the molecule sum rule result :
\beq
{\cal L}_0^c({D^*D}) \equiv f^2_{D^*D}M^8_{D^*D}\, {\rm e}^{-\tau_R M^2_{D^*D}}
\simeq  1.09\times 10^{-5} {\rm GeV}^{10},
\eeq
which we have evaluated at the stability point  $\tau_R\simeq 0.3$ GeV$^{-2}$.  We have used the previous values of the molecule parameters $f_{D^*D}$=140 keV and $M_{D^*D}$=3.91 GeV given in Table\,\ref{tab:res} and the average values of $f_{D}=204(6)$ MeV and  $f_{D^*}=250(8)$ MeV from\,\cite{SNFB15,SNFB15,SNB2}. We have neglected the equal and small contributions  to the two sum rules from the QCD continuum above the continuum threshold taken to be : $t_c\approx 30$ GeV$^2$ which is the mean of the two extremal values delimiting the stability region. 
These results indicate that : 
 
 -- For finite $N_c$, the non-resonant contribution is about one order of magnitude smaller than the one of the resonance molecule (a similar conclusion using an alternative approach has been reached in\,\cite{WANG1}) and disprove the claim of Ref.\,\cite{LUCHA} based on large $N_c$ once an appropriate matching of the two correlators via the $k^2$ factor is done. 
  
-- A posteriori,  the  existence of  the stability region or ``sum rule window" in the LSR analysis where one can extract the (postulated) resonance mass and coupling using the spectral function parametrization within the duality ansatz  ``one resonance" + QCD continuum is  a strong indication of the duality between the QCD-OPE and the phenomenological side of the sum rule,
where the resonace contribution is dominant  over the one of the non-resonant states. This fact does not also support the claim of Ref.\,\cite{LUCHA}. 

\section{Radial excitations}
We present in this section a new estimate of the couplings and masses  of the first radial excitations using the lowest moments ${\cal L}^c_0$ and ${\cal R}_0^c$\,\footnote{We note that the moment ${\cal R}_1^c$ which is expected to be more sensitive to the radial excitation contribution has a behaviour similar to ${\cal R}_0^c$ and will not presented here.}. In so doing, we use a ``two resonance" parametrization of the spectral function.
\subsection{The $(D^*D)_1$ first radial excitation}
\begin{figure}[hbt]
\vspace*{-0.25cm}
\begin{center}
\centerline {\hspace*{-7.5cm} \bf a) }
\includegraphics[width=7cm]{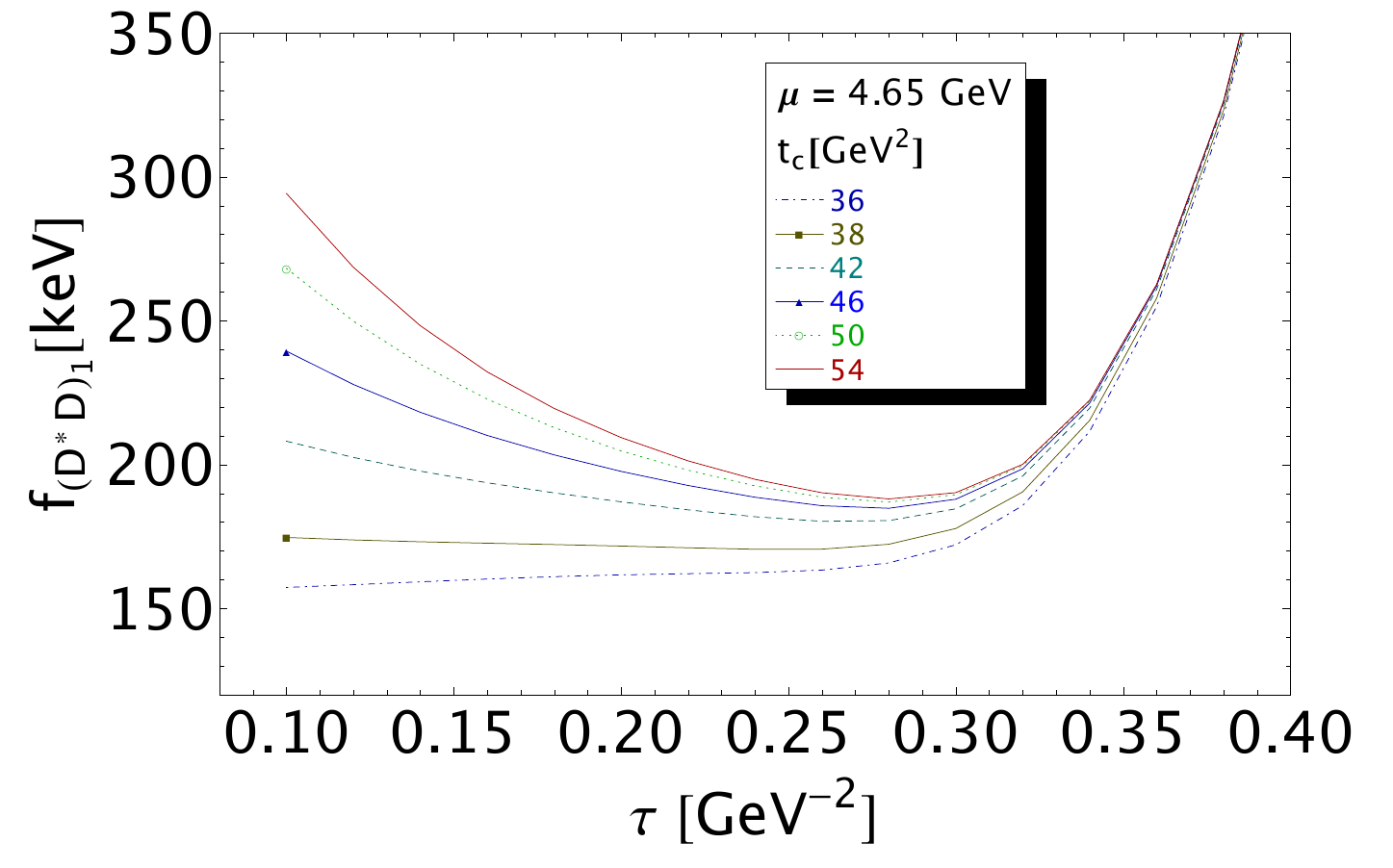}
\vspace{0.25cm}
\centerline {\hspace*{-7.5cm} \bf b) }
\includegraphics[width=7cm]{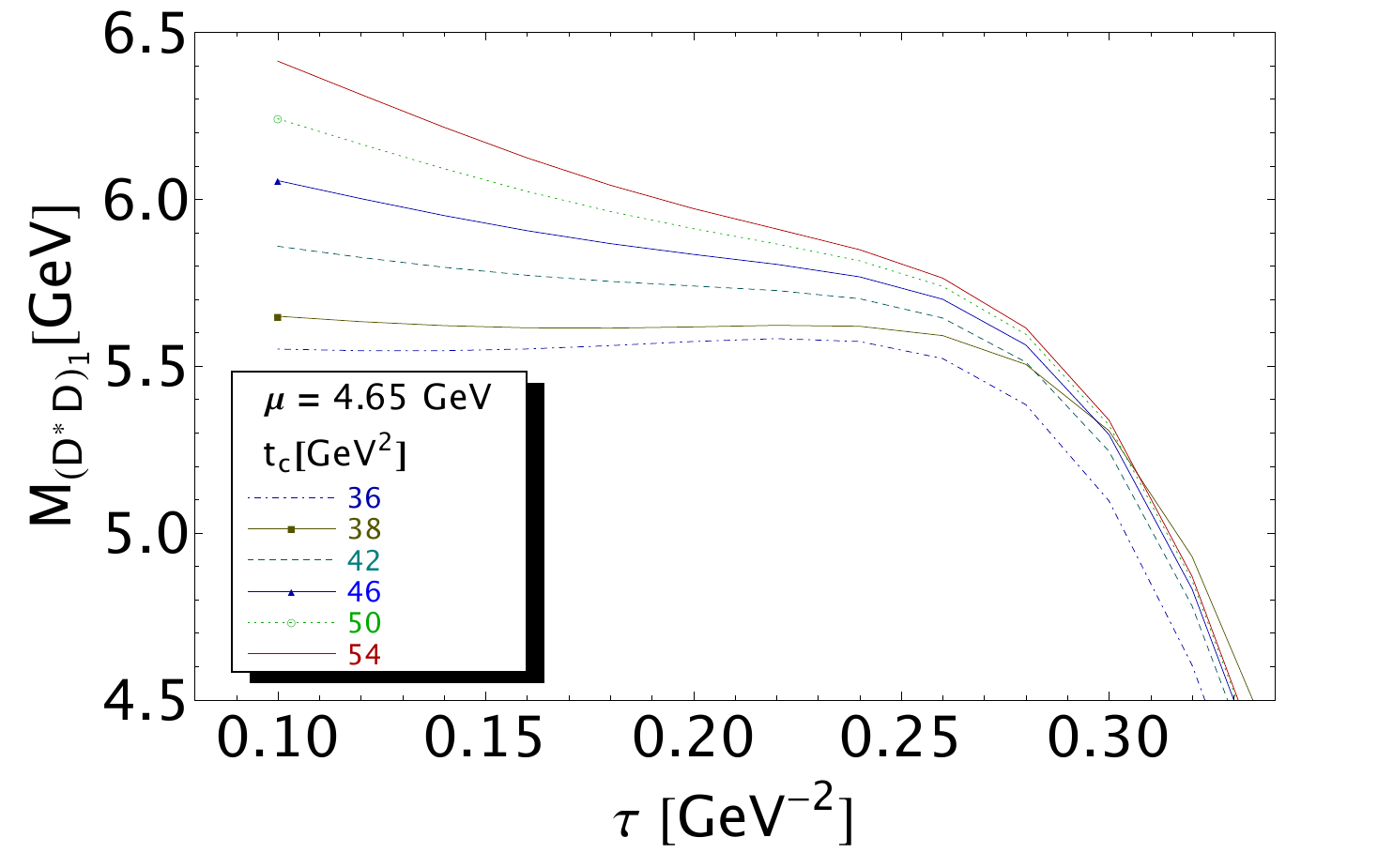}
\vspace*{-0.5cm}
\caption{\footnotesize  $f_{(D^*D)_1}$ and $M_{(D^*D)_1}$ as function of $\tau$ at NLO for different values of $t_c$, for $\mu$=4.65 GeV and for values of $\overline m_{c}(\overline m_{c})$ given in Table\,\ref{tab:param}.} 
\label{fig:radial}
\end{center}
\vspace*{-0.25cm}
\end{figure}

We insert the previously obtained values of the lowest ground state $D^*D$ mass and coupling and study the $\tau$\,- and $t_c$-stability of the sum rules by fixing the subtraction constant $\mu$ as in Eq.\,\ref{eq:mu}. The analysis is shown in Fig.\,\ref{fig:radial} where the stability region is delimited by $t_c$ = 39 and 50 GeV$^2$ to which corresponds respectively the $\tau$-stability of (0.14--0.18) and (0.28--0.29) GeV$^{-2}$.   The results\,:
\beq
f_ {(D^*D)_1}=197(25)~{\rm keV},~~M_ {(D^*D)_1}=5709(70)~{\rm MeV}
\label{eq:rad}
\eeq
are quoted in Table\,\ref{tab:radial}. The set of $(t_c,\tau)$ is compiled in Table\,\ref{tab:lsr-rad}. 

 One can notice that the mass value is roughly about the (expected) one of $\sqrt{t_c}\simeq (4.7-6.2)$ GeV inside the stability region where the lowest ground state mass has been extracted. 

\vspace*{-0.25cm}
\begin{table*}[hbt]
\setlength{\tabcolsep}{0.15pc}
\catcode`?=\active \def?{\kern\digitwidth}
    {\scriptsize
  \begin{tabular*}{\textwidth}{@{}l@{\extracolsep{\fill}}|ccccccccc   c ccccccc  l}
\hline
\hline
 Observables\,&$ \Delta t_c$&$\Delta\tau$&$ \Delta\mu $ &$\Delta \alpha_s$& $\Delta PT$ &$\Delta m_s$ & $\Delta m_c$&$\Delta \bar\psi\psi$&$\Delta \kappa$&$\Delta  G^2 $&$\Delta M_0^2$&$\Delta\bar\psi\psi^2$&$\Delta G^3 $&$\Delta OPE$&$\Delta M_G$& $\Delta f_G$&$\Delta M_{(G)_1}$&Values\\
\hline
\hline
Coupling $f_{G}$ [keV] \\
\cline{0-0} 
$(D^*D)_0$&21&0.7&2.9&11&6.2&--&8.5&8.2&--&0.09&4.5&26.8&0.10&19&14.5&34&--&46(56)\\
\\
$(D^*D)_1$&10.0&0.18&1.35&8.5&0.60&--&5.60&9.30&--&0.06&3.70&7.40&0.07&3.71&1.80&15.1&10.1&197(25)\\
$(D^*_0D_1)_1$&5.2&0.5&0.4&4.3&9.3&--&4.7&19.25&--&0.10&6.30&10.5&0.28&9.10&0.45&16.5&26.0&238(41)\\
$(A_{cd})_1$&18.9&0.12&2.29&11.8&9.60&--&7.31&9.49&--&0.04&6.04&7.43&0.15&6.41&4.05&15.1&17.2&272(38)\\
\\
$(D^{*}_{s}D)_1$ &9.0&0.20&1.50&8.8&0.20&0.95&4.90&8.65&10.5&0.06&2.20&6.9&0.08&6.1&1.75&16.8&8.70&199(29)\\
$(D^{*}_{s}D_s)_1$&6.5&0.60&1.80&9.5&4.9&1.5&4.6&11.7&13.0&0.18&4.80&10.0&0.20&12.4&7.30&16.2&27.1&197(43)\\
\hline
\hline
Mass $M_{G}$ [MeV]\\
\cline{0-0}
$(D^*D)_1$&30.0&38.9&11.8&5.0&0.02&--&16.5&10.0&--&1.10&29.5&6.0&1.15&23.0&18.0&12&--&5709(70)\\
$(D^*_0D_1)_1$&83.0&2.5&0.4&10&14&--&11.8&55.2&--&0.2&34&74.6&0.7&50.6&20.8&55&--&6375(152)\\
$(A_{cd})_1$&59.0&15.2&9.46&6.54&3.81&--&19.1&26.4&--&0.15&22.5&4.96&0.80&14.3&20.7&26.5&--&5717(82)\\
\\
$(D^{*}_{s}D)_1$ &2.0&30.0&16.5&7.5&0.02&2.0&18.5&13.0&6.0&0.85&20.5&4.0&1.40&10.0&19.0&10&--&5725(52)\\
$(D^{*}_{s}D_s)_1$&42.0&49.0&39.5&42.5&1.05&4.5&20.5&38.5&45.5&1.0&35.0&58.0&3.0&40.8&25.5&69.0&--&5786(152)\\
\hline\hline
\end{tabular*}
{\scriptsize
 \caption{Sources of errors and predictions from LSR at NLO and  for the decay constants and masses of the molecules and tetraquark radial excitation states. The errors from the QCD input parameters are from Table\,\ref{tab:param}. We take $|\Delta \tau|= 0.01$ GeV$^{-2}$ and  $\Delta \mu=0.05$ GeV. The quoted errors have been estimated fixing $t_c$ at the mean of the two extremal values delimiting the stability region quoted in Table\,\ref{tab:lsr-rad}. We notice that the relative large values of some individual errors in the case of ${D^*_sD_s}$ and $(D^*_0D_1)_1$ are mainly induced by the shift of the position of the minima compared to the one of the central values. 
 }
 \label{tab:radial}
}
}
\end{table*}
\begin{table*}[hbt]
\setlength{\tabcolsep}{0.pc}
\catcode`?=\active \def?{\kern\digitwidth}
{\scriptsize
\begin{tabular*}{\textwidth}{@{}l@{\extracolsep{\fill}}|ccccccc  }
\hline
\hline
      States         & \multicolumn{6}{c}{ ($1^+$) radial excitations}\\ 

\hline
Parameters          
 & \multicolumn{1}{c}{$(D^*D)_0$}
& \multicolumn{1}{c}{$(D^*D)_1$} 
    & \multicolumn{1}{c}{$(D^{*}_{0}D_{1})_1$} 
        & \multicolumn{1}{c}{$(A_{c})_1$} 
       & \multicolumn{1}{c}{$(D^{*}_{s}D)_1$} 
                 & \multicolumn{1}{c}{{$(D^{*}_{s}D_{s})_1$ }} 
          
                                \\
\hline
$t_c$ [GeV$^2$]&27 -- 46&39 -- 50&48 -- 56&39 -- 50&40 -- 50 &42 -- 52\\
$\tau$ [GeV]$^{-2} 10^2$ &30 ; 34& 13 ; 26& 15, 24 ; 20, 25 &7, 26; 29, 27&9 ; 27&21 ; 29 \\
\hline\hline
\end{tabular*}
}
 \caption{Values of the set of the LSR parameters $(t_c,\tau)$ at the optimization region for the PT series up to NLO and for the OPE truncated at the dimension-six condensates and for $\mu=4.65$ GeV. The first set for $(D^{*}_{0}D_{1})_1$ and $(A_{c})_1$ correspond repsectively to the mass and coupling.}
\label{tab:lsr-rad}
\end{table*}
\subsection{The $(D^*_0D_1)_1$ radial excitation}
\begin{figure}[hbt]
\vspace*{-0.25cm}
\begin{center}
\centerline {\hspace*{-7.5cm} \bf a) }
\includegraphics[width=7.2cm]{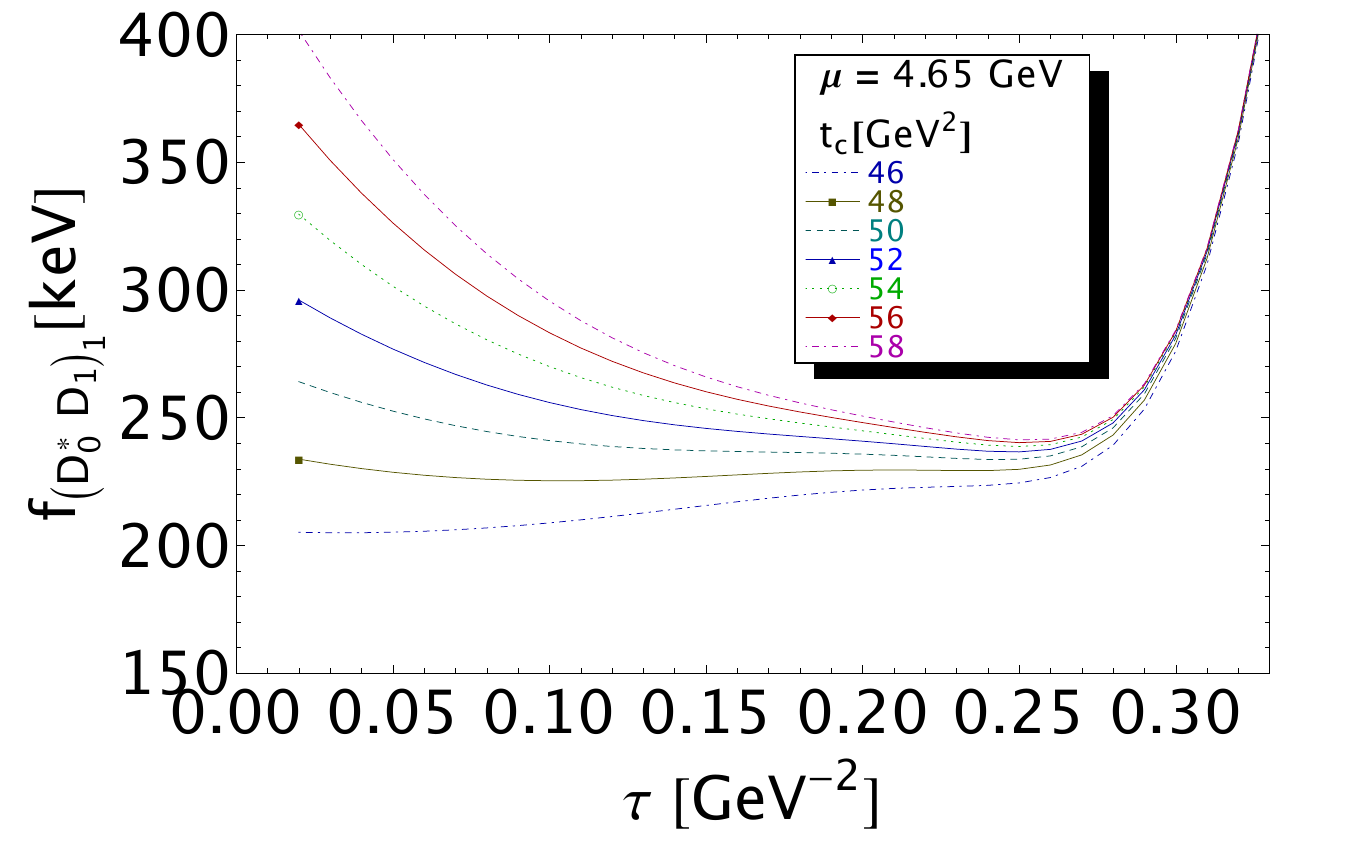}
\vspace{0.25cm}
\centerline {\hspace*{-7.5cm} \bf b) }
\includegraphics[width=6.cm]{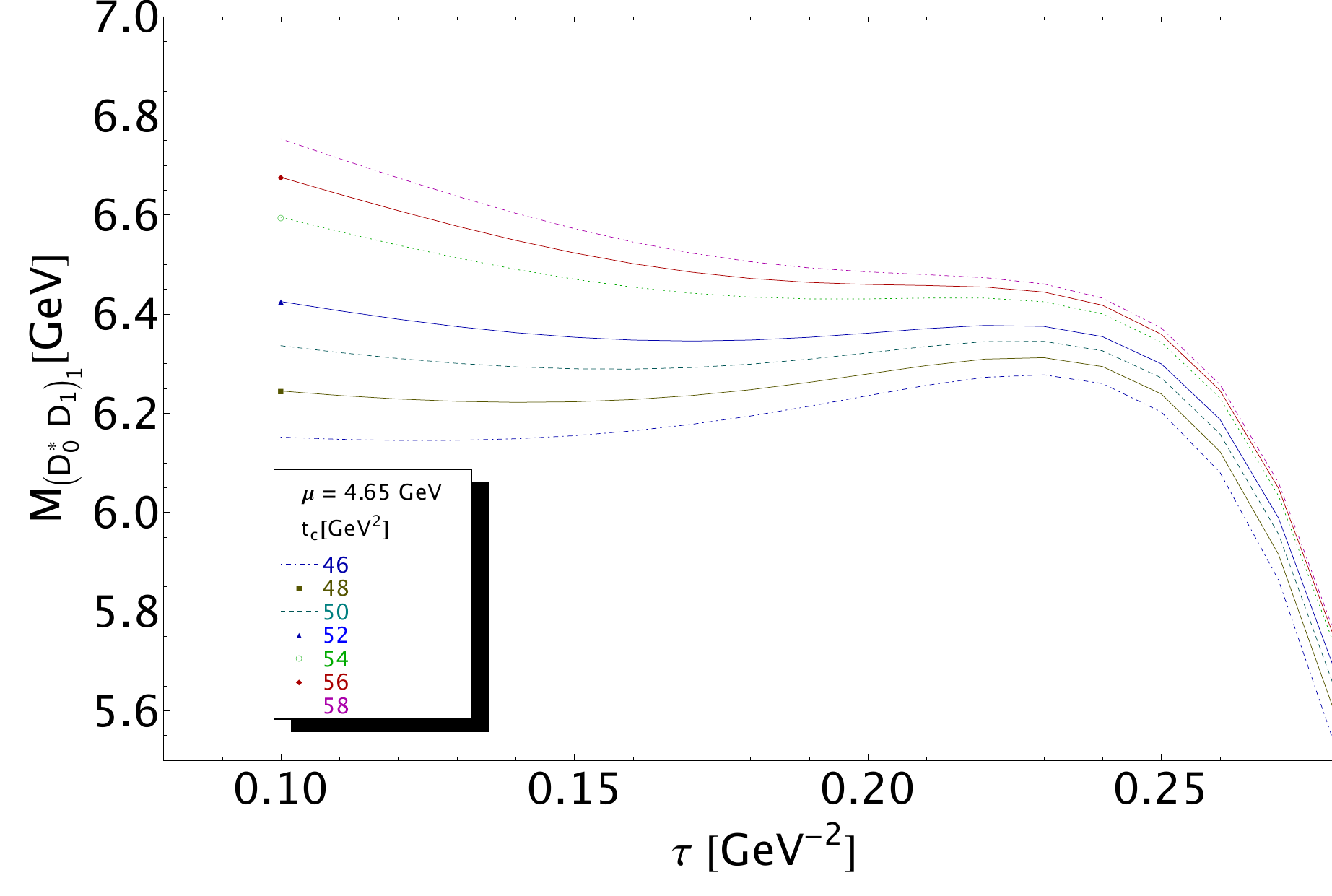}
\vspace*{-0.5cm}
\caption{\footnotesize  $f_{(D^*_0D_1)_1}$ and $M_{(D^*_0D_1)_1}$ as function of $\tau$ at NLO for different values of $t_c$ and for $\mu$=4.65 GeV.} 
\label{fig:dstar0d1-radial}
\end{center}
\vspace*{-0.25cm}
\end{figure}

We show the analysis in Fig.\,\ref{fig:dstar0d1-radial}. The curves have similar behaviour as in the case of $(D^*D)_1$
but the stabilities are reached for higher values of $t_c$ which imply a higher value of the $(D^*_0D_1)_1$ radial excitation mass. The results are quoted in Table\,\ref{tab:radial} and the set of $(t_c,\tau)$ is compiled in Table\,\ref{tab:lsr-rad}. 
\subsection{The $(A_{cd})_1$ radial excitation}
The behaviours of the coupling and mass versus $\tau$ and $t_c$ are similar to the case of the one of $(D^*D)_1$ and will not be repeated here. The results are quoted in Table\,\ref{tab:radial}. 
\subsection{The $(D^*_sD)_1$ radial excitation}
We estimate the mass and coupling of the $(D^*_sD)_1$ radial excitation like in the case of $(D^*D)_1$. The analysis is similar and the curves have the same behaviours. The set of LSR parameters used to get the results are quoted in Table\,\ref{tab:lsr-rad}. We obtain the results quoted in Tables\,\ref{tab:radial}\,:
\beq
f_ {(D^*_sD)_1}=199(29)~{\rm keV},~~M_ {(D^*_sD)_1}=5725(52)~{\rm MeV}~,
\label{eq:rads}
\eeq
where they are also quite high compared to the ones of ordinary mesons.  
\subsection{The first radial excitation $(D^*_sD_s)_1$}
\begin{figure}[hbt]
\vspace*{-0.25cm}
\begin{center}
\centerline {\hspace*{-7.5cm} \bf a) }
\includegraphics[width=6cm]{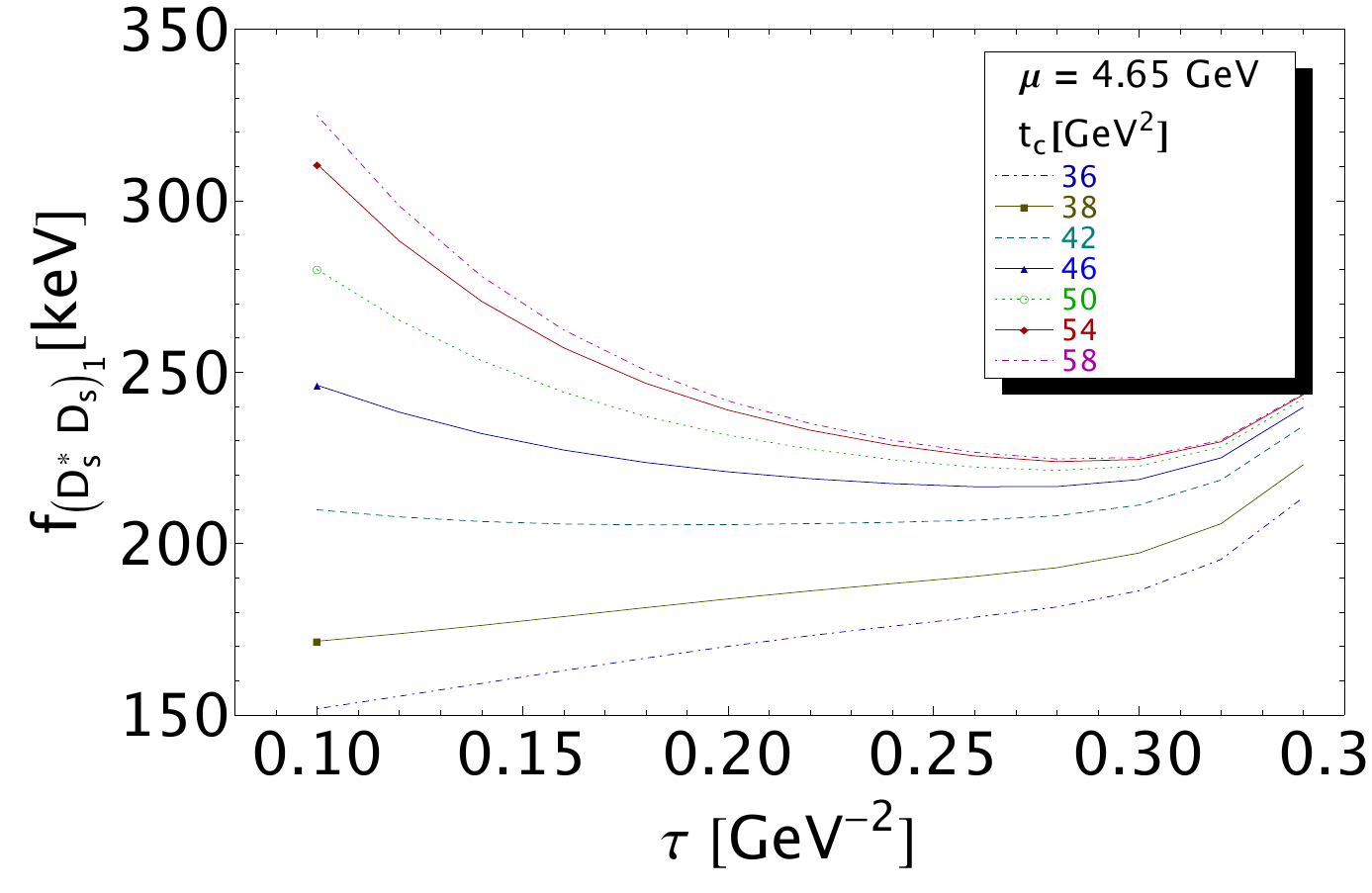}
\vspace{0.25cm}
\centerline {\hspace*{-7.5cm} \bf b) }
\includegraphics[width=6cm]{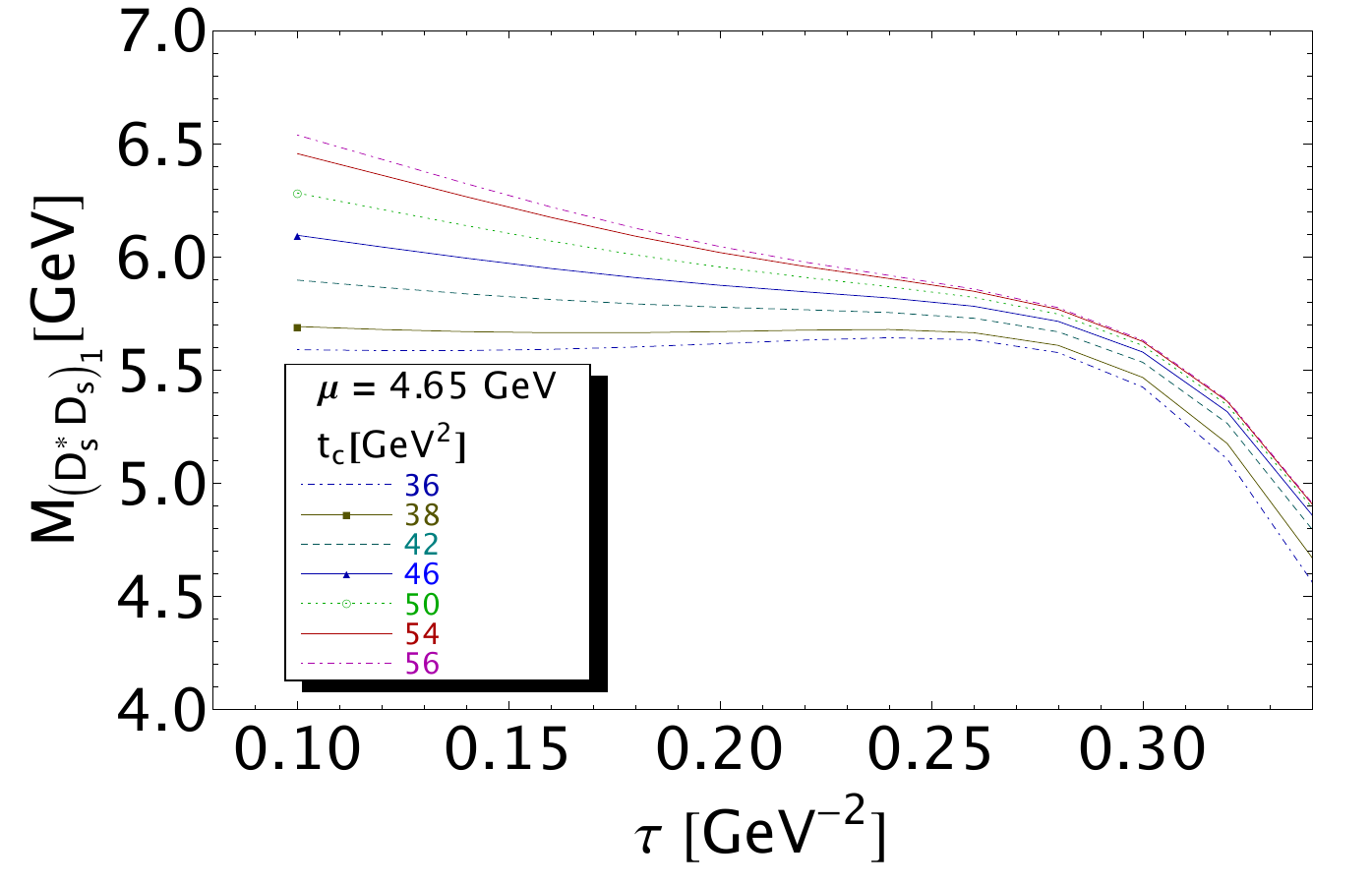}
\vspace*{-0.5cm}
\caption{\footnotesize  $f_{(D^*_sD_s)_1}$ and $M_{(D^*_sD_s)_1}$ as function of $\tau$ at NLO for different values of $t_c$ and for $\mu$=4.65 GeV.} 
\label{fig:dstarsds-radial}
\end{center}
\vspace*{-0.25cm}
\end{figure}

In this subsection, we study the first radial excitation $(D^*_sD_s)_1$.  The $\tau$ and $t_c$  behaviours of its mass and coupling are shown in Fig.\,\ref{fig:dstarsds-radial} for $\mu=4.65$ GeV. The $\tau$ and $t_c$-stabilities are reached for the set $(\tau,t_c)$ =(0.28,42) to (0.30,54) in units of (GeV$^{-2}$, GeV$^2$). We deduce (see Table\,\ref{tab:radial})\,:
\beq 
f_{(D^*_sD_s)_1}=197(43)~{\rm keV},~~M_{(D^*_sD_s)_1}=5786(152)\,{\rm MeV}.
\label{eq:radialss}
\eeq
One can notice that the coupling of the 1st radial excitation is  similar to the previous radial excited states which are relatively large compared to the ones of lowest ground states. This feature differs from the case of ordinary mesons built from bilinear currents. The mass is also found to be relatively high. 
 \subsection{Comments on the radial excitations\label{sec:radial1}}
 We have shown  previously that the couplings of the excited states to the corresponding currents are as large as the one of the ground states (see Table\,\ref{tab:radial})
 which is a new feature compared to the case of ordinary hadrons.  
 
 We have also shown that the mass-splittings between the first radial excitation and the lowest ground state are (see Table\,\ref{tab:radial}) :
\beq
M_{(G)_1}-M_{G}\simeq (1.70\sim 2.35)~{\rm GeV},
\eeq
which is much bigger than the one $(500\sim 600)$ MeV for ordinary mesons but comparable with the one obtained for the $DK$ state in\,\cite{DK}. The authors in Ref.\,\cite{WANGMU} have also noticed a such anomalously large value of the mass of the radial excitation $M_{R1}\approx \sqrt{t_c} \geq 4.5$ GeV in the analysis of the $Z_c(4020,4025)$.  Then, they have concluded that the $Z_c(4020,4025)$ cannot be a tetraquark state as the value of $t_c$ used to extract these masses is much larger than the one expected from the empirical relation : 
\beq
\sqrt{t_c}\approx M_{Z_c}+ 0.5~{\rm GeV},
\label{eq:excited}
\eeq
where the value 0.5~{\rm GeV} has been inspired from ordinary mesons and from the results of\,\cite{LEBED}. 

 From our result, one can already conclude that the $Z_c(4020)$ to  $Z_c(4430)$ are too low to be the radial excitations of the $Z_c(3900)$ unless they couple weakly to the interpolating current such that their effect is tiny in the LSR analysis.  

 \subsection{Can there be a weakly coupled radial excitation\label{sec:radial}}
 If one literally extrapolates the phenomenological observation from ordinary hadrons, one would expect a radial excitation with a mass:
 \beq
 M_{(D^*D)_0}\approx  M_{D^*D} +0.5 ~{\rm GeV}\simeq 4.4~{\rm GeV}~,
 \label{eq:radial}
 \eeq
 which is relatively low compared to the previous prediction in Eq.\,\ref{eq:rad} but seems to fit the $Z_c(4430)$. To understand why it can have been eventually missed in the previous analysis, we shall determine the $Z_c(4430)$ coupling to the current and use as input the experimental
 mass value\,\footnote{No stability region is obtained for an the attempt to determine this mass from the LSR ${\cal R}_0^c$.}.
  
  \begin{figure}[hbt]
\vspace*{-0.25cm}
\begin{center}
\includegraphics[width=6.5cm]{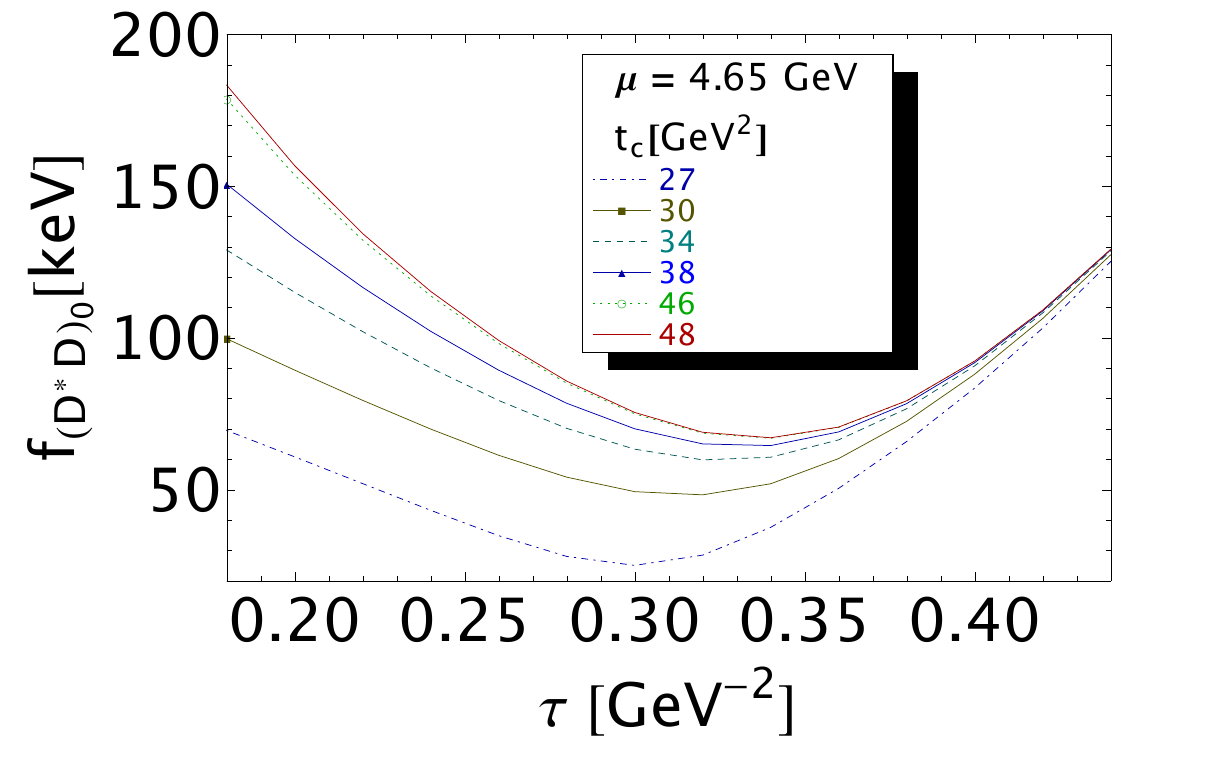}
\vspace{0.25cm}
\vspace*{-0.5cm}
\caption{\footnotesize  The coupling $f_{(D^*D)_0}$ of the $Z_c(4430)$  at NLO as a function of $\tau$ for different values of  $t_c$ and for $\mu=4.65$ GeV.}
\label{fig:dstardradial0}
\end{center}
\vspace*{-0.25cm}
\end{figure} 

In so doing, we reconsider the ${\cal L}^c_0$ moment. We use a two-resonance parametrization of the spectral function and introduce as inputs the previous  values of the $D^*D$ molecule mass and coupling. We use the experimental mass of the $Z_c(4430)$. We include the {\it high-mass} $(D^*D)_1$ radial excitation  obtained previously into the QCD continuum contribution. 

We show the $\tau$-behaviour of the coupling for different $t_c$-values and for fixed $\mu=4.65$ GeV in Fig.\,\ref{fig:dstardradial0}. At the $(\tau,t_c)$ stability regions $(0.30,27)$ to $(0.34,46)$ (GeV$^{-2}$, GeV$^2$), we deduce\,:
\beq
f_{(D^*D)_0}=46(56)~{\rm keV}~,
\eeq
where the different sources of the errors are given in Table\,\ref{tab:radial}. The coupling is indeed relatively small compared to that of $f_{D^*D}=140$ keV. It can even be consistent with zero due to the large errors mainly induced by the coupling of the ground state  and of the dimension-6 condensates. This result may support an eventual radial excitation interpretation of the $Z(4340)$ which couples  very weakly to the current and having a mass much lower than the strongly coupled [$f_{(D^*D)_1}=197$ MeV] radial excitation with a mass 5709 MeV (see Table\,\ref{tab:radial}). This weak coupling disagrees with the one obtained in Ref\,\cite{WANGRAD} and may originate from the fact that the latter analysis has been done at a smaller value of $t_c$ and at the scale $\mu=1.5$ GeV which is too low compared to the optimal choice obtained in the present work. However, as we have already mentioned in Subsection\,\ref{sec:mu}, we do not find any convincing theoretical basis for justifying this low choice of $\mu$. We expect that similar results can be obtained in some other channels.

 
\section{Versus our previous results}

\begin{table*}[hbt]
\setlength{\tabcolsep}{1pc}
\catcode`?=\active \def?{\kern\digitwidth}
{\scriptsize
\begin{tabular*}{\textwidth}{@{}l@{\extracolsep{\fill}}|llllll}
\hline
\hline
      
States      & \multicolumn{3}{c}{Couplings} 
      & \multicolumn{3}{c}{Masses} 
     \\
\cline{2-4}\cline{5-7}
& Ref.\,\cite{MOLE16} & Ref.\,\cite{SU3} &New & Ref.\,\cite{MOLE16} & Ref.\,\cite{SU3} &New \\
\cline{1-1}\cline{2-4}\cline{5-7}
\footnotesize Molecule  $(\bar cd)(c\bar u)$\\
$D^*D$ & 154(7) &--&140(15) &3901(62)&--&3912(61) \\
$D^*_0D_1$ &96(15) &--& 96(23)&4394(164)&3854(182)& 4023(130) \\
\footnotesize Tetraquark  $(\bar c\bar d)(cu)$ \\
$A_{cd}$&176(30)&--&173(17)&3890(130)&--&3889(42)  \\
 \hline
\footnotesize Molecule  $(c\bar s)(\bar cu)$\\
$D^*_sD$ &--& -- &130(15)&-- &--&3986(51) \\
$D^*D_s$ &--&--&133(16)&-- &-&3979(56) \\
$D^*_{s0}D_{1}$ &--&--&86(23)&-- &-&4064(133) \\
$D^*_{0}D_{s1}$ &--&--&89(22)&-- &-&4070(133) \\
\footnotesize  Tetraquark $(\bar c\bar s)(cu)$ \\
$A_{cs}$&--&--&148(17)&--&--&3950(56)  \\
\hline
\footnotesize Molecule $(\bar cs)(c\bar s)$\\
$D^*_sD_s$ &--& 114(13) &--&-- &3901(62)&4091(57) \\
$D^*_{s0}D_{s1}$ &--&79(14) & 70(16)&--&4269(205) & 4198(129) \\
$A_{css}$&--&114(16)&131(14)&--&4209(112)&4014(86)  \\
\hline\hline

\end{tabular*}
}
 \caption{Summary of the couplings and masses predictions of the lowest ground states and comparison with our previous results in Refs.\,\cite{MOLE16,SU3}.}
\label{tab:summary}
\end{table*}
We compare our results with our previous ones from Refs.\,\cite{MOLE16,SU3} in Table\,\ref{tab:summary}.  Notice that in \,\cite{SU3}, the double ratio of moments has been also used to extract directly the $SU(3)$ breaking contributions to the masses and couplings. One can notice a good agreement between the different results. The exception is the central value of $M_{D^*_0D_1}$ which slightly moves in the 3 papers though the results are in agreement within the errors. This is due to the difficult localization of the inflexion point which we have identified in the present work with the minimum of the coupling like in some other channels.  
\section{Confrontation with the data}
As proposed in\,\cite{DK}, we consider that the physical states are superposition of quasi-degenerated hypothetical molecules and tetraquark states having the same quantum numbers $J^{PC}$ and having almost the same coupling strength to the currents. We have denoted these observed states as {\it Tetramoles} ${\cal T}$ (tetraquarks $\oplus$ molecules).
\subsection{$Z_c(3900)$ as a tetramole  state}
 One may define the tetramole ${\cal T}_c$ as a superposition of the $D^*D$ molecule and $A_{cd}$ tetraquark state with the parameters\,:
\beq
M_{{\cal T}_c}=3900(42)~{\rm MeV},~~~~~~f_{{\cal T}_c}=155(11)~{\rm keV} ,
\eeq 
which are the mean of the two previous couplings and masses. We identify this {\it tetramole} state with the $Z_c(3900)$ found by BELLE\,\cite{BELLE1} and BESIII\,\cite{BES0}.
\subsection{$Z_c(4025,4050)$ as a $D^*_0D_1$ molecule}

 The $D^*_0D_1$ molecule with the parameters:
\beq
M_{D^*_0D_1}= 4023(130)~{\rm MeV},~~~~f_{D^*_0D_1}=96(23)~{\rm keV},
\eeq
as given in Tables\,\ref{tab:summary} and \,\ref{tab:res} might be identified with the $Z_c(4025)$ found by BES\,\cite{BES3} or with  the $Z_c(4050)$ found by BELLE\,\cite{BELLE3}. However, one should mention that these states are not well established and have not yet been retained in the PDG Summary Table\,\cite{PDG}. 

\subsection{$Z_c(4430)$ as a weakly coupled radial excitation\,?\label{sec:weak-radial}}
 Pursuing our confrontation with the data, we note that the $Z_c(4430)$\,\cite{BELLE4,LHCb5} is too low to be the strongly coupled ($f_{(D^*D)_1}=196(41)$ MeV)  first radial excitation of the $D^*D$ expected to be in the range $(5.4 \sim 5.8)$ GeV (see Table\,\ref{tab:res}). However, 
 it can be fitted by the weakly coupled ($f_{(D^*D)_0}=46(56)$ MeV) low mass one obtained in Subsection\,\ref{sec:radial}.
\subsection{The $Z_{cs}(3983)$ as a Tetramole}
We consider the tetramole ${\cal T}_s$ as the combination of the $D^*_sD,~D^*D_s$ molecules and $A_{cs}$ tetraquark states  have almost the same mass and same coupling to the currents (see Table\,\ref{tab:res}).  Taking the mean values of these parameters, we obtain:
\beq
M_{{\cal T}_{cs}}=3973(31)~{\rm MeV},~~~~~~f_{{\cal T}_{cs}}=136(9)~{\rm keV} ,
\eeq 
which we can identify with the recent $Z_{cs}(3983)$ found recently by BESIII\,\cite{BES}. 
\subsection{The $Z_{cs}(4100)$ bump as a  $D^*_{s0}D_1\oplus D^*_0D_{s1}$ molecule}
Inspecting our predictions for the $D^*_{s0}D_1$ and $D^*_0D_{s1}$ masses in Tables\,\ref{tab:res} and\,\ref{tab:summary}, we find that they are almost degenerated and have almost the same couplings. Taking their combination having a mean mass :
\bea
M_{D^*_{s0}D_1\oplus D^*_{0}D_{s1}}&=&4067(94)~{\rm MeV},\nnb\\
f_{D^*_{s0}D_1\oplus D^*_{0}D_{s1}}&=&88(16)~{\rm keV},
\eea
we are tempted to identify this state with the $Z_{cs}(4100)$ bump observed by  BESIII\,\cite{BES}
. Its relatively small coupling to the current compared to the one of the ground state $Z_{cs}(3983)$ may indicate that  it can be large using a Golberger-Treiman-type relation argument where the hadronic coupling behaves as the inverse of coupling (see e.g: \cite{FURLAN}). 
\subsection{The future $Z_{css}$ states}
 From our predictions in Table\,\ref{tab:res}, one can also define a tetramole ${\cal T}_{css}$ which is a superposition of the $D^*_sD_s$ molecule and ${A}_{css}$ tetraquark states with\,:
\beq
M_{{\cal T}_{css}}=4068(48)~{\rm MeV},~~~~
f_{{\cal T}_{css}}=114(10)~{\rm keV}.
\eeq

 One also expects to have a $D^*_{s0}D_{s1}$ molecule at a higher mass value:
\beq
M_{D^*_{s0}D_{s1}}= 4198(129)~{\rm MeV},~~f_{D^*_{s0}D_{s1}}=79(14)~{\rm keV},
\eeq
where its coupling to the current is relatively small indicating that it can be relatively large using a Golberger-Treiman-type relation argument where the hadronic coupling behaves as 1/$f_{D^*_{s0}D_{s1}}$\,(see e.g: \cite{FURLAN}).

 These predicted states are expected to be seen in the near future experiments and can be considered as a test of the predictions given in this paper. 
\section{On some other and LSR results}
After the publication of the recent BESIII results on the observation of the $Z_{cs}(3982)$ candidate\,\cite{BES}, many papers using different models appear in the literature\,
\cite{JIN} 
for attempts to explain the true nature of this state. 

Besides the pioneer QSSR estimate of the $D^*D_s$ molecule and tetraquark states\,\cite{NIELSEN1}\,\footnote{For more complete references, see e.g.\,\cite{MOLEREV}.}, some recent papers using LSR come to our attention\,
(see e.g \cite{WAN})
where we notice some common caveats\,:

-- All analysis is done at lowest order (LO) of perturbation theory where the choice of the value of the  $\overline{MS}$ running mass in favour of the pole mass is unjustified because the definitions of the two masses are undistinguishable at this order. Moreover, the calculation of the spectral functions using on-shell renormalization would (a priori) favour the choice of the on-shell mass. The difference on the effect of this choice is explicitly shown in Fig.\,\ref{fig:lo}.  Hopefully, the effects of NLO corrections in the ${\overline MS}$-scheme are tiny (see Fig.\,\ref{fig:lo}) in the LSR analysis confirming (a posteriori) the intuitive choice of the $ {\overline MS}$ running mass at LO. 

-- The value of $t_c\equiv s_0$ used to determine the mass and coupling of about $(18 \sim 23)$ GeV$^2$ is relatively low as it corresponds to the beginning of the $\tau$-stability region (see  previous figures) where one also notice that the predictions increase until the $t_c$-stability value. As a consequence, the absolute value and the error in the extraction of the mass and coupling have been underestimated. 

-- In general, the way how the errors from different sources  has been estimated are not explained in details, raises some doubts on the real size of the quoted errors having in mind that the extraction of the different errors quoted in Tables\,\ref{tab:res} and \,\ref{tab:radial} require some  painful works. 


-- The value of the four-quark condensates from the vacuum saturation assumption is often used. However, though this estimate is correct in the large $N_c$ limit, it has been found phenomenologically from different light-quarks and $\tau$-decay channels\,\,\cite{DOSCH,SNTAU,LNT,JAMI2a,JAMI2c,LAUNERb} that this estimate is largely violated at the realistic case $N_c=3$. Moreover, it is also known that the dimension-six quark condensates which mix under renormalization\,\cite{SNTARRACH}\,(Part VII page 285) does not support the previous assumption. 

-- The previous papers extend the OPE to high-dimension up to $d=10$ vacuum condensates but only for some classes of high-dimension condensate contributions and by assuming the validity of the factorization assumption for estimating their sizes. However, it has been shown in\,\cite{SNTARRACH}\,\cite{SNB1}\,(Part VII page 285) that the structure of these high-dimension condensates are quite complex due to their mixing under renormalization such that their inclusion in the QSSR analysis should deserve more care. 

In addition to these 
caveats,  we also notice that :

-- Differentiating the neutral from the charged $Z_{cs}$  is purely academic from the approach as the two states are almost degenerated within the errors. 


-- In Ref.\,\cite{WANG1,WANGMU}, the author introduces a relation between the PT subtraction scale $\mu$ and the so-called energy bound which is quite obscure to us as discussed in Subsection\,\ref{sec:mu}. 


\section{Summary and Conclusions}
We have\,: 

-- Systematically studied the spectra and couplings of the $(\bar cq)(\bar q'c)$ molecules and $(\bar c\bar q)(q'c)$ tetraquark states where $q,q'\equiv d,s$ are light quarks.

-- Improved our previous predictions obtained in Refs.\,\cite{MOLE16,SU3} by a much better localization of the $t_c,~\tau$ and $\mu$-stability points and by using updated values of some QCD input parameters. 

-- Emphasized that  the localization  of the inflexion point for extracting the values of the masses can be fixed more precisely in most channels by identifying it with the value of $\tau$ corresponding to the minimum of the curve where optimal value of the coupling is extracted. 

-- Provided new predictions of the $(\bar cq)(\bar sc)$ molecules and $(\bar c\bar q)(sc)$ tetraquark states.

-- Introduced the {\it tetramole} states as a superposition of quasi-degenerated molecules and tetraquark states having the same quantum numbers $J^{PC}$ with almost the same couplings strengths to the interpolating currents. It is remarkable to notice that the mass-spliitings of the tetramoles ${\cal T}_c,~ {\cal T}_{cs}$ and ${\cal T}_{css}$ due to $SU(3)$ breakings are successively about $(73\sim 91)$ MeV. 

-- Completed the analysis with new predictions of some first radial excitation masses and couplings.  One can notice that the mass gap of about $(1.7\sim 2.35)$ GeV, between the lowest mass ground state and the 1st radial excitation strongly coupled to the current,  is quite large compared to $(0.5\sim 0.7)$ GeV for ordinary $\bar qq$ mesons.  Similar results have been obtained for the $DK$-like states\,\cite{DK}. However, a weakly coupled radial excitation (see Section\,\ref{sec:radial}) having a lower mass of about 4.4 GeV is not excluded from the approach. These features may signal some new dynamics of these exotic states which can be found in these high-mass regions. 

-- Compared successfully our predictions with the observed $Z_c$ and $Z_{cs}$-spectra. 

-- Given new predictions for the future $Z_{css}$ states.

-- Also shown that a result based on a qualitative $N_c$ counting without taking into account the dynamics from Feynman loop calculation leads to a wrong conclusion. 
\section*{Acknowledgements}
We thank Prof. G. Veneziano for discussions and comments on the manuscript. 
\appendix
\section{QCD spectral functions with $J^P=1^+$}
We shall present in the following the different QCD expressions of the spectral functions related to the molecular and tetraquarks currents,
which come from the evaluation of the two-point correlation function using the $J^{P} = 1^+$
 hadronic currents  given in Table\,\ref{tab:current} where $q\equiv s$. 
 The expressions obtained in the chiral limit $m_s=m_d=0$ can be found in Ref.\,\cite{MOLE16}
 while the one with a double strange quarks has been obtained in\,\cite{SU3}. 
 We have used the expression of the $\la G^3\ra$ condensate contribution obtained in the chiral limit in\,\cite{MOLE16} which will not be given below.  
 One should notice that compared to the QCD expressions given in the literature, the ones which we give below and in the two previous papers\,\cite{MOLE16,SU3} are completely integrated and compact.  
 Hereafter, we define :
 $\rho(t)\equiv \frac{1}{\pi}{\rm Im}\Pi^{(1)}(t)$ where ${\rm Im}\Pi(t)^{(1)}$ is the spectral function defined in Eq.\,\ref{eq:2-pseudo} with :
 \beq
 \rho(t)\simeq \rho^{pert}+ \rho^{\langle \bar{q}q \rangle}+ \rho^{\langle G^2 \rangle}+ \rho^{\langle \bar{q}Gq \rangle}+ \rho^{\langle \bar{q}q \rangle^2}+ \rho^{\langle G^3 \rangle}
\eeq
 where : 
$ \la G^2\ra \equiv \la g^2G^2\ra,~\la \bar qGq\ra\equiv M_0^2\la\bar qq\ra,~\la G^3\ra \equiv \la g^3G^3\ra, $
 and:
$x \equiv  m_c^2/t~,~~~ v=\sqrt{1-4x}~,~~~   \lv=\ln{\frac{(1+v)}{(1-v)}},\nnb\\
\lp=\lid\left(\frac{1+v}{2}\right)-\lid\left(\frac{1-v}{2}\right).$
 $m_c \equiv M_c$ (resp. $m_s$) is the on-shell charm (resp. running strange) quark mass. ; $b$ is the current mixing parameter where its optimal value is found to be $b=0$\,\cite{MOLE16}. For the estimate of the four-quark operator, we introduce the violation of the vacuum saturation estimate qunatified by the factor $\rho\simeq (3\sim 4)$ defined in Table\,\ref{tab:param}. 
 \begin{widetext}
 \subsection{The Molecular  Currents}
{\bf $D_s D^\ast$ molecule }
\footnotesize
\vspace{-0.25cm}
\begin{eqnarray}
  \rho^{pert}(t) &=& 
  \frac{m_c^8}{5\cdot 3 \cdot 2^{15} \pi^6} 
  \bigg[ v \bigg( 840x + 140 + \frac{5248}{x}
  - \frac{1164}{x^2} - \frac{182}{x^3} + \frac{5}{x^4} 
  \bigg) + 120 {\cal L}_v \bigg( 14x^2 + 15 - 18\log(x)\nnb \\&&
  - \frac{32}{x} + \frac{9}{x^2} \bigg) 
  - 4320 {\cal L}_{+} \bigg]
  ~-~ \frac{m_s m_c^7}{5 \cdot 2^{14} \pi^6} \bigg[ 
  v \bigg( 420x + 70 + \frac{1574}{x} - \frac{257}{x^2} 
  - \frac{16}{x^3} \bigg)\nnb \\&& 
  +~ 60{\cal L}_v \bigg( 14x^2 + 4
  - 12\log(x) - \frac{16}{x} + \frac{3}{x^2} \bigg) 
  - 1440 {\cal L}_{+} \bigg]\nnb \\
  \rho^{\langle \bar{q}q \rangle}(t) &=&
  \frac{ m_c^5 \langle \bar{s}s \rangle}{3 \cdot 2^{10} \pi^4} 
  \bigg[ v \bigg( 60x + 10 - \frac{34}{x}
  - \frac{9}{x^2} \bigg) + 24 {\cal L}_v \bigg( 5x^2 - 3 
  + \frac{2}{x} \bigg) \bigg] 
  ~+~ \frac{ m_c^5 \langle \bar{q}q \rangle}{2^{8} \pi^4} 
  \bigg[ v \bigg( 6 - \frac{5}{x} - \frac{1}{x^2} \bigg) \nnb\\&&
  + 6 {\cal L}_v \bigg( 2x - 2 + \frac{1}{x} \bigg) \bigg] 
  ~+~ \frac{ m_s m_c^4 \langle \bar{s}s \rangle}{2^{11} \pi^4} 
  \bigg[ v \bigg( 12x + 2 - \frac{26}{x} + \frac{3}{x^2} \bigg)
  + 24 {\cal L}_v ( x^2 + 1 ) \bigg] \nnb\\&&
  -~ \frac{ 3m_s m_c^4 \langle \bar{q}q \rangle}{2^{7} \pi^4}
  \bigg[ v \bigg( 2 + \frac{1}{x} \bigg) + 4 {\cal L}_v 
  ( x - 1 ) \bigg]  \nnb
  \\ &&\nnb \\
  \rho^{\langle G^2 \rangle}(t) &=& 
  -\frac{m_c^4 \langle G^2 \rangle}{3^2 \cdot 2^{14} \pi^6}
  \bigg[ v \bigg( 60x - 62 + \frac{26}{x}
  + \frac{3}{x^2} \bigg) + 24 {\cal L}_v \bigg( 5x^2 - 6x + 3 
  - \frac{1}{x} \bigg) \bigg] \nnb
 \\ &&\nnb \\
  \rho^{\langle \bar{q}Gq \rangle}(t) &=& 
  - \frac{m_c^3 \langle \bar{s}G s \rangle}{3\cdot2^{10}\pi^4}
  \bigg[ v \bigg( 66x + 11 - \frac{68}{x} \bigg) + 
  6 {\cal L}_v \bigg( 22x^2 + 3 + \frac{4}{x} \bigg) \bigg] 
  ~+~ \frac{3m_c^3 \langle \bar{q}G q \rangle}{2^9 \pi^4}
  \bigg( \frac{v}{x} - 2 {\cal L}_v \bigg)\nnb \\ &&
  -~ \frac{m_s m_c^2 \langle \bar{s}G s \rangle}
  {3\cdot2^{10}\pi^4} \bigg[ 
  v \bigg( 18x - 1 - \frac{8}{x} \bigg) 
  + 18 {\cal L}_v ( 2x^2 + 1) \bigg] ~+~ 
  \frac{3 m_s m_c^2 \langle \bar{q}G q \rangle}{2^8 \pi^4} \:v\nnb
  \\ &&\nnb \\
  \rho^{\langle \bar{q}q \rangle^2}(t) &=& 
  \frac{m_c^2 \langle \bar{q}q \rangle 
  \langle \bar{s}s \rangle}{64 \pi^2} \:v \bigg(
  4 - \frac{m_s m_c \tau}{x} - \frac{m_s}{m_c} \bigg)
\end{eqnarray}

\vspace{-0.5cm}
{\bf $D_s^\ast D$ molecule }
\vspace{-0.5cm}
\begin{eqnarray}
  \rho^{pert}(t) &=& 
  \frac{m_c^8}{5\cdot 3 \cdot 2^{15} \pi^6} 
  \bigg[ v \bigg( 840x + 140 + \frac{5248}{x}
  - \frac{1164}{x^2} - \frac{182}{x^3} + \frac{5}{x^4} 
  \bigg) + 120 {\cal L}_v \bigg( 14x^2 + 15 - 18\log(x) \nnb\\&&
  - \frac{32}{x} + \frac{9}{x^2} \bigg) 
  - 4320 {\cal L}_{+} \bigg]
  ~-~ \frac{m_s m_c^7}{2^{12} \pi^6} \bigg[ 
  v \bigg( 60 + \frac{130}{x} - \frac{18}{x^2} 
  - \frac{1}{x^3} \bigg)\nnb \\&& 
  +~ 12{\cal L}_v \bigg( 10x - 4
  - 6\log(x) - \frac{6}{x} + \frac{1}{x^2} \bigg) 
  - 144 {\cal L}_{+} \bigg]\nnb
  \\ 
  \rho^{\langle \bar{q}q \rangle}(t) &=&
  \frac{ m_c^5 \langle \bar{s}s \rangle}{2^{8} \pi^4} 
  \bigg[ v \bigg( 6 - \frac{5}{x} - \frac{1}{x^2} \bigg) 
  + 6 {\cal L}_v \bigg( 2x - 2 + \frac{1}{x} \bigg) \bigg] 
  ~+~ \frac{ m_c^5 \langle \bar{q}q \rangle}{3\cdot2^{10} \pi^4} 
  \bigg[ v \bigg( 60x + 10 - \frac{34}{x} - \frac{9}{x^2} \bigg)\nnb
  \\&&
  + 24 {\cal L}_v \bigg( 5x^2 - 3 + \frac{2}{x} \bigg) \bigg] 
  ~+~ \frac{ m_s m_c^4 \langle \bar{s}s \rangle}{2^{11} \pi^4} 
  \bigg[ v \bigg( 12x + 2 - \frac{26}{x} + \frac{3}{x^2} \bigg)
  + 24 {\cal L}_v ( x^2 + 1 ) \bigg]\nnb \\&&
  -~ \frac{ 3m_s m_c^4 \langle \bar{q}q \rangle}{2^{7} \pi^4}
  \bigg[ v \bigg( 2 + \frac{1}{x} \bigg) + 4 {\cal L}_v 
  ( x - 1 ) \bigg] \nnb \\ 
  \rho^{\langle G^2 \rangle}(t) &=& 
  -\frac{m_c^4 \langle G^2 \rangle}{3^2 \cdot 2^{14} \pi^6}
  \bigg[ v \bigg( 60x - 62 + \frac{26}{x}
  + \frac{3}{x^2} \bigg) + 24 {\cal L}_v \bigg( 5x^2 - 6x + 3 
  - \frac{1}{x} \bigg) \bigg] \nnb
 \\ 
  \rho^{\langle \bar{q}Gq \rangle}(t) &=& 
  \frac{3m_c^3 \langle \bar{s}G s \rangle}{2^{9}\pi^4}
  \bigg( \frac{v}{x} - 2 {\cal L}_v \bigg) 
  ~-~ \frac{m_c^3 \langle \bar{q}G q \rangle}{3\cdot2^{10}\pi^4}
  \bigg[ v \bigg( 66x + 11 - \frac{68}{x} \bigg) + 
  6 {\cal L}_v \bigg( 22x^2 + 3 + \frac{4}{x} \bigg) \bigg]\nnb \\&&
  -~ \frac{m_s m_c^2 \langle \bar{s}G s \rangle}{2^{10}\pi^4}
  \bigg[ v \bigg( 2x - 25 + \frac{4}{x} \bigg) 
  + 2 {\cal L}_v ( 2x^2 - 24 x + 9) \bigg] ~+~ 
  \frac{3 m_s m_c^2 \langle \bar{q}G q \rangle}{2^8 \pi^4} \:v
  \nnb\\ 
  \rho^{\langle \bar{q}q \rangle^2}(t) &=& 
  \frac{m_c^2 \langle \bar{q}q \rangle 
  \langle \bar{s}s \rangle}{64 \pi^2} \:v \bigg(
  4 - \frac{ m_s m_c \,\tau}{x} \bigg)
 \end{eqnarray}
{\bf $D_{s0}^\ast D_1$ molecule}
\vspace*{-0.5cm}
\begin{eqnarray}
  \rho^{pert}(t) &=& 
  \frac{m_c^8}{5\cdot 3 \cdot 2^{15} \pi^6} 
  \bigg[ v \bigg( 840x + 140 + \frac{5248}{x}
  - \frac{1164}{x^2} - \frac{182}{x^3} + \frac{5}{x^4} 
  \bigg) + 120 {\cal L}_v \bigg( 14x^2 + 15 - 18\log(x) \nnb\\&&
  - \frac{32}{x} + \frac{9}{x^2} \bigg) 
  - 4320 {\cal L}_{+} \bigg]
  ~+~ \frac{m_s m_c^7}{5 \cdot 2^{14} \pi^6} \bigg[ 
  v \bigg( 420x + 70 + \frac{1574}{x} - \frac{257}{x^2} 
  - \frac{16}{x^3} \bigg)\nnb \\&& 
  +~ 60{\cal L}_v \bigg( 14x^2 + 4
  - 12\log(x) - \frac{16}{x} + \frac{3}{x^2} \bigg) 
  - 1440{\cal L}_{+} \bigg]\nnb\\
  \eea
  \bea
  \rho^{\langle \bar{q}q \rangle}(t) &=&
  -~\frac{ m_c^5 \langle \bar{q}q \rangle}{2^{8} \pi^4} 
  \bigg[ v \bigg( 6 - \frac{5}{x} - \frac{1}{x^2} \bigg) 
  + 6 {\cal L}_v \bigg( 2x - 2 + \frac{1}{x} \bigg) \bigg] 
  ~-~ \frac{ m_c^5 \langle \bar{s}s \rangle}{3\cdot2^{10} \pi^4} 
  \bigg[ v \bigg( 60x + 10 - \frac{34}{x} - \frac{9}{x^2} \bigg)
 \nnb \\&&
  +~  24 {\cal L}_v \bigg( 5x^2 - 3 + \frac{2}{x} \bigg) \bigg] 
  ~-~ \frac{ 3m_s m_c^4 \langle \bar{q}q \rangle}{2^{7} \pi^4} 
  \bigg[ v \bigg( 2 + \frac{1}{x} \bigg) 
  + 4 {\cal L}_v (x - 1) \bigg] \nnb\\&&
  +~ \frac{ m_s m_c^4 \langle \bar{s}s \rangle}{2^{11} \pi^4}
  \bigg[ v \bigg( 12x + 2 - \frac{26}{x} + \frac{3}{x^2} \bigg) 
  + 24 {\cal L}_v ( x^2 + 1 ) \bigg] \nnb \\ 
  \rho^{\langle G^2 \rangle}(t) &=& 
  -~ \frac{m_c^4 \langle G^2 \rangle}{3^2 \cdot 2^{14} \pi^6}
  \bigg[ v \bigg( 60x - 62 + \frac{26}{x}
  + \frac{3}{x^2} \bigg) + 24 {\cal L}_v \bigg( 5x^2 - 6x + 3 
  - \frac{1}{x} \bigg) \bigg] \nnb
 \\ 
  \rho^{\langle \bar{q}Gq \rangle}(t) &=& 
  -~ \frac{3m_c^3 \langle \bar{q}G q \rangle}{2^{9}\pi^4}
  \bigg( \frac{v}{x} - 2 {\cal L}_v \bigg) 
  ~+~ \frac{m_c^3 \langle \bar{s}G s \rangle}{3\cdot2^{10}\pi^4}
  \bigg[ v \bigg( 66x + 11 - \frac{68}{x} \bigg) + 
  6 {\cal L}_v \bigg( 22x^2 + 3 + \frac{4}{x} \bigg) \bigg]\nnb \\&&
  +~ \frac{3m_s m_c^2 \langle \bar{q}G q \rangle}{2^{8}\pi^4} \:v
 ~-~ \frac{m_s m_c^2 \langle \bar{s}G s \rangle}
 {3 \cdot 2^{10} \pi^4} \bigg[ 
 v \bigg( 18x - 1 - \frac{8}{x} \bigg) 
 + 18 {\cal L}_v ( 2x^2 + 1) \bigg]\nnb
 \\ 
  \rho^{\langle \bar{q}q \rangle^2}(t) &=& 
  \frac{\langle \bar{q}q \rangle 
  \langle \bar{s}s \rangle}{64 \pi^2} \:v \bigg[
  4m_c^2 + m_s m_c \Big( 1 + \frac{m_c^2 \tau}{x} \Big) \bigg]
\end{eqnarray}

{\bf $D_{0}^\ast D_{s1}$ molecule} 
\begin{eqnarray}
  \rho^{pert}(t) &=& 
  \frac{m_c^8}{5\cdot 3 \cdot 2^{15} \pi^6} 
  \bigg[ v \bigg( 840x + 140 + \frac{5248}{x}
  - \frac{1164}{x^2} - \frac{182}{x^3} + \frac{5}{x^4} 
  \bigg) + 120 {\cal L}_v \bigg( 14x^2 + 15 - 18\log(x)\nnb \\&&
  - \frac{32}{x} + \frac{9}{x^2} \bigg) 
  - 4320 {\cal L}_{+} \bigg]
  ~+~ \frac{m_s m_c^7}{2^{12} \pi^6} \bigg[ 
  v \bigg( 60 + \frac{130}{x} - \frac{18}{x^2} 
  - \frac{1}{x^3} \bigg) \nnb \\&& 
  +~ 12{\cal L}_v \bigg( 10x - 4
  - 6\log(x) - \frac{6}{x} + \frac{1}{x^2} \bigg) 
  - 144{\cal L}_{+} \bigg]\nnb
  \\ 
  \rho^{\langle \bar{q}q \rangle}(t) &=&
  -~ \frac{m_c^5 \langle \bar{q}q \rangle}{3\cdot2^{10} \pi^4} 
  \bigg[ v \bigg( 60x + 10 - \frac{34}{x} - \frac{9}{x^2} \bigg)
  +~ 24 {\cal L}_v \bigg( 5x^2 - 3 + \frac{2}{x} \bigg) \bigg] \nnb
  \\&&
  -~\frac{m_c^5 \langle \bar{s}s \rangle}{2^{8} \pi^4} 
  \bigg[ v \bigg( 6 - \frac{5}{x} - \frac{1}{x^2} \bigg) 
  + 6 {\cal L}_v \bigg( 2x - 2 + \frac{1}{x} \bigg) \bigg] 
  ~-~ \frac{ 3m_s m_c^4 \langle \bar{q}q \rangle}{2^{7} \pi^4} 
  \bigg[ v \bigg( 2 + \frac{1}{x} \bigg) 
  + 4 {\cal L}_v (x - 1) \bigg] \nnb\\&&
  +~ \frac{ m_s m_c^4 \langle \bar{s}s \rangle}{2^{11} \pi^4}
  \bigg[ v \bigg( 12x + 2 - \frac{26}{x} + \frac{3}{x^2} \bigg) 
  + 24 {\cal L}_v ( x^2 + 1 ) \bigg] \nnb \\ 
  \rho^{\langle G^2 \rangle}(t) &=& 
  -~ \frac{m_c^4 \langle G^2 \rangle}{3^2 \cdot 2^{14} \pi^6}
  \bigg[ v \bigg( 60x - 62 + \frac{26}{x}
  + \frac{3}{x^2} \bigg) + 24 {\cal L}_v \bigg( 5x^2 - 6x + 3 
  - \frac{1}{x} \bigg) \bigg] \nnb
 \\ &&\nnb \\
  \rho^{\langle \bar{q}Gq \rangle}(t) &=& 
  \frac{m_c^3 \langle \bar{q}G q \rangle}{2^{10}\pi^4}
  \bigg[ v \bigg( 30x + 5 + \frac{4}{x} \bigg) + 
  6 {\cal L}_v \bigg( 10x^2 - 3 \bigg) \bigg]
  ~-~ \frac{3m_c^3 \langle \bar{s}G s \rangle}{2^{9}\pi^4}
  \bigg( \frac{v}{x} - 2 {\cal L}_v \bigg)\nnb \\&&
  +~ \frac{3m_s m_c^2 \langle \bar{q}G q \rangle}{2^{8}\pi^4}
  \bigg[ 3v + 2{\cal L}_v ( 2x - 1) \bigg]
  ~-~ \frac{m_s m_c^2 \langle \bar{s}G s \rangle}{2^{10} \pi^4} 
  \bigg[ v \bigg( 2x - 1 + \frac{4}{x} \bigg) 
  + 2{\cal L}_v ( 2x^2 - 3) \bigg]\nnb
 \\ 
  \rho^{\langle \bar{q}q \rangle^2}(t) &=& 
  \frac{m_c^2 \langle \bar{q}q \rangle 
  \langle \bar{s}s \rangle}{64 \pi^2} \:v \bigg[
  4 + \frac{m_s m_c \tau}{x} \bigg]
\end{eqnarray}
\subsection{The tetraquark current $A_{cd}$}
\begin{eqnarray}
  \rho^{pert}(t) &=& 
  \frac{m_c^8(1+b^2)}{5\cdot 3^2 \cdot 2^{13} \pi^6} 
  \bigg[ v \bigg( 840x + 140 + \frac{5248}{x}
  - \frac{1164}{x^2} - \frac{182}{x^3} + \frac{5}{x^4} 
  \bigg) + 120 {\cal L}_v \bigg( 14x^2 + 15 - 18\log(x)\nnb \\&&\nnb
  - \frac{32}{x} + \frac{9}{x^2} \bigg) 
  - 4320 {\cal L}_{+} \bigg]
  ~-~ \frac{m_s m_c^7(1-b^2)}{5 \cdot 3\cdot 2^{12} \pi^6} \bigg[ 
  v \bigg( 420x + 70 + \frac{1574}{x} - \frac{257}{x^2} 
  - \frac{16}{x^3} \bigg) \\&& 
  +~ 60{\cal L}_v \bigg( 14x^2 + 4
  - 12\log(x) - \frac{16}{x} + \frac{3}{x^2} \bigg) 
  - 1440{\cal L}_{+} \bigg]\nnb\\
  \eea
  \bea
  \rho^{\langle \bar{q}q \rangle}(t) &=&
  \frac{m_c^5 (1-b^2) \langle \bar{q}q \rangle}
  {3\cdot2^{6} \pi^4} 
  \bigg[ v \bigg( 6 - \frac{5}{x} - \frac{1}{x^2} \bigg) 
  + 6 {\cal L}_v \bigg( 2x - 2 + \frac{1}{x} \bigg) \bigg] 
  ~+~\frac{m_c^5 (1-b^2) \langle \bar{s}s \rangle}
  {3^2 \cdot 2^{8} \pi^4} \bigg[
  v \bigg( 60x + 10 - \frac{34}{x} - \frac{9}{x^2} \bigg)  
\nnb  \\&&
  +~ 24 {\cal L}_v \bigg( 5x^2 - 3 + \frac{2}{x} \bigg) \bigg] 
  ~-~ \frac{ m_s m_c^4 (1+b^2) \langle \bar{q}q \rangle}
  {2^{5} \pi^4} \bigg[ v \bigg( 2 + \frac{1}{x} \bigg) 
  - 4 {\cal L}_v (1 - x) \bigg] \nnb\\&&
  +~ \frac{ m_s m_c^4 (1+b^2) \langle \bar{s}s \rangle}
  {3 \cdot 2^{9} \pi^4}
  \bigg[ v \bigg( 12x + 2 - \frac{26}{x} + \frac{3}{x^2} \bigg) 
  + 24 {\cal L}_v ( x^2 + 1 ) \bigg] \nnb 
  \\ &&\nnb \\
  \rho^{\langle G^2 \rangle}(t) &=& 
  \frac{m_c^4 (1+b^2) \langle G^2 \rangle}
  {3^2 \cdot 2^{11} \pi^6}
  \bigg[ v \bigg( 6 - \frac{5}{x} - \frac{1}{x^2} \bigg) + 
  6 {\cal L}_v \bigg( 2x - 2 + \frac{1}{x} \bigg) \bigg] \nnb
 \\ &&\nnb \\
  \rho^{\langle \bar{q}Gq \rangle}(t) &=& 
  \frac{m_c^3 (1-b^2) \langle \bar{q}G q \rangle}{2^{7}\pi^4}
  \bigg[ \frac{v}{x} - 2{\cal L}_v \bigg]
  ~-~ \frac{m_c^3 (1-b^2) \langle \bar{s}G s \rangle}
  {3^2 \cdot 2^{8}\pi^4}
  \bigg[ v\Big(42x + 7 - \frac{40}{x} \Big) 
  + 6 {\cal L}_v \Big( 14x^2 + 3 + \frac{2}{x} \Big) \bigg] \nnb
  \\&&
  +~ \frac{m_s m_c^2 (1+b^2) \langle \bar{q}G q \rangle}
  {2^{6}\pi^4} \:v ~-~ \frac{m_s m_c^2 (1+b^2) 
  \langle \bar{s}G s \rangle}{3^2 \cdot 2^{9} \pi^4} 
  \bigg[ v \bigg( 18x - 5 - \frac{4}{x} \bigg) 
  + 18{\cal L}_v ( 2x^2 + 1) \bigg]\nnb
 \\ &&\nnb \\
  \rho^{\langle \bar{q}q \rangle^2}(t) &=& 
  \frac{ \langle \bar{q}q \rangle 
  \langle \bar{s}s \rangle}{48 \pi^2} \:v \bigg[
  4m_c^2 (1+b^2) - m_s m_c (1-b^2) 
  \Big( 1 + \frac{m_c^2 \tau}{x} \Big) \bigg]
\end{eqnarray}
 \end{widetext}


\end{document}